\documentclass[11pt, a4paper]{article}
\usepackage[dvipsnames]{xcolor}
\usepackage[utf8]{inputenc}
\usepackage{url}
\usepackage{amsmath}
\usepackage{amsfonts}
\usepackage{amssymb} 
\usepackage{graphicx}
\usepackage{array}
\usepackage{tikz}
\usepackage{tikz-cd}
\usetikzlibrary{arrows.meta,decorations.pathmorphing,positioning,calc}

\usepackage[pdftitle={ModifiedAbelianGaugeTheories},
  pdfauthor={Markus Dierigl, Ruben Minasian, Du\v{s}an Novi\v{c}i\'c},
  pdfsubject={generalized symmetries, anomalies, Swampland},
  bookmarksopen, bookmarksnumbered, bookmarksopenlevel=2, colorlinks=false, linkcolor=blue, citecolor=blue, urlcolor=blue]{hyperref}
  
\usepackage{caption}

\numberwithin{equation}{section}

\usepackage[nosort]{cite}

\begin{document}

\baselineskip=18pt

\vspace*{-2cm}
\begin{flushright}
  \texttt{CERN-TH-2026-027}
\end{flushright}

%% title, authors, affiliation
\vspace*{0.6cm} 
\begin{center}
{\Large{\textbf{Modified Abelian Gauge Theories}}} \\
 \vspace*{1.5cm}
Markus Dierigl$^1$, Ruben Minasian$^2$, Du\v{s}an Novi\v{c}i\'c$^3$\\

{
 \vspace*{1.0cm} 
{\it 
${}^1$ Theoretical Physics Department, CERN, 1211 Geneva 23, Switzerland\\
${}^2$ Institut de Physique Th\'eorique, Universit\'e Paris-Saclay, \\
CNRS, CEA, F-91191, Gif-sur-Yvette, France\\
${}^3$ Max-Planck-Institut f\"ur Physik, Werner-Heisenberg-Institut, \\ 85748 Garching bei M\"unchen, Germany
}}

\vspace*{0.8cm}
\end{center}
\vspace*{.5cm}

\noindent The topological properties of field configurations in gauge theory contain important data about the (generalized) global symmetries of the theory as well as potential inconsistencies in the form of gauge anomalies. In this work we modify the topological classes of Abelian $p$-form fields, generating new global variants of gauge theories. These modifications implement constraints directly on the classifying space of the gauge field and its cohomology classes via homotopy fiber construction. This general approach allows us to investigate the universal effects of the constraints on the conserved global charges encoded in gauge characteristic classes. We further demonstrate that this procedure generically leads to new topological sectors introducing additional global charges and anomalies in the modified gauge theories.

\vspace*{\fill}
\noindent
markus.dierigl@cern.ch, ruben.minasian@ipht.fr, novicic@mpp.mpg.de

\thispagestyle{empty}
\clearpage

\setcounter{page}{1}

\newpage
\tableofcontents

\newpage

\section{Introduction}

Nature's fundamental interactions are very effectively captured by quantum field theory. This involves the appearance of particles that transmit the forces which are the excitations of the underlying gauge field. These gauge theories are further subject to redundancies associated to gauge symmetries, whose breaking would render the theory inconsistent. This remains to be true after the inclusion of quantum corrections and demands the absence of so-called anomalies, that classify the properties of the partition function under gauge transformations. Moreover, the partition function splits into separate sectors which correspond to topologically distinct gauge field configurations. Since it is the topological features of the fields that specify the individual sectors, they are invariant under continuous deformations and can be thought of as defining conserved charges. The goal of this work is to modify the topological properties of gauge theories and study how this affects the topological charges as well as anomalies of the underlying theory.

While 1-form gauge fields can describe Abelian or non-Abelian gauge equivalence, their higher-form variants are generically Abelian. These Abelian $p$-form fields frequently appear in higher-dimensional supergravity theories as well as string theory. Even in the Abelian case the theories are far from trivial and present many interesting phenomena. Many of these are described by the analysis of their generalized (categorical) symmetries \cite{Gaiotto:2014kfa} (see \cite{Schafer-Nameki:2023jdn, Brennan:2023mmt, Bhardwaj:2023kri} for reviews) which have advanced our understanding tremendously within the last years.

One class of generalized symmetries, typically called magnetic (higher-form) symmetries, originates from the fact that the field strength is closed. In case of a $p$-form gauge field the field strength is described by a $(p+1)$-form $F^{(p+1)}$ and defines a conserved current $j \sim \ast F^{(p+1)}$
\begin{equation}
    d\ast j \sim d \ast \big( \ast F^{(p+1)}\big) = \pm d F^{(p+1)} = 0 \,.
\end{equation}
Since the same is true for a power (wedge product) of the field strength this leads to a whole set of (composite) conserved currents, e.g., for $p$ odd one can define the $k$-fold product current, 
\begin{equation}
   d \ast \big( \ast (F^{(p+1)} \wedge \dots \wedge F^{(p+1)}) \big) = 0 \,.
\end{equation}
Each of these currents is associated to a generalized global symmetry, sometimes referred to as Chern-Weil symmetry, \cite{Heidenreich:2020pkc} (see also \cite{Garcia-Valdecasas:2024cqn}). The corresponding conserved charges are given by the integral of $\ast j$ over a sub-manifold of appropriate dimension, which in the case above can be written as
\begin{equation}
    Q \sim\int_{\Sigma_{kp+k}} \big( F^{(p+1)} \wedge \dots \wedge F^{(p+1)} \big) \,.
\end{equation}
This formula suggests a close connection between these generalized higher-form symmetries, their conserved charges, and the characteristic classes of the underlying gauge theory. This follows simply from the fact, that the characteristic classes on a spacetime manifold $X$ are given by elements of $H^{\bullet}(X;\mathbb{Z})$ pulled back from the associated classifying space (see, e.g., \cite{milnor_stasheff_1974, hatcher2002}).\footnote{We work in Euclidean signature.} Since these classes are closed by definition they describe conserved quantities which include the U$(1)$ Chern-Weil symmetries above, but as cohomology classes with integer coefficients they can also include discrete $\mathbb{Z}_k$ torsion charges. Thus, the characteristic classes of gauge theories are an important piece of information for a full understanding of the gauge theory and its (generalized) symmetries. This is further corroborated by the fact that typically the values of these characteristic classes split the gauge theory in topologically distinct sectors which need to be summed over in the full partition function.

This raises the question of whether there can be theories that differ in these symmetry sectors. Such {\it modified Abelian gauge theories} and their construction are the subject of this work. In particular we construct theories where some characteristic classes, i.e., elements of $H^{\bullet}(X;\mathbb{Z})$, are restricted to satisfy certain constraints. This procedure can be understood as a modification of the underlying global higher-form symmetry and leads to (global) variants of Abelian $p$-form theories.

The technique that we use in order to implement such constraints is insensitive to the choice of spacetime manifold $X$ because it directly modifies the classifying space. This is because characteristic classes on $X$, come from (the pull-back of) cohomology classes of the classifying space itself. The concrete implementation uses what is known as a {\it homotopy fiber construction}, see, e.g., \cite{hatcher2002}, which can be understood as gluing additional cells into the classifying space trivializing the wanted (co)homology classes. However, these constructions tend to also introduce additional characteristic classes, which come from the implementation of the constraint. Thus, we find that while we can get rid of certain conserved global charges this often comes at the price of introducing new ones. Interestingly, the homotopy fiber construction also provides a physical interpretation for this since the modified classifying space can be understood as the total space of a fibration, with the original classifying space appearing as the base space. While the fibration can trivialize cohomology classes of the base, the fiber in general introduces new ones, which are precisely the additional classes we encounter. 

While this type of modification might seem exotic, we know several examples where similar constraints are implemented. Most prominently, this appears in the coupling of non-Abelian gauge dynamics to gravity and is also manifested in the Green-Schwarz anomaly cancellation mechanism \cite{Green:1984sg}, where the presence and transformation properties of a $2$-form field trivializes otherwise conserved currents by making them exact, and hence trivial in cohomology. An explicit example would be
\begin{equation}
    d H^{(3)} \sim F^{(2)} \wedge F^{(2)} \,,
\label{eq:Bianchi}
\end{equation}
where $H^{(3)}$ is the (gauge invariant) field strength of a 2-form field. Such an equation is also called Bianchi identity. Since, here, we are working with cohomology classes with integer coefficients we are also sensitive to the proportionality factor in \eqref{eq:Bianchi} as well as torsion classes. In string theory and supergravity examples one further can include characteristic classes of the tangent bundle accounting for a gravitational contribution as is the case for the original Green-Schwarz mechanism in heterotic string theory. Here, we focus on the field theory side without the appearance of the curvature 2-form in the constraints, and we analyze how a modification of the gauge theory affects its characteristic classes and anomalies.

Our main topic is the modification of U$(1)$ $p$-form gauge theories, whose classifying space $B^p \mathrm{U}(1)$ is given by the Eilenberg-MacLane space $K(\mathbb{Z},p+1)$, which is defined to have a single non-trivial homotopy group $\pi_{m}\big( K(A,m) \big) = A$ in positive degree. An analogous approach could be used for discrete $\mathbb{Z}_n$ $p$-form gauge theories with classifying space $K(\mathbb{Z}_n,p)$. Implementing a constraint on a characteristic class in $H^k(X;\mathbb{Z})$ describes the modified classifying space as a fibration of $K(A,k-1)$ over $B^p \mathrm{U}(1)$ (where $A$ is an Abelian group). This suggests that the constraints are enforced by the mixing with higher-form fields with gauge group $A$, very similar to what happenes in the Bianchi identity \eqref{eq:Bianchi} above and leads to a so-called higher-group symmetry (see, e.g., \cite{Kapustin:2013uxa, Cordova:2018cvg, Benini:2018reh}). In this sense the homotopy fiber approach is very physical, but also general in that it does not need the specification of the number of spacetime dimensions, spacetime manifold $X$, or the explicit formulation at the level of a Lagrangian. The resulting theory will have a different set of conserved charges and therefore a different set of (generalized) symmetries.

These modifications have further consequences for the possible gauge anomalies. For perturbative anomalies this can be be seen rather straightforwardly from the fact that the anomaly polynomial, which encodes the non-trivial phase of the partition function under small gauge transformations, is built out of characteristic classes of the gauge theory. Thus, once one restricts the characteristic classes, also the gauge anomalies are modified. Including diffeomorphisms, one can further include the characteristic classes of the tangent bundle and finds additional constraints capturing mixed gravitational anomalies which might also be modified. However, the newly introduced characteristic classes might lead to additional anomalies, which were absent before and demand extra consistency conditions. For non-perturbative anomalies (including diffeomorphisms) the analysis becomes more tricky and potential anomalies are classified by non-trivial bordism classes, \cite{Dai:1994kq, Freed:2004yc, Freed:2014iua, Witten:2015aba, Yonekura:2016wuc, Garcia-Etxebarria:2018ajm, Witten:2019bou}, for which the gauge theory characteristic classes are important input data. That a modification of a gauge theory should have an influence on the anomalies of the underlying theory should not come as a surprise since this is exactly the purpose of the Bianchi identity in the Green-Schwarz mechanism, whose more global versions have been studied in for example \cite{Witten:1999eg, Garcia-Etxebarria:2017crf, Monnier:2018nfs, Tachikawa:2021mvw, Tachikawa:2021mby, Debray:2021vob, Yonekura:2022reu, Lee:2022spd, Dierigl:2022zll, Basile:2023knk, Torres:2024sbl, Dierigl:2025rfn}. The application of the homotopy fiber construction for fermion anomaly cancellation was further discussed in \cite{Saito:2025idl}, from which we adopt many of the relevant techniques.

Since the mathematics involved can be quite heavy (see \cite{mccleary2001, hatcherspectral, beaudry2018guidecomputingstablehomotopy, Garcia-Etxebarria:2018ajm}), we try to illustrate the general principles in various explicit examples. In particular, we will be interested in $D$-dimensional Maxwell theory, i.e., with U$(1)$ 1-form field, whose first non-trivial characteristic class, the first Chern class, $c_1 \in H^2(X;\mathbb{Z})$, remains unmodified, but whose composite class $c_1^2$ is subject to some constraints. For the constraints we primarily focus on a `mod $n$' constraint with $c_1^2$~mod~$n \in H^4(X;\mathbb{Z}_n)$ being trivial; and a `times $n$' constraint with $n c_1^2 \in H^4(X;\mathbb{Z})$ being trivial. The effect on the local dynamics of these theories depends on the way they are implemented at the level of differential cohomology and goes beyond the topological analysis of the homotopy fiber in the main part of this work. We comment on this issue in Appendix~\ref{app:homBianchi}. Nevertheless, even on the topological level we find interesting new characteristic classes and modifications of the gauge anomalies of these global modifications of Maxwell theory.

The rest of this work is organized as follows: In Section~\ref{sec:modgauge} we discuss the topological sectors of gauge theories with focus on U$(1)$ $p$-form fields, and their relation to characteristic classes. We will also introduce the two restrictions; `times $n$' and `mod $n$', which we will implement throughout this work. In Section~\ref{sec:homfiber} we explain the realization of the constraints using the homotopy fiber construction and demonstrate the evaluation of the characteristic classes of the new theory $H^{\bullet}(Q;\mathbb{Z})$ in various interesting examples. We further explore what happens in case one wants to implement more than one constraint. In Section~\ref{sec:bordism} we further include gravitational information for some of the discussed examples and explore the oriented and Spin bordism groups of the modified gauge theories which classify the potential anomalies of the modified theory. We point out various applications and generalizations of the techniques described here in Section~\ref{sec:concl}. In Appendix~\ref{app:homBianchi}, we connect Bianchi identities and the homotopy fiber construction and further discuss different ways to implement the constraints beyond the topological level. We provide the cohomology classes of Eilenberg-MacLane space of the form $K(\mathbb{Z},m)$ and $K(\mathbb{Z}_2,m)$ in Appendix~\ref{app:EMachom} and determine some bordism groups we use in the main text in Appendix~\ref{app:bordism}.

\section{Topological sectors of Abelian gauge theories}
\label{sec:modgauge}

In this section we describe the topological sectors of Abelian $p$-form gauge fields. In particular, we describe how the different gauge configurations are distinguished by their realization of characteristic classes. These characteristic classes can be understood, by pulling back the non-trivial cohomology classes from the classifying space $BG$ to the spacetime manifold $X$. We then describe two possible modifications of the topological sectors, which can be understood as a modification of the magnetic, sometimes also called solitonic, (higher-form) symmetries. While we focus on Abelian, especially U$(1)$ gauge fields, an analogous approach also works in other cases.

\subsection{Abelian gauge fields and differential cohomology}
\label{subsec:difcohom}

A convenient way to describe U$(1)$ $p$-form fields uses differential cohomology, see,  e.g., \cite{Cheeger1985DifferentialCA, Hopkins:2002rd} and \cite{Hsieh:2020jpj, Moore:2025tmt} for reviews. A U$(1)$ $p$-form field is fully described by a differential character, i.e., by an element of $\check{H}^{p+1}(X)$. This differential character contains information about both the topological features of the configurations as well as the continuous information encoded for example in the connection or field strength. 

For our purpose it will be useful to consider one of the short exact sequences describing $\check{H}^{p+1}(X)$, given by
\begin{equation}
    0 \longrightarrow \frac{\Omega^p(X)}{\Omega^p_{\mathbb{Z}}(X)}\longrightarrow \check{H}^{p+1}(X) \longrightarrow H^{p+1} (X;\mathbb{Z})\longrightarrow 0 \,.
    \label{eq:partfuncsplit}
\end{equation}
Here, $\Omega^p(X)$ ($\Omega^p_{\mathbb{Z}}(X)$) denote differential $p$-forms (with integer periods). The first entry, can be understood as the topologically trivial fields, where the U$(1)$ gauge field is defined as a global differential form $\Omega^p(X)$, up to (large) gauge transformations $\Omega^p_{\mathbb{Z}}(X)$. The last entry, $H^{p+1}(X;\mathbb{Z})$, captures the topological class $[N^{(p+1)}]$ of the gauge configuration in integer cohomology, whose representative co-cycle is given by $N^{(p+1)}$. For a U$(1)$ $p$-form field, this decomposition of a differential character $\check{A} \in \check{H}^{(p+1)}(X)$ also motivates a split of the path integral into a discrete sum over topological sectors, labeled by elements in $H^{p+1}(X;\mathbb{Z})$, and a continuous integral over topologically trivial differential characters $\check{a} \in \check{H}^{p+1}(X)$ which we write as:
\begin{equation}
    \int D\check{A} \rightarrow \sum_i \int D\check{a} \,.
\end{equation}
In the following we will be interested in modifying the discrete sum by restricting the allowed topological classes of the gauge theory. However, we will not only modify the topological class in $H^{p+1}(X;\mathbb{Z})$, but also more general characteristic classes in higher degree, e.g., powers of the topological class $(N^{(p+1)})^k$. In the case of a U$(1)$ 1-form field described by an element in $\check{H}^2(X)$ we will also use the common notation 
\begin{equation}
    [N^{(2)}] = c_1 \in H^2(X;\mathbb{Z}) \,,
\end{equation}
in terms of the first Chern class of the bundle.

The fate of the continuous integral over $\check{a}$ above, goes beyond the topological data of the classifying space and depends on the particular implementation of the modification, as we argue in Appendix~\ref{app:homBianchi}. In the following we concentrate on the topological features that modify the discrete sum.

\subsection{Characteristic classes and classifying spaces}
\label{subsec:classspace}

Every topological class of a gauge configuration with gauge group $G$ on the spacetime manifold $X$ can be identified with a homotopy class of maps from $X$ into the classifying space $BG$ (see, e.g., \cite{milnor_stasheff_1974, HatcherKtheory}),
\begin{equation}
    f: X \rightarrow BG \,.
\end{equation}
For that reason $f$ is called the classifying map. Note however, that there is more physical data in the gauge configuration, e.g., the choice of connection, which requires information beyond the topological class. Denoting the homotopy class of a map from $X$ to $BG$, by $[X,BG]$ each separate class defines a topologically distinct gauge configuration. Since the topological sectors of the gauge theory are distinguished by characteristic classes, these must be part of the information of the classifying space $BG$.

\begin{figure}
    \centering
    \includegraphics[width = \textwidth]{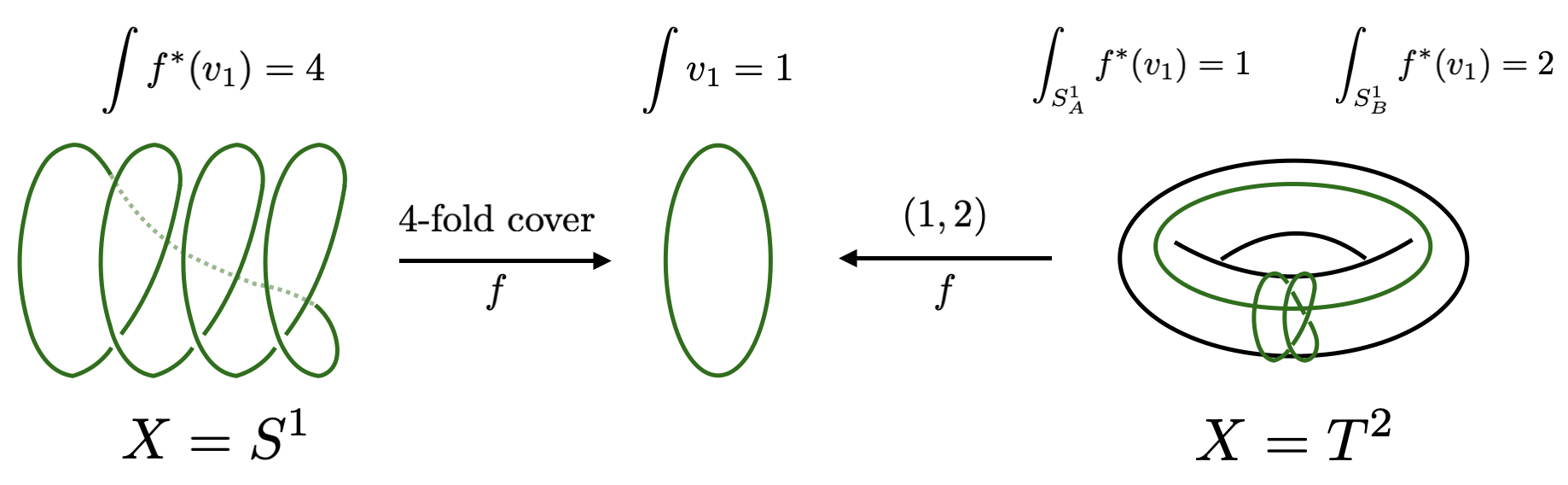}
    \caption{Illustration of classifying maps $f$ from $X = S^1$ and $X = T^2$ to the classifying space of a periodic scalar field, $K(\mathbb{Z},1) \simeq S^1$.}
    \label{fig:classmap}
\end{figure}

Indeed, the possible characteristic classes of the gauge theory can be identified with non-trivial elements of the cohomology of the classifying space, sometimes also referred to as group cohomology, $H^{\bullet}(BG;\mathbb{Z})$. The characteristic classes of a particular gauge background on $X$, specified by the homotopy class of its classifying map $f$, are then obtained by pulling back, $f^{\ast}$, the cohomology classes $H^{\bullet}(BG;\mathbb{Z})$ to spacetime. Since these are integer-valued classes, they remain constant under continuous deformations of the map $f$, exactly as we would expect since only the homotopy class $[X,BG]$ enters. Let us discuss this in some simple examples: 

One can understand a periodic scalar field $\varphi \sim \varphi + 2\pi$ as a 0-form U$(1)$ gauge field (the `large gauge transformations' make the scalar periodic). The associated classifying space is given by $S^1$, the circle, which is also an Eilenberg-MacLane space, $K(\mathbb{Z},1)$ (see \cite{hatcher2002}). The topological classes of this periodic scalar field are then classified by homotopy classes of maps from $X$ to $S^1$. Since in positive degree the only non-trivial cohomology group of the circle is given by $H^1(S^1;\mathbb{Z}) = \mathbb{Z}$, the topological classes of periodic scalars are fully classified by specifying the pull-back of this class under the classifying map $f$, i.e. specifying the corresponding element in $H^1(X;\mathbb{Z})$. In the case that spacetime $X$ is also given by a circle, we know that homotopy classes of maps from $X$ to $BG \simeq S^1$ are fully specified by an integer $n$, which describes how often $X$ winds around $BG$, see Figure~\ref{fig:classmap}. This defines the value of $\int_X N^{(1)}_{\varphi}$, i.e., the winding of the periodic scalar around the spacetime circle.\footnote{We avoid the use of $d \varphi$ since $\varphi$ is not globally well-defined due to large gauge transformations.}
Denoting the generator of the cohomology group by $v_1 \in H^1(S^1;\mathbb{Z})$, one has
\begin{equation}
    \int_{X} f^{\ast} (v_1) = \int_{f_{\ast} X} v_1 = \int_{nS^1} v_1 = n \int_{S^1} v_1 =  n \,.
\end{equation}
Similar, considerations can be done for higher-dimensional $X$, where one can determine the winding number of $\varphi$ around the non-trivial 1-cycles of $X$ in the same way. For the torus, $T^2$, for instance, the homotopy classes of maps $T^2 \rightarrow S^1$ are specified by two integers, which characterize the winding numbers of the periodic scalar around a basis of 1-cycles (with representatives $S^1_A$ and $S^1_B$), Figure~\ref{fig:classmap}.

Next, let us move to U$(1)$ 1-form gauge fields, i.e., Maxwell theory. In this case the classifying space is infinite-dimensional and given by the infinite-dimensional complex projective space
\begin{equation}
    B\mathrm{U}(1) \simeq \mathbb{CP}^{\infty} \simeq K(\mathbb{Z},2) \,,
\end{equation}
which again can be identified with an Eilenberg-MacLane space. While this seems to present a significant complication when compared to the periodic scalar field the logic proceeds analogously. The different topological sectors are described by homotopy classes $[X,BG]$, and the characteristic classes on $X$ can be obtained via pull-back. For example, the first Chern class $c_1 \in H^2(X;\mathbb{Z})$ is obtained by pulling back the generator of $H^2(\mathbb{CP}^{\infty};\mathbb{Z}) = \mathbb{Z}$ which we denote by $v_2$. For $X = S^2 \simeq \mathbb{CP}^1$, there is again the analogue of a `winding number' which here characterizes the magnetic flux through $S^2$, since homotopy classes of maps $[S^2; \mathbb{CP}^{\infty}]$ are described by $H^2(S^2;\mathbb{Z}) = \mathbb{Z}$, which can be identified as
\begin{equation}
    \int_{S^2} c_1 = \int_{S^2} f^{\ast}(v_2) \,.
\end{equation}
However, as opposed to the circle above, $\mathbb{CP}^{\infty}$ has more non-trivial cohomology classes, which correspond to powers of the generator $v_2$, since $H^{\bullet}(\mathbb{CP}^{\infty};\mathbb{Z})=\mathbb{Z}[v_2]$. For a four-dimensional spacetime manifold $X$, and using naturality of the cup product, one can define 
\begin{equation}
f^{\ast} (v_2 \cup v_2) = f^{\ast}(v_2) \cup f^{\ast}(v_2) = c_1 \cup c_1 \in H^4(X;\mathbb{Z}) \,,
\end{equation}
a characteristic class of higher degree.

The classifying space of a U$(1)$ $p$-form field is given by
\begin{equation}
    B^p \mathrm{U}(1) \simeq K(\mathbb{Z},p+1) \,,
\end{equation}
which again can be identified with an Eilenberg-MacLane space. We see that the topological classes of the higher-form gauge field can be identified with homotopy classes $[X,K(\mathbb{Z},p+1)]$ from spacetime into the classifying space, which are isomorphic to $H^{p+1}(X;\mathbb{Z})$. This is exactly as expected from our discussion of differential cohomology in Section~\ref{subsec:difcohom}, where the topological class was determined by $[N^{(p+1)}] \in H^{p+1}(X;\mathbb{Z})$. More general characteristic classes can then be obtained by pulling back cohomology classes of the Eilenberg-MacLane spaces $K(\mathbb{Z},p+1)$ of higher degree to $X$.

Our goal is to restrict the allowed values of characteristic classes on $X$, for which we need to modify the classifying spaces $BG$.

\subsection{Modifications of topological sectors}\label{subsec:modofopsec}

In this work we explore how to describe a modification of the sum over topological sectors of Abelian U$(1)$ $p$-form gauge theories. For U$(1)$ global symmetries associated to characteristic classes valued in $\mathbb{Z}$, we consider the following two modifications
\begin{itemize}
    \item{The `{\it times $n$}' constraint: $n [N^{(p+1)}]^k = 0 \in H^{kp+k}(X;\mathbb{Z})$:  The case $n = 1$ enforces the characteristic class to be topologically trivial. It can be understood as gauging the full associated U$(1)$ symmetry. For $n \neq 1$, this still presents a gauging of the full associated U$(1)$ global symmetry\footnote{This can also be understood at the level of symmetry generator (see for example \cite{Gaiotto:2014kfa}) $\text{exp}\big( \tfrac{i \alpha}{2 \pi} \int [N^{(p+1)}]^k \big) = \text{exp}\big( \tfrac{i \alpha}{2 \pi n} \int n [N^{(p+1)}]^k \big)$ which is trivial for U$(1)$ parameter $\alpha$.}, but there is a discrete remnant in terms of potential torsion classes in $H^{kp+k}(X;\mathbb{Z})$. These can be understood as the topological classes of a $(kp+k-1)$-form $\mathbb{Z}_n$ gauge field, since these correspond to the image of $H^{kp+k-1}(X;\mathbb{Z}_n)$ in $H^{kp+k}(X;\mathbb{Z})$, via the Bockstein map $\beta_{\mathbb{Z}}$ explained below.}
    \item{The `{\it mod $n$}' constraint: $\mathfrak{m}_n \big( [N^{(p+1)}]^k \big) = 0 \in H^{kp+k} (X;\mathbb{Z}_n)$: This constraint implements that the `mod $n$' reduction $\mathfrak{m}_n$ of a certain characteristic class vanishes and can be regarded as the gauging of a $\mathbb{Z}_n$ subgroup of the associated U$(1)$ global symmetry.}
\end{itemize}
That indeed we are dealing with a non-trivial gauging becomes clear in later sections, using spectral sequences techniques, where the trivialized characteristic classes become the target of a differentials making them exact. For continuous U$(1)$ symmetries this precisely corresponds to the definition of gauging in \cite{Heidenreich:2020pkc}, in which the associated charge density becomes exact and does not lead to a conserved charge when integrated over closed sub-manifolds.

The two constraints above are closely related as can be seen by considering the long exact sequence in cohomology associated to the short exact sequence
\begin{equation}
    0 \longrightarrow \mathbb{Z} \overset{\cdot n}\longrightarrow \mathbb{Z} \overset{\mathfrak{m}_n}{\longrightarrow} \mathbb{Z}_n \longrightarrow 0 \,,
    \label{eq:modnSES}
\end{equation}
given by
\begin{equation}
    ... \longrightarrow H^{r-1} (X;\mathbb{Z}_n)\overset{\beta_{\mathbb{Z}}}{\longrightarrow} H^{r} (X;\mathbb{Z}) \overset{\cdot n}{\longrightarrow} H^r(X;\mathbb{Z}) \overset{\mathfrak{m}_n}{\longrightarrow} H^r(X;\mathbb{Z}_n) \longrightarrow H^{r+1} (X; \mathbb{Z}) \longrightarrow ... \,.
    \label{eq:lesbockz}
\end{equation}
Here, $\beta_{\mathbb{Z}}$ denotes the Bockstein homomorphism associated to \eqref{eq:modnSES} (see \cite{hatcher2002}). 

The `times $n$' constraint demands that the modified characteristic class is in the kernel of $\cdot n$, which by exactness means that it is in the image of $\beta_{\mathbb{Z}}$, i.e., there is a class $x_n \in H^{kp+k-1} (X;\mathbb{Z}_n)$ such that
\begin{equation}
    \beta_{\mathbb{Z}} ( x_n ) = [N^{(p+1)}]^k \,. 
\end{equation}
The `mod $n$' constraint on the other hand means that the modified characteristic class is in the kernel of $\mathfrak{m}_n$ and hence by exactness in the image of $\cdot n$, i.e., there exists an $x \in H^{kp+k} (X;\mathbb{Z})$ such that
\begin{equation}
    nx = [N^{(p+1)}]^k \,.
\end{equation}
Note that these two properties are not mutually exclusive, as we will also see in Section~\ref{subsec:multconst}.

Let us now describe how we can implement these constraints very efficiently in a model-independent way.

\section{The homotopy fiber construction}
\label{sec:homfiber}

In this section we implement the modification of the Abelian gauge theories discussed above, using a homotopy fiber construction, see \cite{hatcher2002} [Section 4.3] as well as its applications in \cite{Saito:2025idl}. This procedure does not require the explicit implementation at the level of the Lagrangian, whose connection to Bianchi identities we discuss in Appendix~\ref{app:homBianchi}, but rather proceeds via a modification of the classifying space of the underlying gauge theory.

\subsection{The homotopy fiber}

Since our goal is to restrict the `allowed' characteristic classes on $X$, we can achieve that by modifying the classifying space of the underlying gauge theory. In particular, we want a new space $Q$, which has a map $p$ into the original $BG$, but has restricted homotopy classes of maps $[X,Q]$ as compared to $[X,BG]$. To be more precise, we want that a map $[X,BG]$ only lifts to a map $[X,Q]$, if the restriction on the characteristic classes is satisfied. Such a space can be constructed as the fiber in a fibration of the form
\begin{equation}
    Q \overset{p}\longrightarrow BG \overset{\alpha}{\longrightarrow} Y \,,
\label{eq:fiberseq}
\end{equation}
where the space $Y$ and the map $\alpha$ are chosen to implement the constraint. This construction guarantees that a map $f$ from $X$ to $BG$ lifts to a map $f_Q$ from $X$ to $Q$ if and only if
\begin{equation}
    \alpha \circ f = 0 \,,
\end{equation}
meaning that the map is in the trivial class $[X,Y]$. The homotopy fiber $Q$ constructed in such a way is unique (up to homotopy).

Next, let us consider how to choose $Y$ and $\alpha$ for the type of constraints defined above. For that we note that we aim for a restriction of characteristic classes on $X$, i.e., elements of $H^m (X;A)$. Thus, to implement such a constraint we ideally would like $Y$ to be sensitive to that, but only that information. Indeed, these spaces are known and given precisely by the Eilenberg-MacLane spaces $K(A,m)$.\footnote{One can alternatively choose the base space to be $Y=E_m$ where $E_m$ is a classifying space for the generalized cohomology $E^m(X)$. This would correspond to a modification of the so called solitonic symmetries corresponding to the generalized cohomology theories, see \cite{Chen:2023czk}.} To be more specific, the homotopy classes of maps from $X$ to $K(A,m)$ are in one-to-one correspondence with cohomology classes $H^m(X;A)$
\begin{equation}
    [X;K(A,m)] \simeq H^m(X;A) \,.
\end{equation}
In some sense Eilenberg-MacLane spaces can be understood as very simply spectra, that define cohomology theories in exactly the way described above. They can also be understood as the minimal choice, since the homotopy classes of maps exclusively depend on this particular cohomology element.

Having identified the spaces $K(A,m)$ as natural choice for $Y$, we now turn to the maps $\alpha$. Due to the properties just discussed, we know that homotopy classes of $\alpha$ are fully determined by
\begin{equation}
    [BG,K(A,m)] \simeq H^m(BG;A) \,.
\end{equation}
Pulling back these elements to $X$, we choose $\alpha$ in such a way that it corresponds to the characteristic class that should be trivialized. With $BG$, $Y$, and $\alpha$ fixed, the space $Q$ naturally appears as fiber in the sequence \eqref{eq:fiberseq} above. Let us again try to illustrate this in several simple examples.
\begin{figure}
    \centering
    \includegraphics[width = 0.75 \textwidth]{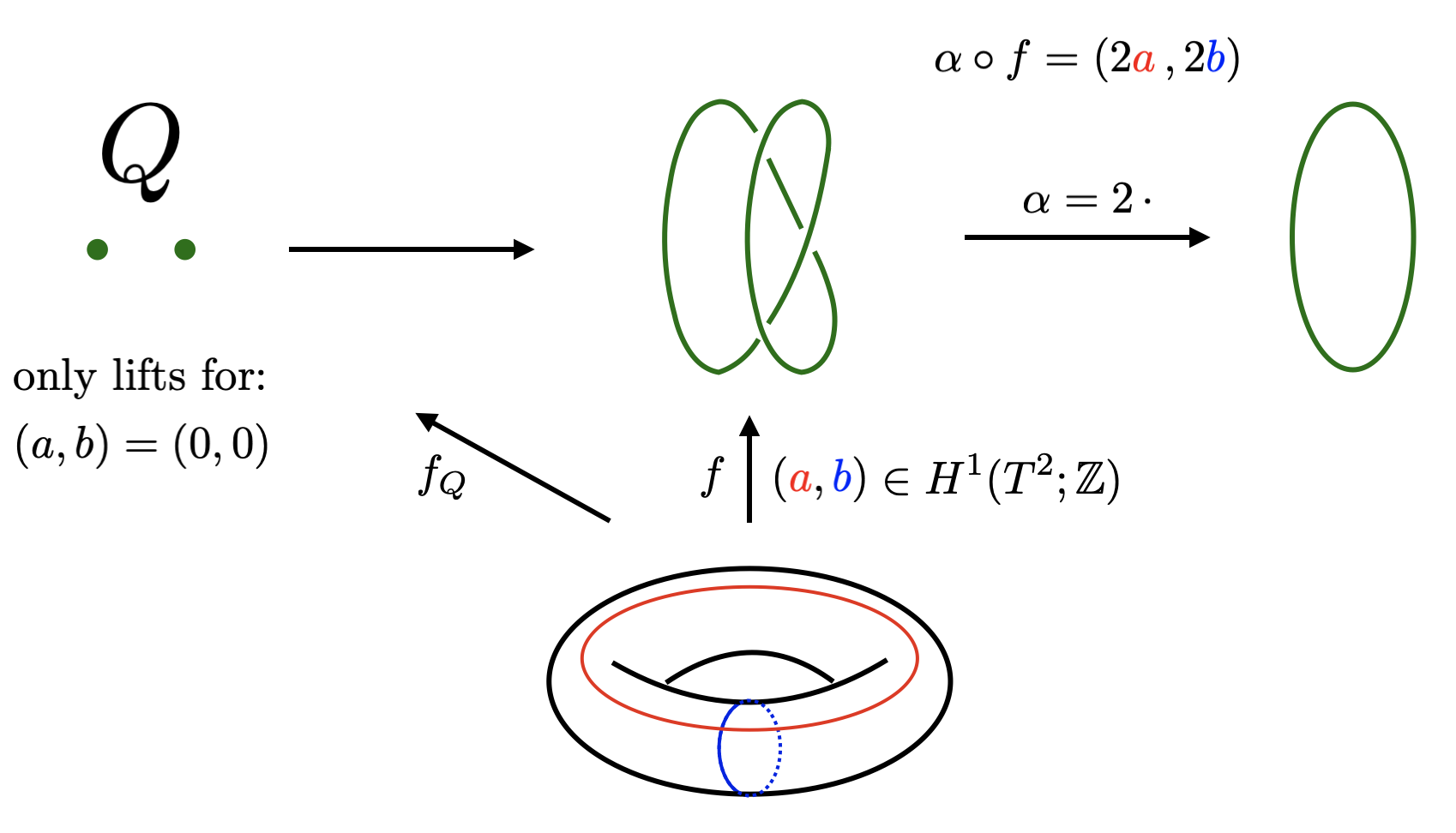}
    \caption{Illustration of the homotopy fiber construction of the `times $n$' constraint for a periodic scalar field (for $X = T^2$ and $\alpha$ a double cover). Here, $f^{\ast}(v_1) = (\textcolor{red}{a}, \textcolor{blue}{b}) \in H^1(T^2;\mathbb{Z})$ denotes the pull-back of the fundamental class of the circle, $v_1$, with respect to $f$. }
    \label{fig:homfiber_timesn}
\end{figure}

\vspace{0.5cm}

\noindent Let us once more consider the periodic scalar field $\varphi$ with associated classifying space $S^1$. Since the only cohomology class in positive dimension is given by $H^1(S^1;\mathbb{Z}) = \mathbb{Z}$, the only spaces $Y$ we want to consider are of the form $K(A,1)$. And to be more precise we choose $A$ to be a free $\mathbb{Z}$ or cyclic $\mathbb{Z}_n$ group, leading to
\begin{equation}
    K(\mathbb{Z},1) \simeq S^1 \,, \quad K(\mathbb{Z}_n,1) \simeq L^{\infty}_n \simeq B\mathbb{Z}_n \,.
\end{equation}
The first choice is appropriate to implement the `times $n$' constraint, for which we choose $\alpha$ in the class
\begin{equation}
    n v_1 \in H^1(S^1;\mathbb{Z}) \simeq [S^1,S^1] \,.
\end{equation}
Thus, $\alpha$ is an $n$-fold covering. The fiber is thus given by $\mathbb{Z}_n$, which is 0-dimensional. From that we learn that, for any $X$, implementing the constraint $n f^{\ast}(v_1) = 0 \in H^1(X;\mathbb{Z})$ trivializes all topological sectors, see Figure~\ref{fig:homfiber_timesn}.

\begin{figure}
    \centering
    \includegraphics[width = 0.75 \textwidth]{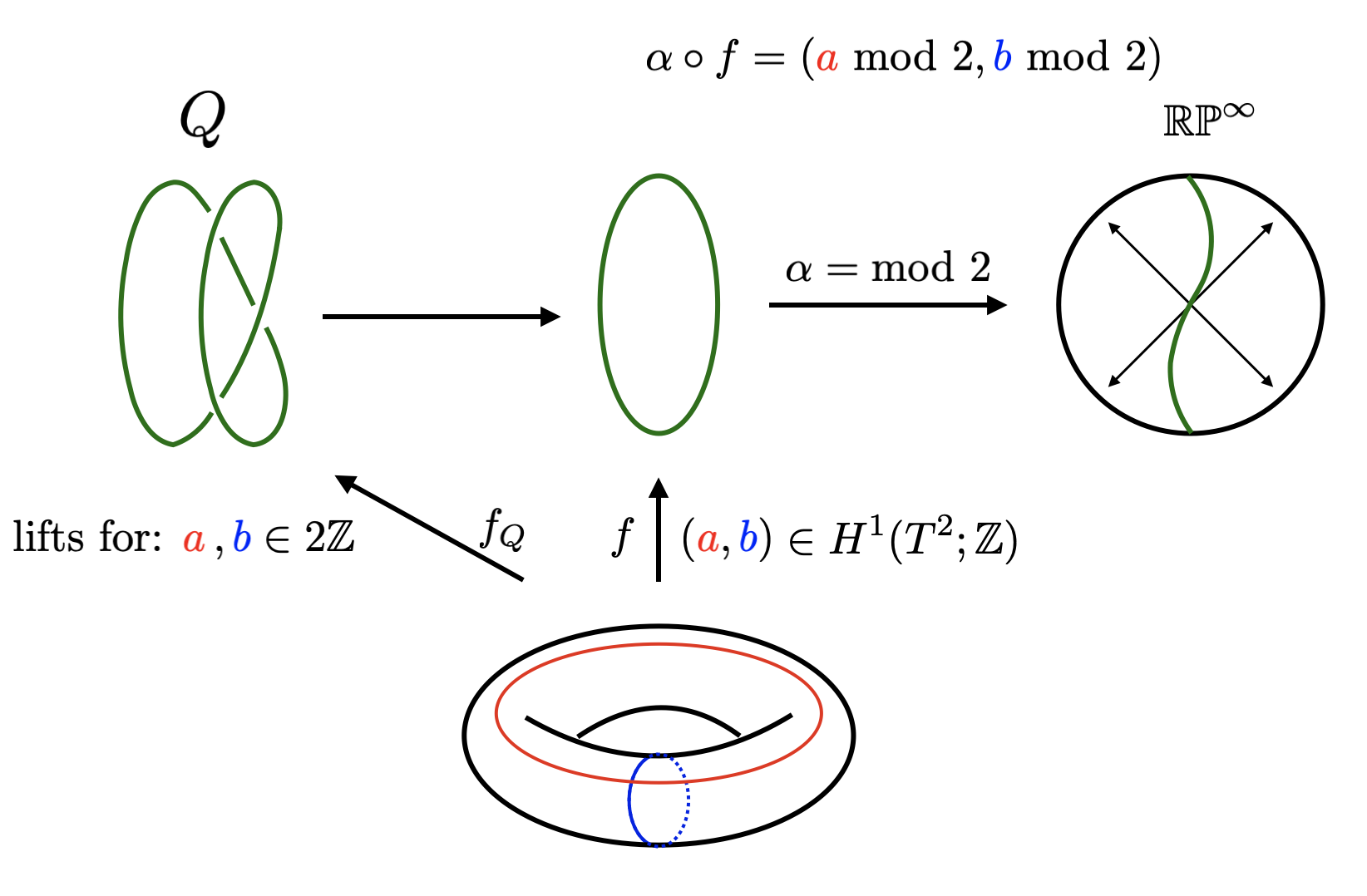}
    \caption{Illustration of the homotopy fiber construction of the `mod $n$' constraint for a periodic scalar field (for $X = T^2$ and $\alpha$ the non-trivial element in $H^1(S^1;\mathbb{Z}_2)$, i.e., $n=2$). }
    \label{fig:homfiber_modn}
\end{figure}

The second choice of $Y$ allows us to implement the `mod $n$' constraint, in particular we have that $\alpha$ is described by a class in
\begin{equation}
    [S^1, B\mathbb{Z}_n] \simeq H^1(S^1;\mathbb{Z}_n) \simeq \mathbb{Z}_n \,.
\end{equation}
Choosing it to be given by $e^{2 \pi i/n} \in \mathbb{Z}_n$, will successfully restrict $f^{\ast}(v_1) \in H^1(X;\mathbb{Z})$ to vanish in $H^1(X;\mathbb{Z}_n)$.\footnote{One could also choose other elements $e^{2 \pi i k/n}$ with $k$ co-prime to $n$.} Here, the homotopy fiber $Q$ can be interpreted as an $n$-fold covering of $S^1$, since those map to the trivial class in $[S^1,B\mathbb{Z}_n]$. From this construction we find that $f^{\ast}(v_1) \in H^1(X;\mathbb{Z})$ needs to vanish mod $n$, in order to lift to $Q$, exactly as intended, see Figure~\ref{fig:homfiber_modn}.

\vspace{0.5cm}

\noindent Below we will study more interesting examples where we modify the characteristic classes of U$(1)$ $p$-form fields, but before that let us describe $Q$ in a way that will be more convenient for us. The sequence \eqref{eq:fiberseq} can be extended to the left, leading to a description of $Q$ itself given as the total space of a fibration 
\begin{equation}
    Z \longrightarrow Q \overset{p}\longrightarrow BG \,.
\end{equation}
The fiber $Z$ is homotopically equivalent to $\Omega Y$ the loop space of $Y$. Luckily, the Eilenberg-MacLane spaces have a nice behavior under $\Omega$ and one has
\begin{equation}
    \Omega K(A,m) \simeq K(A,m-1) \,.
\end{equation}
With $Q$ given as a total space of a fibration and the knowledge of the cohomology of the fiber and base, we can use spectral sequences techniques in order to extract the cohomology classes
\begin{equation}
    H^{\bullet} (Q;\mathbb{Z}) \,,
\end{equation}
which correspond to the new set of characteristic classes of the modified gauge theory (after pull-back to $X$ via $f^{\ast}_Q$).\footnote{An explicit way to construct $Q$ is to consider a path fibration of $K(A,m)$ i.e. $K(A,m-1) \rightarrow \mathcal{P}K(A, m) \rightarrow K(A,m)$ where $\mathcal{P}K(A,m)$ is the space of paths in $K(A,m)$ starting at the based point and which is homotopically trivial. $Q$ can be constructed as a pull-back of this fibration with the map $\alpha$, i.e. $Q \simeq \alpha^*\mathcal{P}K(A, m)$ and one automatically obtains the fibration structure \cite{MosherTangora1968}.} In the following we will find that while the restrictions are successfully implemented, there can be new characteristic classes, i.e., new conserved global charges, originating from the $K(A,m-1)$ fiber.

\begin{figure}
\centering
\includegraphics[width = 0.8 \textwidth]{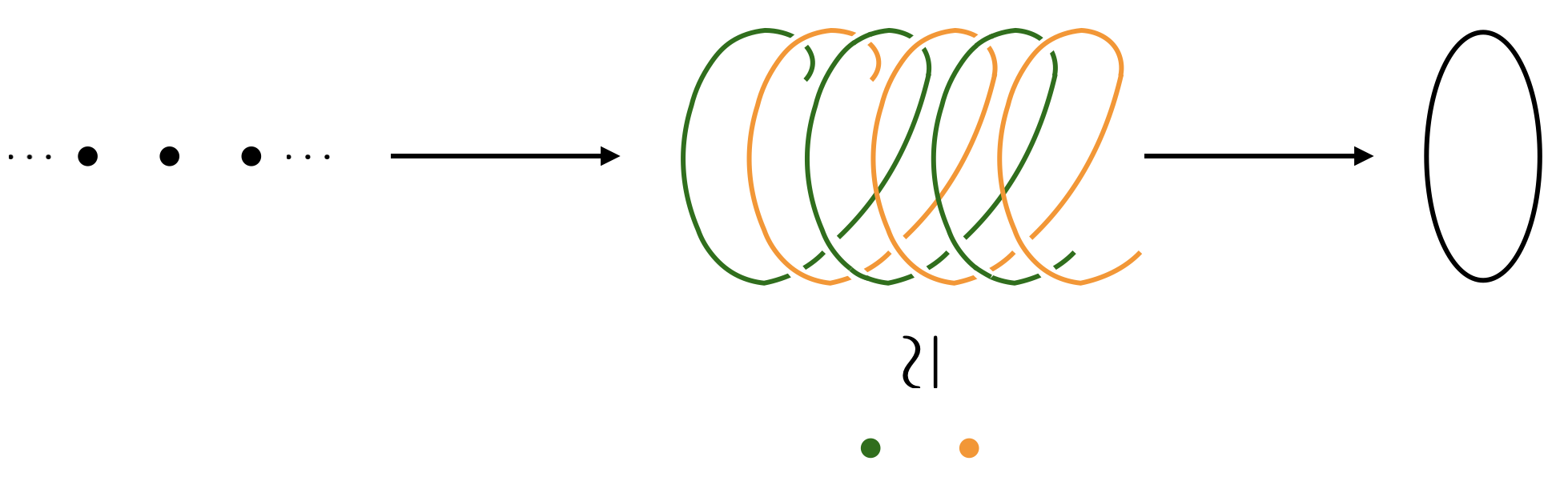}
\caption{Illustration of $Q$ in `times $n$' constraint as $\mathbb{Z}$ fibered over $S^1$}
\label{fig:fibration_timesn}
\end{figure}

Even though the low degrees make the examples above somewhat special, let us try to apply this strategy. One has
\begin{equation}
    \Omega K(\mathbb{Z},1) = K(\mathbb{Z},0) \simeq \mathbb{Z} \,,
\end{equation}
This means that we obtain $\mathbb{Z}_n$ as a fibration of $\mathbb{Z}$ over $S^1$. This can be made more explicit by noting that $\text{pt} \simeq \mathbb{R}$ and the fact that fibering $\mathbb{Z}$ over $S^1$ can turn it into a line. The fibration is defined such that going around the base circle relates points in $\mathbb{Z}$ which are $n$ steps apart. One therefore obtains $\mathbb{R}^{\oplus n} \simeq \text{pt}^{\oplus n} \simeq \mathbb{Z}_n$ as the total space, see Figure~\ref{fig:fibration_timesn}.
Similarly, for the `mod $n$' constraint one has
\begin{equation}
    \Omega K(\mathbb{Z}_n,1) = K (\mathbb{Z}_n, 0) \simeq \mathbb{Z}_n \,,
\end{equation}
and the fibration of $Q$, being the $n$-fold cover of the base circle has indeed a fiber given by $\mathbb{Z}_n$, reproducing the fibration structure depicted in Figure~\ref{fig:homfiber_timesn}. This relation of the two different fibration structures once more suggests the relation between the `times $n$' and `mod $n$' constraint.

Let us now move to the discussion of U$(1)$ 1-form gauge fields.

\subsection{Warm-up example: Modification of \texorpdfstring{$c_1$}{c1}}

As a warm-up example we consider the modification of the first Chern class in a U$(1)$ gauge theory as described in Section~\ref{subsec:modofopsec}.

\vspace{0.5cm}

\noindent For the `times $n$' constraint, we construct the homotopy fiber similar to the construction for the periodic scalar above\footnote{Even though $B\mathrm{U}(1) \simeq K(\mathbb{Z},2)$, we use $B\mathrm{U}(1)$ to distinguish between the base and the total space.}
\begin{equation}\label{eq:fibrationfornc_1}
    Q \longrightarrow B\mathrm{U}(1) \overset{\alpha}{\longrightarrow} K(\mathbb{Z},2) \,,
\end{equation}
where the map $\alpha$ is `multiplication by $n$' in the following sense. From the discussion in Section~\ref{subsec:classspace}, we know that the homotopy class of $\alpha$ is determined by
\begin{equation}
    [B\mathrm{U}(1),K(\mathbb{Z},2)] \simeq H^2 \big( B\mathrm{U}(1);\mathbb{Z}\big) \simeq \mathbb{Z} \,.
\end{equation}
Let us denote the generator of $H^2 \big( B\mathrm{U}(1);\mathbb{Z}\big)$ by $v_2$, then we can identify the different maps $\alpha$ with an element $n v_2$, which is obtained via pull-back
\begin{equation}
    \alpha^{\ast} (v_2) = n v_2 \in H^2\big(B\mathrm{U}(1);\mathbb{Z}\big) \,.
\end{equation}
Thus, to lift to a map into $Q$, one needs 
\begin{equation}
    f^{\ast}(n v_2) = n f^{\ast}(v_2) = n c_1 = 0 \in H^2(X;\mathbb{Z}) \,,
\end{equation}
to be trivial in cohomology, as desired.

Let us start by considering $n = 1$, in which case we can choose $\alpha$ to be the identity map. However, this means that $Q$ itself is a point, whose cohomology classes $H^{\bullet}(\mathrm{pt};\mathbb{Z})$ in positive degree vanish identically. Indeed, this means that the new gauge theory with classifying space $Q$ has no non-trivial characteristic classes. One nevertheless has a non-trivial theory since one needs to integrate over topologically trivial gauge fields, which in particular do not need to be flat ($F=0$).\footnote{In fact, in Euclidean space $\mathbb{R}^d$ one could not distinguish between the two versions of U$(1)$ gauge theory, since the cohomology of spacetime is trivial.} Taking loop spaces one can rephrase $Q$ as
\begin{equation}
    K(\mathbb{Z},1) \simeq S^1 \longrightarrow Q \longrightarrow B\mathrm{U}(1) \,,
    \label{eq:trivialexample}
\end{equation}
i.e., as a circle fibration over $B\mathrm{U}(1)$ and another way to interpret $Q$ is given by the universal bundle $E\mathrm{U}(1)$ which is contractible and therefore can be identified with a point. Since further  $S^1\simeq B\mathbb{Z}$ and $Q\simeq B\mathbb{R}$ one can think of this fibration as a fibration of classifying spaces corresponding to the exact sequence of groups
\begin{equation}
    0\rightarrow \mathbb{Z} \hookrightarrow \mathbb{R} \rightarrow \mathrm{U}(1) \rightarrow 0 \,,
\end{equation}
with an intriguing relation to gauge fields with gauge group $\mathbb{R}$, which recently featured in the description of symmetry topological field theories for continuous symmetries, see, e.g., \cite{Brennan:2024fgj, Antinucci:2024zjp}.

\begin{figure}
\centering
\begin{tikzpicture}[scale=0.65]
    % Axes
    \draw[->] (0,0) -- (6.2,0) node[right] {$p$};
    \draw[->] (0,0) -- (0,3.2) node[above] {$q$};
    % Grid (optional)
    \draw[step=1cm,gray!20] (0,0) grid (6,3);
    % Labels
    \foreach \x in {0,1,...,5}
        \node[below] at (\x+0.5, 0) {\small $\x$};
    \foreach \y in {0,1,2}
        \node[left] at (0,\y+0.5) {\small $\y$};
    \node(00) at (0.5, 0.5) {$\mathbb{Z}$};
\node(20) at (2.5, 0.5) {$\mathbb{Z}$};
\node(40) at (4.5, 0.5) {$\mathbb{Z}$};
\node(01) at (0.5, 1.5) {$\mathbb{Z}$};
\node(21) at (2.5, 1.5) {$\mathbb{Z}$};
\node(41) at (4.5, 1.5) {$\mathbb{Z}$};
\node(60) at (6.5, 0.5) {};
\draw[->] (01)-- node[above]{$\cdot n$} (20) ;
\draw[->] (21)-- node[above]{$\cdot n$} (40) ;
\draw (41)-- node[above]{$\cdot n$} (6, 0.7) ;
     % Axes
    \draw[->] (8,0) -- (14.2,0) node[right] {$p$};
    \draw[->] (8,0) -- (8,3.2) node[above] {$q$};
    % Grid (optional)
    \draw[step=1cm,gray!20] (8,0) grid (14,3);
    % Labels
    \foreach \x in {0,1,...,5}
        \node[below] at (\x+8.5, 0) {\small $\x$};
    \foreach \y in {0,1,2}
        \node[left] at (8,\y+0.5) {\small $\y$};
    \node at (8.5, 0.5) {$\mathbb{Z}$};
\node at (10.5, 0.5) {$\mathbb{Z}_n$};
\node at (12.5, 0.5) {$\mathbb{Z}_n$};
\end{tikzpicture}
\caption{$E^{p,q}_2$ and $E^{p,q}_{\infty}$ page of $Q$ implementing $n c_1 = 0$.}
\label{fig:E2nc1}
\end{figure}
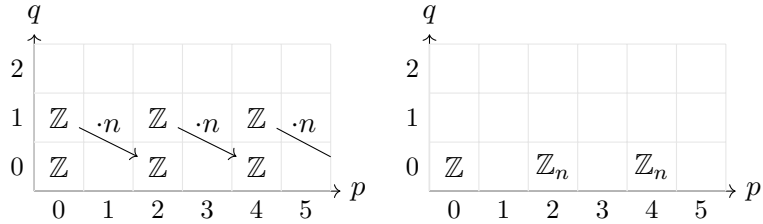

Now consider $n > 1$, extending to the left we again find
\begin{equation}\label{eq:nc1examplefibrat}
    K(\mathbb{Z},1) \simeq S^1 \longrightarrow Q \longrightarrow B \mathrm{U}(1) \,,
\end{equation}
but with different fibration structure as compared to \eqref{eq:trivialexample}.
With the cohomology of $S^1$ and $B\mathrm{U}(1)$ known, we can apply the Leray-Serre spectral sequence (LSSS), \cite{hatcherspectral, mccleary2001} and \cite{Saito:2025idl, Christensen:2025ktc} for recent applications, whose second page is given by
\begin{equation}
    E_2^{p,q} = H^p \big( B \mathrm{U}(1); H^q(S^1;\mathbb{Z})\big) \Rightarrow H^{p+q} (Q;\mathbb{Z}) \,,
\end{equation}
and summarized in Figure \ref{fig:E2nc1}.
We also draw the differentials which indicate the non-trivial fibration and trivialize $n c_1$. Since there can be no other differentials the $E^{p,q}_{\infty}$ page is fully determined and also depicted in Figure \ref{fig:E2nc1}. There are no extension problems and one finds the integer cohomology of $Q$ to be given by
\begin{equation}
    H^k (Q;\mathbb{Z}) = \begin{cases} \mathbb{Z} \,, \quad k = 0 \,, \\ \mathbb{Z}_n \,, \quad k>0 \,, \text{even} \,, \\ 0 \,, \quad \text{otherwise} \,. \end{cases}
\end{equation}
This precisely matches our expectations that requiring $n c_1$ to be trivial also restricts the higher topological classes $c_1^k$ to respect
\begin{equation}
    n c_1^k = (n c_1) c_1^{k-1} = 0 \in  H^{2k}(X;\mathbb{Z}) \,.
\end{equation}
In fact, we see that the cohomology groups of $Q$ coincide with that of $K(\mathbb{Z}_n,1) \simeq B \mathbb{Z}_n$ and the two theories have the same classes of topological sectors.\footnote{The fact that the cohomology groups agree is not sufficient to show that the two spaces are homotopy equivalent, but is necessary if they are. One way to see that these spaces agree follows from the fact that $B\mathbb{Z}_n$ is part of the fibration $B\mathbb{Z}_n\rightarrow BU(1)\rightarrow BU(1)$ induced by the exact sequence of groups $0\rightarrow\mathbb{Z}_n\rightarrow U(1)\rightarrow U(1)\rightarrow 0$, so it can be realized as a total space of the same fibration as in \ref{eq:nc1examplefibrat}. Since this fibration is fully classified by the Postnikov class $\alpha\in H^2(BU(1);\mathbb{Z})\cong\mathbb{Z}$, i.e. by a choice of an integer $m\in\mathbb{Z}$, we see from the computation above that obtaining the desired cohomology one needs to take $m=n$ meaning that $B\mathbb{Z}_n$ is the total space of the same fibration as our $Q$ so they must be equal up to homotopy.} 

The intuition behind this is as follows; the constraint $n c_1 = 0$, requires the image of $c_1$ in the free part of $H^2(X;\mathbb{Z})$ to be trivial. While this does not mean that the gauge configurations are flat, it means that each gauge configuration is continuously connected to a flat gauge field, which one can take as the representative of the topological sector. Thus, the topological data is fully specified by an element in $H^1\big(X;\mathrm{U}(1) \big)$. After specifying $n$ the relevant data reduces to the image of $H^1 (X;\mathbb{Z}_n)$ in $H^2(X;\mathbb{Z})$ under the Bockstein map $\beta_{\mathbb{Z}}$, see \eqref{eq:lesbockz}, precisely the topological class encoded by a $\mathbb{Z}_n$ gauge field.

\vspace{0.5cm}

\begin{figure}
\centering
\begin{tikzpicture}[scale=0.65]
    % Axes
    \draw[->] (0,0) -- (6.2,0) node[right] {$p$};
    \draw[->] (0,0) -- (0,6.2) node[above] {$q$};
    % Grid (optional)
    \draw[step=1cm,gray!20] (0,0) grid (6,6);
    % Labels
    \foreach \x in {0,1,...,5}
        \node[below] at (\x+0.5, 0) {\small $\x$};
    \foreach \y in {0,1,..., 5}
        \node[left] at (0,\y+0.5) {\small $\y$};
\node(00) at (0.5, 0.5) {$\mathbb{Z}$};
\node(20) at (2.5, 0.5) {$\mathbb{Z}$};
\node(40) at (4.5, 0.5) {$\mathbb{Z}$};
\node(02) at (0.5, 2.5) {$\mathbb{Z}_n$};
\node(22) at (2.5, 2.5) {$\mathbb{Z}_n$};
\node(42) at (4.5,2.5) {$\mathbb{Z}_n$};
\node(04) at (0.5, 4.5) {$\mathbb{Z}_n$};
\node(24) at (2.5, 4.5) {$\mathbb{Z}_n$};
\node(44) at (4.5, 4.5) {$\mathbb{Z}_n$};
\end{tikzpicture}
\caption{$E^{p,q}_2=E^{p,q}_{\infty}$ page of $Q$ implementing $\mathfrak{m}_n(c_1) = 0$.}
\label{fig:E2c1modn}
\end{figure}
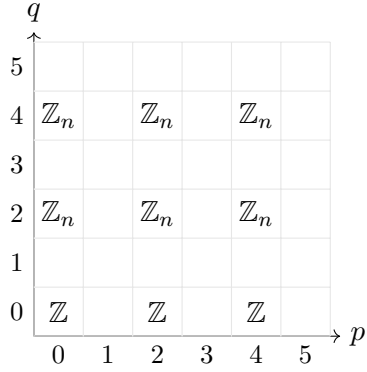

\noindent Let us move to the `mod $n$' constraint, i.e., the trivialization of $\mathfrak{m}_n (c_1) \in H^2(X;\mathbb{Z}_n)$ . Here, we use a homotopy fiber constructed as
\begin{equation}\label{eq:fibrationforc_1modn}
    Q \longrightarrow B\mathrm{U}(1) \longrightarrow K(\mathbb{Z}_n,2) \,.
\end{equation}
The second map is defined such that it pulls back the generator $u_2 \in H^2\big( K(\mathbb{Z}_n,2); \mathbb{Z} \big) \simeq \mathbb{Z}_n$ to the generator $\mathfrak{m}_n (v_2)$ in $H^2(B \mathrm{U}(1); \mathbb{Z}_n) \simeq \mathbb{Z}_n$. Extending to the left we find
\begin{equation}\label{eq:fibrBZn}
    K(\mathbb{Z}_n,1) \simeq B \mathbb{Z}_n \longrightarrow Q \longrightarrow B \mathrm{U}(1) \,,
\end{equation}
which suggests that the constraint involves the topological classes of a $\mathbb{Z}_n$ 1-form field, whose classifying space appears as the fiber. Again, we know the cohomology groups of both base and fiber and can apply the LSSS, whose second page coincides with the $\infty$-page and is provided in Figure \ref{fig:E2c1modn}.

\begin{figure}
\centering
\begin{tikzpicture}[scale=0.65]
    % Axes
    \draw[->] (0,0) -- (6.2,0) node[right] {$p$};
    \draw[->] (0,0) -- (0,6.2) node[above] {$q$};
    % Grid (optional)
    \draw[step=1cm,gray!20] (0,0) grid (6,6);
    % Labels
    \foreach \x in {0,1,...,5}
        \node[below] at (\x+0.5, 0) {\small $\x$};
    \foreach \y in {0,1,..., 5}
        \node[left] at (0,\y+0.5) {\small $\y$};
\node(00) at (0.5, 0.5) {$\mathbb{Z}_n$};
\node(20) at (2.5, 0.5) {$\mathbb{Z}_n$};
\node(40) at (4.5, 0.5) {$\mathbb{Z}_n$};
\node(01) at (0.5, 1.5) {$\mathbb{Z}_n$};
\node(21) at (2.5, 1.5) {$\mathbb{Z}_n$};
\node(41) at (4.5, 1.5) {$\mathbb{Z}_n$};
\node(02) at (0.5, 2.5) {$\mathbb{Z}_n$};
\node(22) at (2.5, 2.5) {$\mathbb{Z}_n$};
\node(42) at (4.5,2.5) {$\mathbb{Z}_n$};
\node(03) at (0.5, 3.5) {$\mathbb{Z}_n$};
\node(23) at (2.5, 3.5) {$\mathbb{Z}_n$};
\node(43) at (4.5, 3.5) {$\mathbb{Z}_n$};
\node(04) at (0.5, 4.5) {$\mathbb{Z}_n$};
\node(24) at (2.5, 4.5) {$\mathbb{Z}_n$};
\node(44) at (4.5, 4.5) {$\mathbb{Z}_n$};
\node(05) at (0.5, 5.5) {$\mathbb{Z}_n$};
\node(25) at (2.5, 5.5) {$\mathbb{Z}_n$};
\node(45) at (4.5, 5.5) {$\mathbb{Z}_n$};
\node(60) at (6.5, 0.5) {};
\node(62) at (6.5, 2.5) {};
\node(64) at (6.5, 4.5) {};
\draw[->](01)-- node{} (20);
\draw[->](21)-- node{} (40);
\draw(41)-- node{} (6,0.8);
\draw[->](03)-- node{} (22);
\draw[->](23)-- node{} (42);
\draw(43)-- node{} (6,2.8);
\draw[->](05)-- node{} (24);
\draw[->](25)-- node{} (44);
\draw(45)-- node{} (6,4.8);
\draw[->] (8,0) -- (14.2,0) node[right] {$p$};
\draw[->] (8,0) -- (8,6.2) node[above] {$q$};2
    % Grid (optional)
    \draw[step=1cm,gray!20] (8,0) grid (14,6);
    % Labels
    \foreach \x in {0,1,...,5}
        \node[below] at (\x+8.5, 0) {\small $\x$};
    \foreach \y in {0,1,..., 5}
        \node[left] at (0,\y+0.5) {\small $\y$};
\node(80) at (8.5, 0.5) {$\mathbb{Z}_n$};
\node(82) at (8.5, 2.5) {$\mathbb{Z}_n$};
\node(84) at (8.5, 4.5) {$\mathbb{Z}_n$};
\end{tikzpicture}
\caption{$E^{p,q}_2$ and $E^{p,q}_{\infty}$ of $Q$ implementing $\mathfrak{m}_n (c_1) = 0$ with $\mathbb{Z}_n$ coefficients.}
\label{fig:E2andEinfc1modn}
\end{figure}
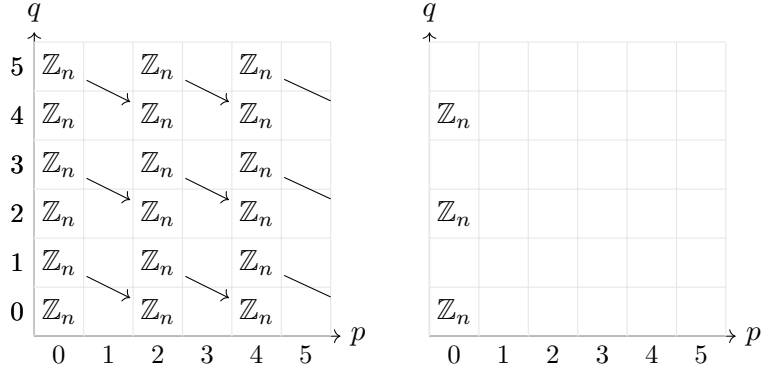

There can be no differentials and the only thing one needs to resolve are extension problems, since for example one has
\begin{equation}
    0 \longrightarrow \mathbb{Z} \longrightarrow H^2(Q;\mathbb{Z}) \longrightarrow \mathbb{Z}_n \longrightarrow 0 \,,
\end{equation}
and there are several options, including $\mathbb{Z}$ and $\mathbb{Z} \oplus \mathbb{Z}_d$, where $d$ divides $n$. To resolve this, we can run the same spectral sequence, but this time for cohomology with $\mathbb{Z}_n$ coefficients.
For that, consider the LSSS with the second page:
\begin{equation}
    E_2^{p,q}=H^p \big( B \mathrm{U}(1) ; H^q\big(K(\mathbb{Z}_n, 1); \mathbb{Z}_n \big) \big) \Rightarrow H^{p+q}(Q; \mathbb{Z}_n) \,,
\end{equation}
given by Figure \ref{fig:E2andEinfc1modn}.
Because we want $H^2(B\mathrm{U}(1);\mathbb{Z}_n)=E^{2,0}_{2} \twoheadrightarrow E_{\infty}^{2,0}\subset H^2(Q; \mathbb{Z}_n)$ to be zero we require the differential $d_2:E_2^{0,1}\rightarrow E_2^{2,0}$ to be an isomorphism. To determine other differentials we can use the ring structure of $H^\bullet \big(K(\mathbb{Z}_n,1); \mathbb{Z}_n\big)$ given by \ref{eq:BZnringstrodd} or \ref{eq:BZnringstreven} and the fact that differentials respect the ring structure. To be more specific, with $x$ and $y$ denoting the generators of $H^{\bullet}(B\mathbb{Z}_n;\mathbb{Z}_n)$ in degree 1 and 2, respectively, we require
\begin{equation}
d_2 x = \mathfrak{m}_n(v_2) \,,
\end{equation}
from which we see that $d_2(x v_2) = \mathfrak{m}_n (v_2^2)$, which means that $d_2 y = 0$. Thus, we obtained the full action of $d_2$, which enables us to determine the third page $E_3$, which in this case is the $\infty$-page, see Figure~\ref{fig:E2andEinfc1modn}. From there we read off the cohomology groups:
\begin{equation}
     H^k (Q; \mathbb{Z}_n) = \begin{cases} \mathbb{Z}_n \,, \quad k \geq 0 \,, \text{even} \,, \\ 0 \,, \quad \text{ otherwise} \,.\end{cases}
\end{equation}
Now using the universal coefficient theorem for cohomology, \cite{fomenko-fuchs-2016},
\begin{equation}
    H^{2k-1}(Q;\mathbb{Z}_n)\simeq H^{2k-1}(Q; \mathbb{Z})\otimes\mathbb{Z}_n\oplus\text{Tor}(H^{2k}(Q;\mathbb{Z});\mathbb{Z}_n)\simeq 0
    \label{eq:coeffcohom}
\end{equation}
we can solve the extension problem in Figure \ref{fig:E2c1modn} by realizing that if there were any finite part in $H^{2k}(Q; \mathbb{Z})$ the torsion functor in the universal coefficient theorem would be non-trivial, leading to a contradiction with the computation above. The conclusion is that one has the maximal non-trivial extension and the correct cohomology groups are
\begin{equation}
    H^k (Q; \mathbb{Z}) = \begin{cases} \mathbb{Z} \,, \quad k \geq 0 \,, \text{even} \,, \\ 0 \,, \quad \text{ otherwise} \,.\end{cases}
\end{equation}
Now, these seem to be identical to $H^\bullet\big( B \mathrm{U}(1); \mathbb{Z}\big)$ but the extensions lead to a different normalization. While pulled back from $B\mathrm{U}(1)$ the topological class $c_1^k$ takes arbitrary integer values they are divisible by $n^k$ when lifting through $Q$. Thus, the $\mathbb{Z}$-valued topological classes for $Q$ are schematically given by $\tfrac{1}{n^k} c_1^k$ making the non-trivial extensions apparent. The fibration \ref{eq:fibrBZn} can be interpreted as corresponding to the non-trivial extension of gauge groups:
\begin{equation}
    0\rightarrow \mathbb{Z}_n\hookrightarrow \mathrm{U}(1)\xrightarrow{\cdot  n} \mathrm{U}(1)\rightarrow 0 \,.
\end{equation}
This extended U$(1)$ allows for smaller electric charges of the form $q \in \tfrac{1}{n} \mathbb{Z}$. Dirac quantization, then requires the magnetic fluxes to be divisible by $n$, precisely as imposed by the constraint $\mathfrak{m}_n (c_1) =0$.

Now that we have some more familiarity with the necessary techniques we can go to more interesting examples, where we modify powers of $c_1$.

\subsection{Modification of \texorpdfstring{$c_1^2$}{square of c1}}

In this next set of examples we still remain within the realm of U$(1)$ gauge theories, but instead of restricting $c_1$, we restrict $c_1^2$, see also \cite{Fiorenza:2012ec, Fiorenza:2016oki} for similar constructions. In analogy to the treatment above, we focus on the two situations
\begin{equation}
\begin{split}
    \text{times }n:& \quad n c_1^2 = 0 \in H^4(X;\mathbb{Z}) \,, \\
    \text{mod }(n=2):& \quad \mathfrak{m}_2(c_1^2) = 0 \in H^4(X;\mathbb{Z}_2) \,.
\end{split}
\end{equation}
Here, we restrict the discussion of the mod $n$ restriction to the case $n =2$, since more is known about the associated Eilenberg-MacLane spaces, but a similar treatment would also be possible for different $n$.

\vspace{0.5cm}

\begin{figure}
\centering
\begin{tikzpicture}[scale=0.5]
    % Axes
    \draw[->] (0,0) -- (11.2,0) node[right] {$p$};
    \draw[->] (0,0) -- (0,11.2) node[above] {$q$};
    % Grid (optional)
    \draw[step=1cm,gray!20] (0,0) grid (11,11);
    % Labels
    \foreach \x in {0,1,...,10}
        \node[below] at (\x+0.5, 0) {\small $\x$};
    \foreach \y in {0,1,..., 10}
        \node[left] at (0,\y+0.5) {\small $\y$};
\node(00) at (0.5, 0.5) {$\mathbb{Z}$};
\node(03) at (0.5, 3.5) {$\mathbb{Z}$};
\node(06) at (0.5, 6.5) {$\mathbb{Z}_2$};
\node(08) at (0.5, 8.5) {$\mathbb{Z}_3$};
\node(09) at (0.5, 9.5) {$\mathbb{Z}_2$};
\node(010) at (0.5, 10.5) {$\mathbb{Z}_2$};
\node(20) at (2.5, 0.5) {$\mathbb{Z}$};
\node(23) at (2.5, 3.5) {$\mathbb{Z}$};
\node(26) at (2.5, 6.5) {$\mathbb{Z}_2$};
\node(28) at (2.5, 8.5) {$\mathbb{Z}_3$};
\node(29) at (2.5, 9.5) {$\mathbb{Z}_2$};
\node(40) at (4.5, 0.5) {$\mathbb{Z}$};
\node(43) at (4.5, 3.5) {$\mathbb{Z}$};
\node(46) at (4.5, 6.5) {$\mathbb{Z}_2$};
\node(60) at (6.5, 0.5) {$\mathbb{Z}$};
\node(63) at (6.5, 3.5) {$\mathbb{Z}$};
\node(66) at (6.5, 6.5) {\phantom{$\mathbb{Z}_2$}};
\node(80) at (8.5, 0.5) {$\mathbb{Z}$};
\node(83) at (8.5, 3.5) {$\mathbb{Z}$};
\node(100) at (10.5, 0.5) {$\mathbb{Z}$};
\draw[->] (03) -- node[above]{$n$} (40);
\draw[->] (23) -- node[above]{$n$} (60);
\draw[->] (43) -- node[above]{$n$} (80);
\draw[->] (63) -- node[above]{$n$}(100);
\draw[->] (09) -- (46);
\draw (29) -- (5,7.65);
\draw (83) -- (11,1.65);
%
%
    % Axes
    \draw[->] (13,0) -- (24.2,0) node[right] {$p$};
    \draw[->] (13,0) -- (13,11.2) node[above] {$q$};
    % Grid (optional)
    \draw[step=1cm,gray!20] (13,0) grid (24,11);
    % Labels
    \foreach \x in {0,1,...,10}
        \node[below] at (\x+13.5, 0) {\small $\x$};
    \foreach \y in {0,1,..., 10}
        \node[left] at (13,\y+0.5) {\small $\y$};
\node at (13.5, 0.5) {$\mathbb{Z}$};
\node at (13.5, 6.5) {$\mathbb{Z}_2$};
\node at (13.5, 8.5) {$\mathbb{Z}_3$};
\node at (13.5, 9.5) {$\xi$};
\node at (17.5, 6.5) {$\xi$};
\node at (13.5, 10.5) {$\mathbb{Z}_2$};
\node at (15.5, 0.5) {$\mathbb{Z}$};
\node at (15.5, 6.5) {$\mathbb{Z}_2$};
\node at (17.5, 0.5) {$\mathbb{Z}_n$};
\node at (19.5, 0.5) {$\mathbb{Z}_n$};
\node at (21.5, 0.5) {$\mathbb{Z}_n$};
\node at (23.5, 0.5) {$\mathbb{Z}_n$};
\node at (15.5, 8.5) {$\mathbb{Z}_3$};
%
%
    % Axes
    \draw[->] (0,-13) -- (11.2,-13) node[right] {$p$};
    \draw[->] (0,-13) -- (0,-1.8) node[above] {$q$};
    % Grid (optional)
    \draw[step=1cm,gray!20] (0,-13) grid (11,-2);
    % Labels
    \foreach \x in {0,1,...,10}
        \node[below] at (\x+0.5, -13) {\small $\x$};
    \foreach \y in {0,1,..., 10}
        \node[left] at (0,\y+0.5-13) {\small $\y$};
\node at (0.5, -2.5) {$\mathbb{Z}_2$};
\node at (0.5, -12.5) {$\mathbb{Z}_2$};
\node(a) at (0.5, -9.5) {$\mathbb{Z}_2$};
\node at (0.5, -7.5) {$\mathbb{Z}_2$};
\node at (0.5, -6.5) {$\mathbb{Z}_2$};
\node at (0.5, -4.5) {$\mathbb{Z}_2$};
\node(h) at (0.7, -3.5) {$\mathbb{Z}_2^{\oplus 2}$};
\node at (2.5, -12.5) {$\mathbb{Z}_2$};
\node(c) at (2.5, -9.5) {$\mathbb{Z}_2$};
\node at (2.5, -7.5) {$\mathbb{Z}_2$};
\node at (2.5, -6.5) {$\mathbb{Z}_2$};
\node at (2.5, -4.5) {$\mathbb{Z}_2$};
\node at (2.7, -3.5) {$\mathbb{Z}_2^{\oplus 2}$};
\node(b) at (4.5, -12.5) {$\mathbb{Z}_2$};
\node(e) at (4.5, -9.5) {$\mathbb{Z}_2$};
\node at (4.5, -7.5) {$\mathbb{Z}_2$};
\node(i) at (4.5, -6.5) {$\mathbb{Z}_2$};
\node(d) at (6.5, -12.5) {$\mathbb{Z}_2$};
\node(g) at (6.5, -9.5) {$\mathbb{Z}_2$};
\node at (6.5, -7.5) {$\mathbb{Z}_2$};
\node(f) at (8.5, -12.5) {$\mathbb{Z}_2$};
\node at (8.5, -9.5) {$\mathbb{Z}_2$};
\node at (10.5, -12.5) {$\mathbb{Z}_2$};
\draw[->] (a) --node[above]{$n$} (b);
\draw[->] (c) --node[above]{$n$} (d);
\draw[->] (e) --node[above]{$n$} (f);
\draw[->] (g) --node[above]{$n$} (9.8,-12);
\draw[->] (1.2,-4) -- (i);
\draw (3.2,-4) -- (5,-5.35);
\draw[->] (1.2,-5) -- (3.8,-7);
\draw[->] (3.2,-5) -- (5.8,-7);
\draw (9.2, -10) -- (11, -11.35);
%
%
    % Axes
    \draw[->] (13,-13) -- (24.2,-13) node[right] {$p$};
    \draw[->] (13,-13) -- (13,-1.8) node[above] {$q$};
    % Grid (optional)
    \draw[step=1cm,gray!20] (13,-13) grid (24,-2);
    % Labels
    \foreach \x in {0,1,...,10}
        \node[below] at (\x+13.5, -13) {\small $\x$};
    \foreach \y in {0,1,..., 10}
        \node[left] at (13,\y+0.5-13) {\small $\y$};
\node at (13.5, -12.5) {$\mathbb{Z}_2$};
\node at (13.5, -9.5) {$\xi$};
\node at (13.5, -7.5) {$\mathbb{Z}_2$};
\node at (13.5, -6.5) {$\mathbb{Z}_2$};
\node at (13.5, -4.5) {$\xi$};
\node at (13.5, -3.5) {$\kappa'$};
\node at (15.5, -3.5) {$\kappa'$};
\node at (15.5, -12.5) {$\mathbb{Z}_2$};
\node at (15.5, -9.5) {$\xi$};
\node at (15.5, -7.5) {$\mathbb{Z}_2$};
\node at (15.5, -6.5) {$\mathbb{Z}_2$};
\node at (15.5, -4.5) {$\xi$};
\node at (17.5, -12.5) {$\xi$};
\node at (17.5, -9.5) {$\xi$};
\node at (17.5, -7.5) {$\xi$};
\node at (17.5, -6.5) {$\xi$};
\node at (19.5, -12.5) {$\xi$};
\node at (19.5, -9.5) {$\xi$};
\node at (21.5, -12.5) {$\xi$};
\node at (13.5, -2.5) {$\mathbb{Z}_2$};
\node at (23.5, -12.5) {$\xi$};
%
%
 % Axes
    \draw[->] (0,-26) -- (11.2,-26) node[right] {$p$};
    \draw[->] (0,-26) -- (0,-14.8) node[above] {$q$};
    % Grid (optional)
    \draw[step=1cm,gray!20] (0,-26) grid (11,-15);
    % Labels
    \foreach \x in {0,1,...,10}
        \node[below] at (\x+0.5, -26) {\small $\x$};
    \foreach \y in {0,1,..., 10}
        \node[left] at (0,\y+0.5-26) {\small $\y$};
\node  at (0.5, -25.5) {$\mathbb{Z}_3$};
\node at (2.5, -25.5) {$\mathbb{Z}_3$};
\node (a1) at (4.5, -25.5) {$\mathbb{Z}_3$};
\node (a2) at (6.5, -25.5) {$\mathbb{Z}_3$};
\node (a3) at (8.5, -25.5) {$\mathbb{Z}_3$};
\node (a4) at (10.5, -25.5) {$\mathbb{Z}_3$};
\node (b1) at (0.5, -22.5) {$\mathbb{Z}_3$};
\node (b2) at (2.5, -22.5) {$\mathbb{Z}_3$};
\node (b3) at (4.5, -22.5) {$\mathbb{Z}_3$};
\node (b4) at (6.5, -22.5) {$\mathbb{Z}_3$};
\node  at (8.5, -22.5) {$\mathbb{Z}_3$};
\node  at (0.5, -18.5) {$\mathbb{Z}_3$};
\node at (2.5, -18.5) {$\mathbb{Z}_3$};
\node (c1) at (4.5, -18.5) {$\mathbb{Z}_3$};
\node  at (0.5, -17.5) {$\mathbb{Z}_3$};
\node at (2.5, -17.5) {$\mathbb{Z}_3$};
\node (d1) at (0.5, -15.5) {$\mathbb{Z}_3$};
\draw[->] (b1) --node[above]{$n$} (a1);
\draw[->] (b2) --node[above]{$n$} (a2);
\draw[->] (b3) --node[above]{$n$} (a3);
\draw[->] (b4) --node[above]{$n$} (a4);
\draw[->] (d1) --node[above]{$n$} (c1);
\draw (9.2, -23) -- (11, -24.35);
%
%
    % Axes
    \draw[->] (13,-26) -- (24.2,-26) node[right] {$p$};
    \draw[->] (13,-26) -- (13,-14.8) node[above] {$q$};
    % Grid (optional)
    \draw[step=1cm,gray!20] (13,-26) grid (24,-15);
    % Labels
    \foreach \x in {0,1,...,10}
        \node[below] at (\x+13.5, -26) {\small $\x$};
    \foreach \y in {0,1,..., 10}
        \node[left] at (13,\y+0.5-26) {\small $\y$};
\node  at (13.5, -25.5) {$\mathbb{Z}_3$};
\node at (15.5, -25.5) {$\mathbb{Z}_3$};
\node  at (17.5, -25.5) {$\zeta$};
\node  at (19.5, -25.5) {$\zeta$};
\node  at (21.5, -25.5) {$\zeta$};
\node  at (23.5, -25.5) {$\zeta$};
\node  at (13.5, -22.5) {$\zeta$};
\node at (15.5, -22.5) {$\zeta$};
\node  at (17.5, -22.5) {$\zeta$};
\node  at (19.5, -22.5) {$\zeta$};
\node  at (13.5, -18.5) {$\mathbb{Z}_3$};
\node at (15.5, -18.5) {$\mathbb{Z}_3$};
\node  at (13.5, -17.5) {$\mathbb{Z}_3$};
\node at (15.5, -17.5) {$\mathbb{Z}_3$};
\node  at (13.5, -15.5) {$\zeta$};
\node at (17.5, -18.5) {$\zeta$};
\end{tikzpicture}
\caption{Relevant entries for $E^{p,q}_2$ and $E^{p,q}_{\infty}$ of $Q$ implementing $n c_1^2 = 0$ (only entries with $p+q \leq 10$ are indicated). Upper diagrams are for $\mathbb{Z}$ coefficients, middle ones for $\mathbb{Z}_2$ coefficients (where we denote $\mathbb{Z}_{(n,2)}$ by $\xi$ and $\mathbb{Z}_{(n,2)} \oplus \mathbb{Z}$ by $\kappa'$) and lower ones for $\mathbb{Z}_3$ coefficients (with $\mathbb{Z}_{(n,3)}$ labeled by $\zeta$).}
\label{fig:E2c1sqn}
\end{figure}
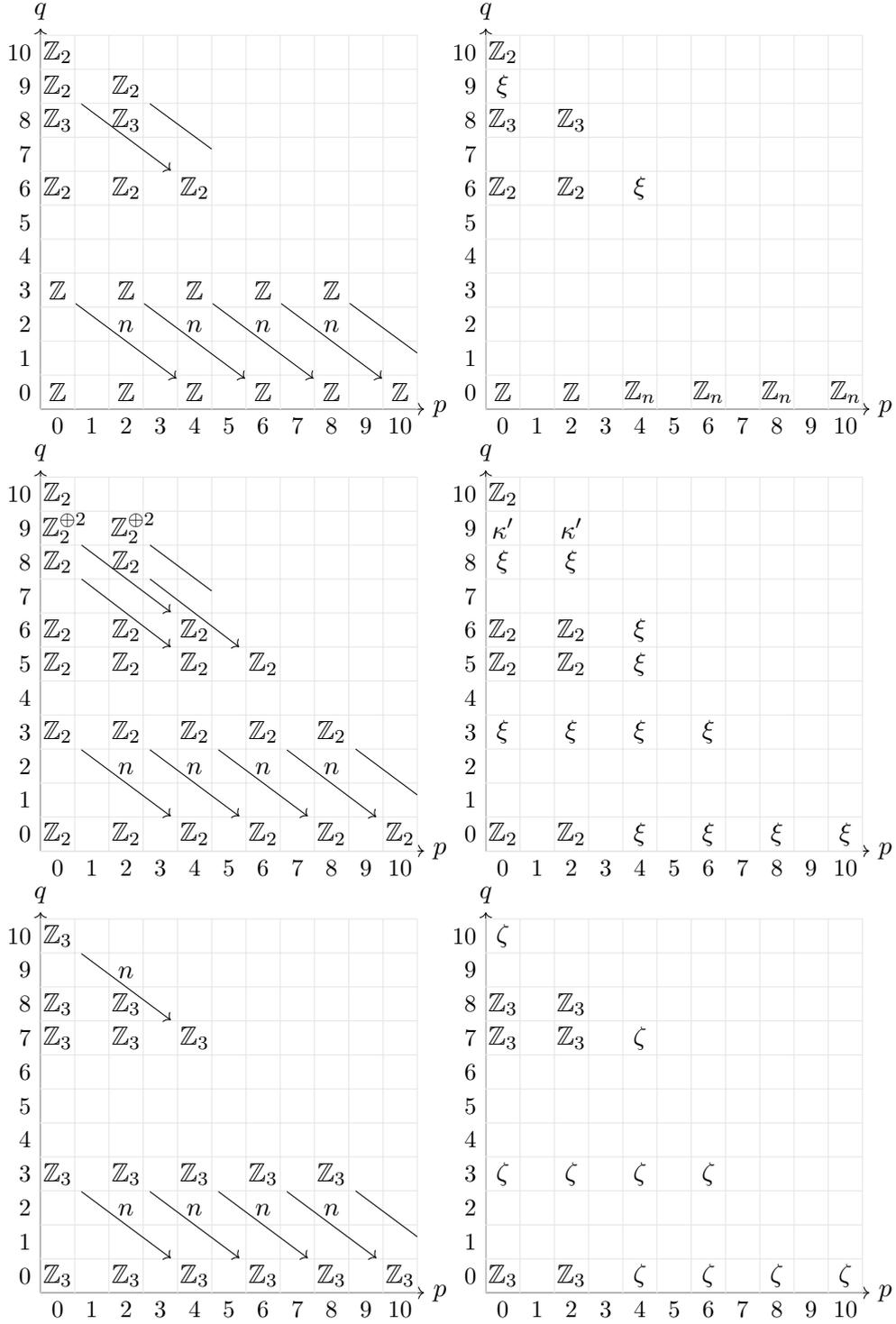

\noindent In the `times $n$' case we consider a homotopy fiber of the form
\begin{equation}
    Q \longrightarrow B \mathrm{U}(1) \overset{\alpha}{\longrightarrow} K(\mathbb{Z},4) \,.
\end{equation}
Form the discussion above we know that the homotopy classes of maps $\alpha$ are determined by an element in $H^4(B\mathrm{U}(1);\mathbb{Z})$ which in the case above is given by $n v_2^2$ which is trivialized in $Q$. In particular, this means that pulling back the generator $v_4 \in H^4\big( K(\mathbb{Z},4); \mathbb{Z}\big)$ to $B\mathrm{U}(1)$ one has $\alpha^*(v_4) = n v_2^2$. Extending to the left we find
\begin{equation}
K(\mathbb{Z},3) \longrightarrow Q \longrightarrow B \mathrm{U}(1) \,,
\end{equation}
which suggests the involvement of a U$(1)$ 2-form field, with classifying space $K(\mathbb{Z},3)$ appearing as the fiber. The integer cohomology of $K(\mathbb{Z},3)$ is known, see e.g., \cite{2013arXiv1312.5676B}, and summarized in Appendix \ref{app:EMachom}. From this we can find the second page of the LSSS, see Figure \ref{fig:E2c1sqn}.

The non-triviality of the fibration is indicated by a differential $d_4$, which trivializes $n v_2^2$ and higher order topological classes. We also know that the $\mathbb{Z}_2$ generator in $E^{0,9}_2$ is given by the product of the element in $E^{0,6}_2$ and $E^{0,3}_2$, see \cite{2013arXiv1312.5676B, Saito:2025idl}, and with that we can argue that there is another differential acting as indicated. Note also that this action is only non-trivial for $n$ odd (we denote the greatest common divisor of $r$ and $s$, by $(r,s)$). With this we can draw the relevant portion of $E^{p,q}_{\infty}$ to determine the cohomology groups of $H^{\bullet} (Q;\mathbb{Z})$, which we also show in Figure \ref{fig:E2c1sqn}. 

The only remaining questions are extension problems for the cohomology groups $H^6(Q;\mathbb{Z})$, $H^8(Q;\mathbb{Z})$, and $H^{10}(Q;\mathbb{Z})$. For that we can re-run the same spectral sequence but for cohomology with $\mathbb{Z}_2$ and  $\mathbb{Z}_3$ coefficients. The first advantage is that there are no extension problems. The second is that it still can distinguish between the various possibilities for $H^{\bullet}(Q;\mathbb{Z})$. The third is that by knowing the differentials on the lower degrees, using the ring and Steenrod algebra structure one can obtain all the differentials more easily. 

For example, for $n$ even, the two options\footnote{For $n$ odd these options are identical, since $\mathbb{Z}_n \oplus \mathbb{Z}_2 \cong \mathbb{Z}_{2n}$ ($n$ odd).} for $H^6(Q;\mathbb{Z})$ are $\mathbb{Z}_2 \oplus \mathbb{Z}_n$ or $\mathbb{Z}_{2n}$; in the first case one would produce $H^6(Q;\mathbb{Z}_2) = \mathbb{Z}_2 \oplus \mathbb{Z}_2$, but in the second case one would find a single $\mathbb{Z}_2$ factor. For $\mathbb{Z}_2$ coefficients we find the second page of the LSSS sequence in Figure \ref{fig:E2c1sqn}. There is a non-trivial differential $d_4:E_4^{0,3}=E_2^{0,3}\rightarrow E_4^{0, 4}=E_2^{0, 4}$ only in the case of $n$ odd. From this one can argue that for $n$ even all relevant differentials are trivial. Denoting the generator of $H^3\big( K(\mathbb{Z},3); \mathbb{Z}_2 \big) \simeq \mathbb{Z}_2$ by $\tilde{v}_3$, in the odd case, the new differential appears, since the element in $E^{0,8}_2$, $\tilde{v}_3 Sq^2 \tilde{v}_3$, is the product of the element in $E^{0,5}_2$ ($Sq^2 \tilde{v}_3$) and $E^{0,3}_2$ ($\tilde{v}_3$). We see that there is a difference depending whether $n$ is even or odd and we have, for example
\begin{equation}
    H^6(Q;\mathbb{Z}_2) = \mathbb{Z}_2 \oplus \mathbb{Z}_{(n,2)} \,,
\end{equation}
which shows that
\begin{equation}
    H^6(Q;\mathbb{Z}) = \mathbb{Z}_n \oplus \mathbb{Z}_2 \,,
\end{equation}
and solves the extension problem in degree 6. Similarly, we have
\begin{equation}
    H^8(Q;\mathbb{Z}_2) =\mathbb{Z}_2\oplus \mathbb{Z}_{(n, 2)} \oplus \mathbb{Z}_{(n, 2)}\,.
\end{equation}
Tracking the factors, through various applications of the universal coefficient theorem demonstrates that $\mathbb{Z}_2$ appears as a separate factor in $H^8(Q;\mathbb{Z})$ and the remaining extension problem is the $\mathbb{Z}_3$ in $E^{0,8}_{\infty}$, which is only relevant if $n$ is divisible by 3. To resolve the last piece of the extension problem we can use the spectral sequence for cohomology with $\mathbb{Z}_3$ coefficients, see Figure~\ref{fig:E2c1sqn}. Analogous to the strategy with $\mathbb{Z}_2$ coefficients one can use the extra structure on the cohomology groups $H^\bullet\big(B\mathrm{U}(1);\mathbb{Z}_3\big)$ and $H^\bullet\big(K(\mathbb{Z}, 3);\mathbb{Z}_3\big)$, namely of $\mathcal{A}_3$-modules, where $\mathcal{A}_3$ is a mod-3 Steenrod algebra.\footnote{The mod-3 Steenrod algebra is generated by a Bockstein map $\beta:H^k(X; \mathbb{Z}_3)\rightarrow H^{k+1}(X; \mathbb{Z}_3) $ which correspond to a short exact sequence of groups $0\rightarrow \mathbb{Z}_3\rightarrow \mathbb{Z}_{9}\rightarrow \mathbb{Z}_3\rightarrow 0$ and Steenrod powers $P^i:H^k(X; \mathbb{Z}_3)\rightarrow H^{k+4i}(X; \mathbb{Z}_3)$, which satisfy the analogous properties to Steenrod squares (for more details see \cite{hatcher2002}).} As for $\mathbb{Z}_2$ coefficients, cohomology groups of Eilenberg-MacLane spaces with $\mathbb{Z}_3$ coefficients are freely generated graded-commutative algebras generated by the fundamental class and the admissible sequence of Steenrod operations acting on it \cite{mccleary2001}. In our case;
\begin{equation}
\begin{split}
    H^\bullet \big(B\mathrm{U(1)};\mathbb{Z}_3\big) &= \mathbb{Z}_3 [v'_2] \,, \\
    H^\bullet \big(K(\mathbb{Z}, 3);\mathbb{Z}_3\big) &= \mathbb{Z}_3[v'_3, P^1 v'_3, \beta P^1 v'_3,...] \,,
\end{split}
\end{equation}
where $v_k'$ denotes the mod $3$ reduction of the generator $v_k \in H^k \big(K(\mathbb{Z},k); \mathbb{Z}\big)$
Since $d_4(v_3')=n (v_2')^2$ we have a non-trivial differential only in the case of $n$ not being divisible by 3 and as above, one can argue that in this case all differentials are trivial. Otherwise there is an extra relevant non-trivial differential $d_4(v_3' P^1 v_3')= n (v_2')^2 P^1 v_3'$. One can just read off the $H^\bullet(Q; \mathbb{Z}_3)$ from the infinity page. Most importantly one obtains 
\begin{equation}
    H^7(Q, \mathbb{Z}_3)=\mathbb{Z}_3\oplus \mathbb{Z}_{(3, n)}.
\end{equation}
Since from the universal coefficient theorem for cohomology
\begin{equation}
    H^7(Q;\mathbb{Z}_3)=H^7(Q; \mathbb{Z})\otimes \mathbb{Z}_3\oplus \text{Tor}\big(H^8(Q; \mathbb{Z}); \mathbb{Z}_3\big)
\end{equation}
one sees that we get a trivial extension of $\mathbb{Z}_3$ with $\mathbb{Z}_n$ in the cohomology of degree 8, and similarly in degree 10. Finally, we obtain the cohomology groups up to degree 10:
\begin{equation}
\begin{array}{c | c c c c c c c c c c c}
    k & 0 & 1 & 2 & 3 & 4 & 5 & 6 & 7 & 8 & 9 & 10 \\ \hline
    H^k(Q;\mathbb{Z}) & \mathbb{Z} & 0 & \mathbb{Z} & 0 & \mathbb{Z}_n & 0 & \mathbb{Z}_n \oplus \mathbb{Z}_2 & 0 & \mathbb{Z}_n \oplus \mathbb{Z}_3 \oplus \mathbb{Z}_2 & \mathbb{Z}_{(n,2)} & \mathbb{Z}_n \oplus\mathbb{Z}_3 \oplus \mathbb{Z}_2 \oplus \mathbb{Z}_{(n,2)}
\end{array}
\end{equation}
Also, since all the extensions were trivial we have that under the pullback map
\begin{equation}
    k \geq 2: \quad \mathbb{Z}\big[v_2^k \big]\cong H^{2k}(B \mathrm{U}(1); \mathbb{Z})\rightarrow H^{2k}(Q; \mathbb{Z})\cong \mathbb{Z}_n \oplus \text{extra} \,,
\end{equation}
the generator $v_2^k$, with $k \geq 2$, is sent to the generator of $\mathbb{Z}_n\subset H^{2k}(Q;\mathbb{Z})$. We can also understand the extra contributions, in particular the factors of $\mathbb{Z}_2$ and $\mathbb{Z}_3$ above are simply inherited from the fiber $K(\mathbb{Z},3)$ and thus describe topologically non-trivial configurations of a U$(1)$ 2-form field. The factors $\mathbb{Z}_{(n,2)}$ are also inherited from the fiber, but only survive the differentials in case $n$ is even. One therefore sees, that in these low degrees the characteristic classes can be associated to the modified U$(1)$ classes and that of a U$(1)$ 2-form gauge field. Further note that even in the case $n =1$ the presence of the fiber generates new characteristic classes.

\vspace{0.5cm}

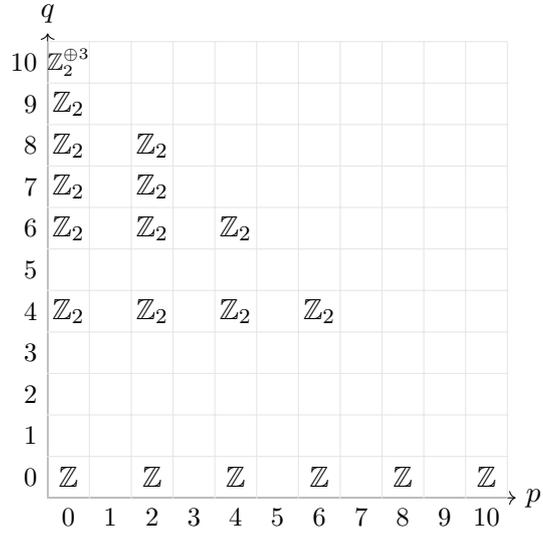
\begin{figure}
\centering
\begin{tikzpicture}[scale=0.55]
    % Axes
    \draw[->] (0,0) -- (11.2,0) node[right] {$p$};
    \draw[->] (0,0) -- (0,11.2) node[above] {$q$};
    % Grid (optional)
    \draw[step=1cm,gray!20] (0,0) grid (11,11);
    % Labels
    \foreach \x in {0,1,...,10}
        \node[below] at (\x+0.5, 0) {\small $\x$};
    \foreach \y in {0,1,..., 10}
        \node[left] at (0,\y+0.5) {\small $\y$};
\node(00) at (0.5, 0.5) {$\mathbb{Z}$};
\node(04) at (0.5, 4.5) {$\mathbb{Z}_2$};
\node(06) at (0.5, 6.5) {$\mathbb{Z}_2$};
\node(07) at (0.5, 7.5) {$\mathbb{Z}_2$};
\node(08) at (0.5, 8.5) {$\mathbb{Z}_2$};
\node(09) at (0.5, 9.5) {$\mathbb{Z}_2$};
\node(20) at (2.5, 0.5) {$\mathbb{Z}$};
\node(24) at (2.5, 4.5) {$\mathbb{Z}_2$};
\node(26) at (2.5, 6.5) {$\mathbb{Z}_2$};
\node(27) at (2.5, 7.5) {$\mathbb{Z}_2$};
\node(28) at (2.5, 8.5) {$\mathbb{Z}_2$};
\node(40) at (4.5, 0.5) {$\mathbb{Z}$};
\node(44) at (4.5, 4.5) {$\mathbb{Z}_2$};
\node(46) at (4.5, 6.5) {$\mathbb{Z}_2$};
\node(60) at (6.5, 0.5) {$\mathbb{Z}$};
\node(64) at (6.5, 4.5) {$\mathbb{Z}_2$};
\node(80) at (8.5, 0.5) {$\mathbb{Z}$};
\node (010) at (0.5, 10.5) {{\footnotesize $\mathbb{Z}_2^{\oplus 3}$}};
\node(100) at (10.5, 0.5) {$\mathbb{Z}$};
\end{tikzpicture}
\caption{$E^{p,q}_2$ page of $Q$ implementing $\mathfrak{m}_2 (c_1^2) = 0$, for $p+q \leq 10$.}
\label{fig:E2c1sqmodn}
\end{figure}

\begin{figure}
\centering
\begin{tikzpicture}[scale=0.55]
    % Axes
    \draw[->] (0,0) -- (11.2,0) node[right] {$p$};
    \draw[->] (0,0) -- (0,11.2) node[above] {$q$};
    % Grid (optional)
    \draw[step=1cm,gray!20] (0,0) grid (11,11);
    % Labels
    \foreach \x in {0,1,...,10}
        \node[below] at (\x+0.5, 0) {\small $\x$};
    \foreach \y in {0,1,..., 10}
        \node[left] at (0,\y+0.5) {\small $\y$};
\node(00) at (0.5, 0.5) {$\mathbb{Z}_2$};
\node(03) at (0.5, 3.5) {$\mathbb{Z}_2$};
\node(04) at (0.5, 4.5) {$\mathbb{Z}_2$};
\node(05) at (0.5, 5.5) {$\mathbb{Z}_2$};
\node(06) at (0.5, 6.5) {\footnotesize{ $\mathbb{Z}_2^{\oplus 2}$}};
\node(07) at (0.5, 7.5) {\footnotesize{$\mathbb{Z}_2^{\oplus 2}$}};
\node(08) at (0.5, 8.5) {\footnotesize{$\mathbb{Z}_2^{\oplus 2}$}};
\node(09) at (0.5, 9.5) {\footnotesize{$\mathbb{Z}_2^{\oplus 4}$}};
\node(010) at (0.5, 10.5) {\footnotesize{$\mathbb{Z}_2^{\oplus 5}$}};
\node(20) at (2.5, 0.5) {$\mathbb{Z}_2$};
\node(23) at (2.5, 3.5) {$\mathbb{Z}_2$};
\node(24) at (2.5, 4.5) {$\mathbb{Z}_2$};
\node(25) at (2.5, 5.5) {$\mathbb{Z}_2$};
\node(26) at (2.5, 6.5) {\footnotesize{$\mathbb{Z}_2^{\oplus 2}$}};
\node(27) at (2.5, 7.5) {\footnotesize{$\mathbb{Z}_2^{\oplus 2}$}};
\node(28) at (2.5, 8.5) {\footnotesize{$\mathbb{Z}_2^{\oplus 2}$}};
\node(40) at (4.5, 0.5) {$\mathbb{Z}_2$};
\node(43) at (4.5, 3.5) {$\mathbb{Z}_2$};
\node(44) at (4.5, 4.5) {$\mathbb{Z}_2$};
\node(45) at (4.5, 5.5) {$\mathbb{Z}_2$};
\node(46) at (4.5, 6.5) {\footnotesize{$\mathbb{Z}_2^{\oplus 2}$}};
\node(60) at (6.5, 0.5) {$\mathbb{Z}_2$};
\node(63) at (6.5, 3.5) {$\mathbb{Z}_2$};
\node(64) at (6.5, 4.5) {$\mathbb{Z}_2$};
\node(80) at (8.5, 0.5) {$\mathbb{Z}_2$};
\node(100) at (10.5, 0.5) {$\mathbb{Z}_2$};
\draw[->] (03) -- (40);
\draw[->] (23) --  (60);
\draw[->] (43) --  (80);
\draw[->] (63) --(100);
\draw[->] (07) -- (44);
\draw[->] (27) -- (64);
\draw[->] (08) --  (45);
\draw (28) --  (5.7,6.1);
\draw[->] (09) --  (46);
\draw (010) --  (3.7,8.1);
%
%
    % Axes
    \draw[->] (13,0) -- (24.2,0) node[right] {$p$};
    \draw[->] (13,0) -- (13,11.2) node[above] {$q$};
    % Grid (optional)
    \draw[step=1cm,gray!20] (13,0) grid (24,11);
    % Labels
    \foreach \x in {0,1,...,10}
        \node[below] at (\x+13.5, 0) {\small $\x$};
    \foreach \y in {0,1,..., 10}
        \node[left] at (13,\y+0.5) {\small $\y$};
\node at (13.5, 0.5) {$\mathbb{Z}_2$};
\node at (13.5, 4.5) {$\mathbb{Z}_2$};
\node at (13.5, 5.5) {$\mathbb{Z}_2$};
\node at (13.5, 6.5) {\footnotesize{$\mathbb{Z}_2^{\oplus 2}$}};
\node at (13.5, 7.5) {$\mathbb{Z}_2$};
\node at (13.5, 8.5) {$\mathbb{Z}_2$};
\node at (13.5, 9.5) {\footnotesize{$\mathbb{Z}_2^{\oplus 2}$}};
\node at (13.5, 10.5) {\footnotesize{$\mathbb{Z}_2^{\oplus 4}$}};
\node at (15.5, 0.5) {$\mathbb{Z}_2$};
\node at (15.5, 4.5) {$\mathbb{Z}_2$};
\node at (15.5, 5.5) {$\mathbb{Z}_2$};
\node at (15.5, 6.5) {\footnotesize{$\mathbb{Z}_2^{\oplus 2}$}};
\node at (15.5, 7.5) {$\mathbb{Z}_2$};
\node at (15.5, 8.5) {$\mathbb{Z}_2$};
\end{tikzpicture}
\caption{$E_2^{p,q}$ and $E^{p,q}_{\infty}$ of $Q$ implementing $\mathfrak{m}_2 (c_1^2) = 0$ with $\mathbb{Z}_2$ coefficients and $p+q\leq 10$.}
\label{fig:c1sqmod2Z2coef}
\end{figure}
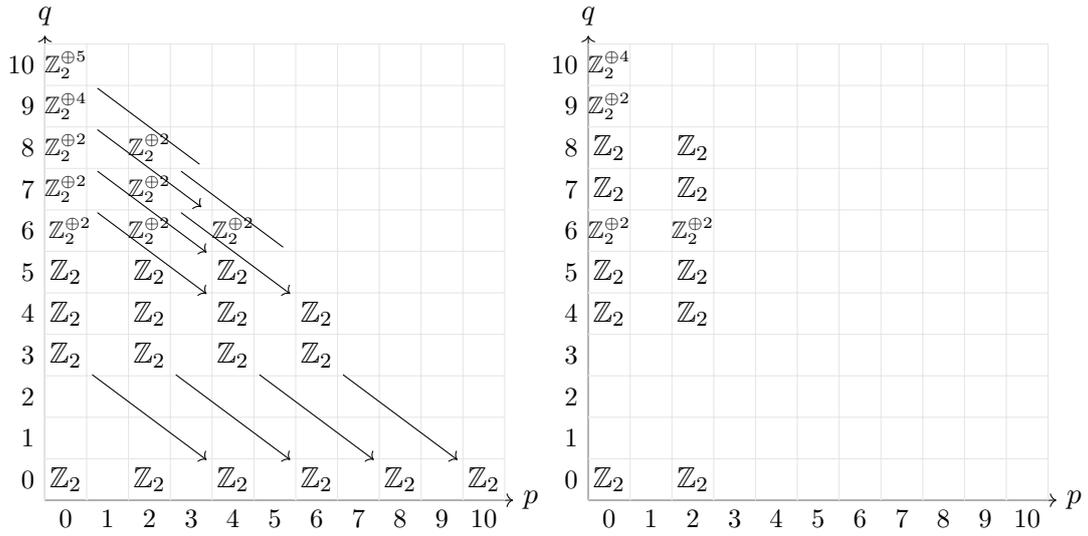

\noindent Now let us move to the `mod $n$' constraint for the particular case $n = 2$, for which the homotopy fiber reads
\begin{equation}
    Q \longrightarrow B\mathrm{U}(1) \longrightarrow K(\mathbb{Z}_2, 4) \,,
\end{equation}
where the generator $u_4 \in H^4\big( K(\mathbb{Z}_2,4); \mathbb{Z}\big)$ pulls back to $\mathfrak{m}_2 (v_2^2) \in H^4\big(B\mathrm{U}(1);\mathbb{Z}_2\big)$. Again, we can extend to the left to find a description of $Q$ as fibration
\begin{equation}\label{eq:fibrationforc1sqmod2}
    K(\mathbb{Z}_2,3) \longrightarrow Q \longrightarrow B \mathrm{U}(1) \,.
\end{equation}
Here, we find the classifying space for a $\mathbb{Z}_2$ 3-form field as a fiber. With the cohomology groups in \cite{Clementthesis} and summarized in Appendix \ref{app:EMachom}, we can write down the second page of the LSSS, see Figure \ref{fig:E2c1sqmodn}.
We can extract useful information about differentials and extensions from the spectral sequence with $\mathbb{Z}_2$ coefficients, which we draw in Figure \ref{fig:c1sqmod2Z2coef}. We will not be able to fully pin down all the relevant differentials, but we will nevertheless manage to determine all the cohomology groups of $Q$ in degree 10 and below as well as their relations to the characteristic classes of $B\mathrm{U}(1)$.
We use Serre's description of the cohomology $H^{\bullet}\big(K(\mathbb{Z}_2, 3 ); \mathbb{Z}_2\big)$ as a polynomial algebra given in terms of Steenrod operations \cite{hatcherspectral, mccleary2001}. To be more precise, in terms of the generator $\tilde{u}_3 \in H^3\big( K(\mathbb{Z}_2,3) ; \mathbb{Z}_2 \big)$ one has:
\begin{equation}
\begin{split}
    H^{\bullet}(K(\mathbb{Z}_2, 3);\mathbb{Z}_2)=\mathbb{Z}_2 \big[ & \tilde{u}_3,\,Sq^1 \tilde{u}_3,\, Sq^2 \tilde{u}_3,\, Sq^2Sq^1 \tilde{u}_3, \\ & Sq^3Sq^1 \tilde{u}_3,\, Sq^4Sq^2 \tilde{u}_3,\,Sq^4Sq^2Sq^1 \tilde{u}_3,\dots \big] \,.
\end{split}
\end{equation}
By design, there is a non-trivial differential $d_4(\tilde{u}_3)= \mathfrak{m}_n (v_2^2)$. Using that $d_4$ is a differential with respect to the ring structure determines all the non-trivial differentials\footnote{For that, note that $Sq^i$ commutes with the transgression map $\tau=d_r: E^{0,r-1}_r \rightarrow E^{r,0}_r$, which is the name for all differentials that take a class coming from the fiber and sends it to a class of the base \cite{hatcherspectral, mccleary2001}. From that one finds $\tau \big( Sq^i \tilde{u}_3\big) = Sq^i \big( \tau (\tilde{u}_3)\big) = Sq^i \big( \mathfrak{m}_2 (v_2^2)\big)$ which vanishes for $Sq^1$ and $Sq^2$, which in our case is enough to show the triviality of the differential on the remaining elements in $H\big( K(\mathbb{Z}_2,3); \mathbb{Z}_2 \big)$, for $k \leq 10$.} in the Figure~\ref{fig:c1sqmod2Z2coef}. From the $\infty$-page we read out the cohomology groups with $\mathbb{Z}_2$ coefficients:
\begin{equation}
     \begin{array}{c | c c c c c c c c c c c }
     k & 0 & 1 & 2 & 3 & 4 & 5 & 6 & 7 & 8 & 9 & 10 \\ \hline
     H^k(Q;\mathbb{Z}_2) & \mathbb{Z}_2 & 0 & \mathbb{Z}_2 & 0 & \mathbb{Z}_2 & \mathbb{Z}_2 & \mathbb{Z}_2^{\oplus 3} & \mathbb{Z}_2^{\oplus 2} & \mathbb{Z}_2^{\oplus 3} & \mathbb{Z}_2^{\oplus 3} & \mathbb{Z}_2^{\oplus 5}
     \end{array}
\end{equation}
Comparing this result with the LSSS for $\mathbb{Z}$ coefficients in combination with \eqref{eq:coeffcohom}, we can resolve some of the extension problems, similar to our strategy above. We find that the groups with integer coefficients have to be of the form
\begin{equation}\label{eq:potentialgroups}
\begin{split}
    H^4(Q;\mathbb{Z}) &= \mathbb{Z} \,, \enspace H^6(Q;\mathbb{Z}) = \mathbb{Z} \oplus \mathbb{Z}_{2^{\ell_1}} \,, \enspace H^8(Q;\mathbb{Z}) = \mathbb{Z} \oplus \mathbb{Z}_{2^{\ell_2}} \,, \\
    \quad H^9 (Q;\mathbb{Z}) &= \mathbb{Z}_{2^{\ell_3}} \,, \quad H^{10}(Q;\mathbb{Z}) = \mathbb{Z} \oplus \mathbb{Z}_{2^{\ell_4}} \oplus \mathbb{Z}_{2^{\ell_5}} \,,
\end{split}
\end{equation}
and all that is left to determine are the explicit extensions encoded in the $\ell_i$.

Since $H^4(Q;\mathbb{Z}) = \mathbb{Z}$ there is a non-trivial extension. Denoting the generator of the free factor in $H^{2k}(Q;\mathbb{Z})$ by $\alpha_{2k}$, this means 
\begin{equation}
    p^* (v_2^2) = 2 \alpha_4 \,,
\end{equation}
where $p$ is given by projection onto the base in \eqref{eq:fibrationforc1sqmod2}. In degree 6, there are two possibilities, given by $\mathbb{Z} \oplus \mathbb{Z}_2$ and $\mathbb{Z} \oplus \mathbb{Z}_4$, the second case, however is inconsistent with the ring structure, since it would imply $p^* (v_2^3) = \alpha_6$ but one has
\begin{equation}
     p^\ast (v_2^3) = p^* (v_2) \cup p^*(v_2^2) = 2 \alpha_2 \alpha_4 \,,
\end{equation}
which is clearly divisible by $2$. Thus, there is a non-trivial extension, which is inherited from degree 4 and implements $p^* (v_2^3) = 2 \alpha_6 = 2 \alpha_2 \alpha_4$. Using similar arguments one can show that, again, there are non-trivial extensions in degree 8, which can be inferred from
\begin{equation}
    p^* (v_2^4) = p^*(v_2^2) \cup p^*(v_2^2) = 4 \alpha_4^2 \,,
    \label{eq:extfactor4}
\end{equation}
which needs extensions with two $\mathbb{Z}_2$ factors and singles out 
\begin{equation}
    H^8 (Q;\mathbb{Z}) = \mathbb{Z} \oplus \mathbb{Z}_2 \,,
\end{equation}
as the correct cohomology group.

In degree 9 we  encounter yet another phenomenon; the appearance of a non-trivial differential in the LSSS with integer coefficients, which is not determined by the ring structure above. This differential takes care of one of the $\mathbb{Z}_2$ entries in the second page and one finds $H^9 (Q;\mathbb{Z}) = \mathbb{Z}_2$. The other possible option would have been that one has a non-trivial enhancement to $\mathbb{Z}_4$. In both cases the universal coefficient theorem \eqref{eq:coeffcohom} leads to contributions  for $H^8(Q;\mathbb{Z}_2)$ and $H^9(Q;\mathbb{Z}_2)$. However, since $Sq^1$ can be understood as the Bockstein homomorphism associated to the short exact sequence
\begin{equation}
    0 \longrightarrow \mathbb{Z}_2 \longrightarrow \mathbb{Z}_4 \longrightarrow \mathbb{Z}_2 \longrightarrow 0 \,,
\end{equation}
its action is different, since one has
\begin{equation}
    ... \longrightarrow H^8(Q;\mathbb{Z}_2) \longrightarrow H^9 (Q;\mathbb{Z}_2) \longrightarrow H^9(Q;\mathbb{Z}_4) \longrightarrow ... \,.
\end{equation}
Denoting the relevant element in $H^8(Q;\mathbb{Z}_2)$ by $b$, on sees that for $H^9(Q;\mathbb{Z}) = \mathbb{Z}_4$, $Sq^1 (b)$ needs to be trivial by exactness. For $H^9(Q;\mathbb{Z}) = \mathbb{Z}_2$ on the other hand, multiplication by $2$ acts trivial and $Sq^1(b)$ is non-trivial. The problem is that, while we have the Steenrod algebra for $B\mathrm{U}(1)$ and $K(\mathbb{Z}_2,3)$, we do need the action of the Steenrod operations on the cohomology elements of $Q$. For our particular problem, here, we can use the spectral Steenrod operations\footnote{Spectral sequence Steenrod operations are operations defined on every page of a spectral sequence related to the fibration $F\rightarrow E\rightarrow B$ such that $Sq^k:E_r^{p,q}\rightarrow E_{r}^{p,q+k}$ if $k\leq q$ and $Sq^k:E_r^{p,q}\rightarrow E_{t}^{p+k-q,2q}$ if $q\leq k$ and $t$ is an integer typically larger then $r$ making $Sq^k$ well defined, however in the case of $r=2$ one can take $t=2$. These operations commute with the differentials, meaning that by defining them on the second page the action on the later pages is directly induced. On the second page they are defined as $(1\otimes Sq^k):H^p(B;\mathbb{Z}_2)\otimes H^q(F;\mathbb{Z}_2)\rightarrow H^p(B;\mathbb{Z}_2)\otimes H^{q+k}(F;\mathbb{Z}_2)$ if $k\leq q$ and $Sq^{k-q}\otimes Sq^q:H^p(B;\mathbb{Z}_2)\otimes H^q(F;\mathbb{Z}_2)\rightarrow H^{p+k-q}(B;\mathbb{Z}_2)\otimes H^{2q}(F;\mathbb{Z}_2)$, where $Sq^i\otimes Sq^j$ are standard Steenrod operations acting on the cohomology of the base and the fiber, respectively. These spectral Steenrod operations do not fully determine the action of the Steenrod algebra on the cohomology of the total space, but converges to it in the sense that the spectral Steenrod operations on the infinity page are equal to those induced from the Steenrod operations on the total space, e.g., $Sq^k:E_\infty^{p,q}\cong\frac{F^pH^{p+q}(E;\mathbb{Z}_2)}{F^{p+1}H^{p+q}(E;\mathbb{Z}_2)}\rightarrow E_{\infty}^{p,q+r}\cong\frac{F^pH^{p+q+r}(E;\mathbb{Z}_2)}{F^{p+1}H^{p+q+r}(E;\mathbb{Z}_2)}$. For more details see \cite{Araki1959Steenrod}.} developed by Araki \cite{Araki1959Steenrod}. While this only fully determines the Steenrod operations on the associated graded groups induced by the filtration of $E^{p,q}_{\infty}$, it is enough for us. Let us denote the generator of $H^4\big( K(\mathbb{Z}_2,3); \mathbb{Z}_2\big)$ by $\tilde u_3$, then one has
\begin{equation}
    Sq^1: \quad E_2^{2,6} \rightarrow E_2^{2,7} \quad \text{with} \quad (Sq^2 Sq^1 \tilde u_3) \tilde v_2 \mapsto (Sq^3 Sq^1 \tilde u_3) \tilde v_2 \,,
\end{equation}
inducing an isomorphism. The spectral Steenrod operation make sure that this remains non-trivial on the $\infty$-page, i.e., there is a non-trivial map
\begin{equation}
    Sq^1: \quad E^{2,6}_{\infty} \rightarrow E^{2,7}_{\infty} \,.
\end{equation}
Since these factors are the building blocks of $H^{\bullet}(Q;\mathbb{Z}_2)$ there is a non-trivial action also here. However, due to non-trivial extensions encoded in the fibration structure the action
\begin{equation}
    Sq^1: \quad H^8(Q;\mathbb{Z}_2) \rightarrow H^9 (Q;\mathbb{Z}_2) \,,
\end{equation}
might also have contributions from other $E^{9-i,i}_{\infty}$, which here is only $E^{0,9}_{\infty}$. What remains, is that the non-trivial $Sq^1$ action demands that
\begin{equation}
    H^9 (Q;\mathbb{Z}) = \mathbb{Z}_2 \,,
\end{equation}
due to an additional (hidden) differential. This will also trivialize a $\mathbb{Z}_2$ entry in of $E^{10-i,i}_{\infty}$ and hence modify $H^{10}(Q;\mathbb{Z})$ accordingly. One can determine the cohomology in degree ten in the same way. At the level of spectral sequence there is a relevant non-trivial action of the Steenrod operation:
\begin{equation}
    Sq^1:\quad E_2^{0,9}\rightarrow E_2^{0,10} \quad\text{with} \quad Sq^4Sq^2\tilde u_3\rightarrow Sq^5Sq^2\tilde u_3 \,, \quad Sq^1\tilde u_3Sq^2\tilde u_3\rightarrow \tilde u_3^2Sq^1\tilde u_3 \,,
\end{equation}
which survives to the infinity page and implies the nontrivial action of the Steenrod operation
\begin{equation}
    Sq^1:H^9(Q;\mathbb{Z}_2)\rightarrow H^{10}(Q;\mathbb{Z}_2) \,,
\end{equation}
on two $\mathbb{Z}_2$ summands inside of $H^9(Q;\mathbb{Z}_2)$. With the same argument as for degree 9, there are two $\mathbb{Z}_2$ summands in $H^{10}(Q;\mathbb{Z})$ and when compared with \ref{eq:potentialgroups} one fixes the cohomology to
\begin{equation}
    H^{10}(Q;\mathbb{Z})=\mathbb{Z}\oplus\mathbb{Z}_2\oplus\mathbb{Z}_2
\end{equation}
Thus, one has
\begin{equation}
    \begin{array}{c | c c c c c c c c c c c}
    k & 0 & 1 & 2 & 3 & 4 & 5 & 6 & 7 & 8 & 9 & 10 \\ \hline
    H^k(Q;\mathbb{Z}) & \mathbb{Z} & 0 & \mathbb{Z} & 0 & \mathbb{Z} & 0 & \mathbb{Z} \oplus \mathbb{Z}_2 & \mathbb{Z}_2 & \mathbb{Z} \oplus \mathbb{Z}_2 & \mathbb{Z}_2 & \mathbb{Z} \oplus \mathbb{Z}_2 \oplus \mathbb{Z}_2
    \end{array} \,,
    \label{eq:Qmodn_intcohom}
\end{equation}
describing the cohomology groups of $Q$ up to degree $10$. 

As for $B\mathrm{U}(1)$ we find factors of $\mathbb{Z}$ in even degrees, however, now these are normalized differently as is encoded in the non-trivial extensions. For example consider the factor of $\mathbb{Z}$ in degree 8, which is extended by a factor of $4$, \eqref{eq:extfactor4}, and the generator in terms of the original first Chern class of the U$(1)$ bundle is given by
\begin{equation}
    \tfrac{1}{4} c_1^4 \in H^8 (Q;\mathbb{Z}) \,.
\end{equation}
We would expect the same factor $4$ for the generator of $\mathbb{Z}$ in degree 10, which however depends on the existence of a non-trivial differential from elements in $E_{j}^{11-r,r}$ which go beyond our discussion here. 

Beyond this change of normalization as compared to the characteristic classes of $B\mathrm{U}(1)$ we also find additional factors of $\mathbb{Z}_2$. From the second page of the LSSS, Figure~\ref{fig:E2c1sqmodn}, these can be traced back to characteristic classes of the fiber $K(\mathbb{Z}_2,3)$, as well as combinations of classes from fiber and base. Once more we find that the modification of the gauge theory leads to desired redefinition of characteristic classes in combination with the introduction of new ones.

\subsection{Modification of classes of higher-form fields}

Since in the discussion above we have focused on the modification of U$(1)$ 1-form fields, we want to illustrate the applicability of the homotopy fiber construction also for higher-form fields, we do so for the modification of a U$(1)$ 3-form field with $[N^{(4)}]$ as its topological class, in particular we consider the constraints
\begin{equation}
\begin{split}
    \text{times } n:& \quad n [N^{(4)}]^2 = 0 \in H^8(X;\mathbb{Z}) \,, \\
    \text{mod } (n = 2):& \quad \mathfrak{m}_2\big([N^{(4)}]^2\big) = 0 \in H^8(X;\mathbb{Z}_2) \,,
\end{split}
\end{equation}
in analogy to the cases above.

\vspace{0.5cm}

\begin{figure}
\centering
\begin{tikzpicture}[scale=0.5]
    % Axes
    \draw[->] (0,0) -- (13.2,0) node[right] {$p$};
    \draw[->] (0,0) -- (0,13.2) node[above] {$q$};
    % Grid (optional)
    \draw[step=1cm,gray!20] (0,0) grid (13,13);
    % Labels
    \foreach \x in {0,1,...,12}
        \node[below] at (\x+0.5, 0) {\small $\x$};
    \foreach \y in {0,1,...,12}
        \node[left] at (0,\y+0.5) {\small $\y$};
\node at (0.5, 0.5) {$\mathbb{Z}$};
\node(a) at (0.5, 7.5) {$\mathbb{Z}$};
\node at (0.5, 10.5) {$\mathbb{Z}_2$};
\node at (1.2, 12.5) {{\footnotesize $\mathbb{Z}_2 \oplus \mathbb{Z}_3$}};
\node at (4.5, 0.5) {$\mathbb{Z}$};
\node(c) at (4.5, 7.5) {$\mathbb{Z}$};
\node at (7.5, 0.5) {$\mathbb{Z}_2$};
\node(b) at (8.5, 0.5) {$\mathbb{Z}$};
\node at (9.5, 0.5) {$\mathbb{Z}_3$};
\node at (11.5, 0.5) {{\footnotesize $\mathbb{Z}_2^{\oplus 2}$}};
\node(d) at (12.5, 0.5) {$\mathbb{Z}$};
\draw[->] (a) -- (b);
\draw[->] (c) -- (d);
\end{tikzpicture}
\caption{$E^{p,q}_2$ page of $Q$ implementing $n [N^{(4)}]^2 = 0$ in the range $p+q\leq 12$.}
\label{fig:E2N4sqn}
\end{figure}
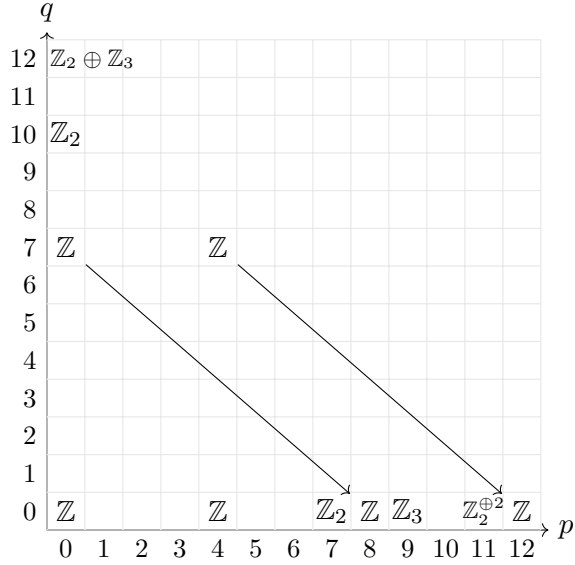

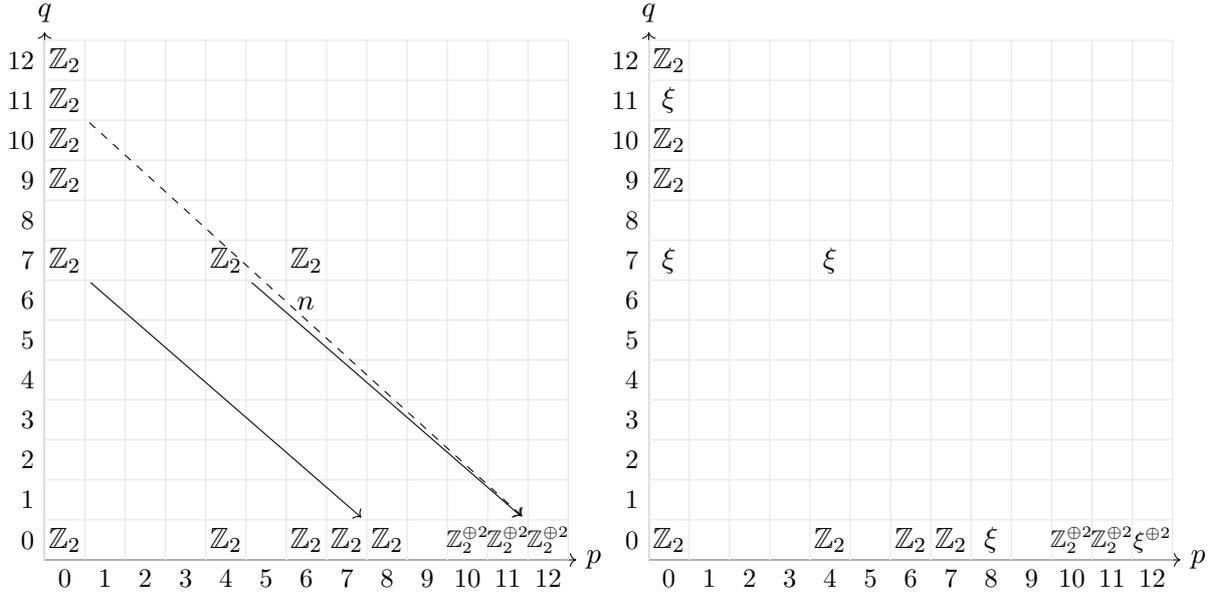
\begin{figure}
\centering
\begin{tikzpicture}[scale=0.53]
    % Axes
    \draw[->] (0,0) -- (13.2,0) node[right] {$p$};
    \draw[->] (0,0) -- (0,13.2) node[above] {$q$};
    % Grid (optional)
    \draw[step=1cm,gray!20] (0,0) grid (13,13);
    % Labels
    \foreach \x in {0,1,...,12}
        \node[below] at (\x+0.5, 0) {\small $\x$};
    \foreach \y in {0,1,..., 12}
        \node[left] at (0,\y+0.5) {\small $\y$};
\node(00) at (0.5, 0.5) {$\mathbb{Z}_2$};
\node(07) at (0.5, 7.5) {$\mathbb{Z}_2$};
\node(09) at (0.5, 9.5) {$\mathbb{Z}_2$};
\node(010) at (0.5, 10.5) {$\mathbb{Z}_2$};
\node(011) at (0.5, 11.5) {$\mathbb{Z}_2$};
\node(012) at (0.5, 12.5) {$\mathbb{Z}_2$};
\node(40) at (4.5, 0.5) {$\mathbb{Z}_2$};
\node(47) at (4.5, 7.5) {$\mathbb{Z}_2$};
\node(60) at (6.5, 0.5) {$\mathbb{Z}_2$};
\node(67) at (6.5, 7.5) {$\mathbb{Z}_2$};
\node(70) at (7.5, 0.5) {$\mathbb{Z}_2$};
\node(80) at (8.5, 0.5) {$\mathbb{Z}_2$};
\node(100) at (10.5, 0.5) {\footnotesize{$\mathbb{Z}_2^{\oplus 2}$}};
\node(110) at (11.5, 0.5) {\footnotesize{$\mathbb{Z}_2^{\oplus 2}$}};
\node(120) at (12.5, 0.5) {\footnotesize{$\mathbb{Z}_2^{\oplus 2}$}};
\draw[->] (07)--(80);
\draw[->] (47)--(120);
\draw[->, dashed] (011)--node[above]{$n$}(120);
%
%
    % Axes
    \draw[->] (15,0) -- (28.2,0) node[right] {$p$};
    \draw[->] (15,0) -- (15,13.2) node[above] {$q$};
    % Grid (optional)
    \draw[step=1cm,gray!20] (15,0) grid (28,13);
    % Labels
    \foreach \x in {0,1,...,12}
        \node[below] at (\x+15.5, 0) {\small $\x$};
    \foreach \y in {0,1,..., 12}
        \node[left] at (15,\y+0.5) {\small $\y$};
\node at (15.5, 0.5) {$\mathbb{Z}_2$};
\node at (15.5, 7.5) {$\xi$};
\node at (15.5, 9.5) {$\mathbb{Z}_2$};
\node at (15.5, 10.5) {$\mathbb{Z}_2$};
\node at (15.5, 11.5) {$\xi$};
\node at (15.5, 12.5) {$\mathbb{Z}_2$};
\node at (19.5, 0.5) {$\mathbb{Z}_2$};
\node at (19.5, 7.5) {$\xi$};
\node at (21.5, 0.5) {$\mathbb{Z}_2$};
\node at (22.5, 0.5) {$\mathbb{Z}_2$};
\node at (23.5, 0.5) {$\xi$};
\node at (25.5, 0.5) {\footnotesize{$\mathbb{Z}_2^{\oplus 2}$}};
\node at (26.5, 0.5) {\footnotesize{$\mathbb{Z}_2^{\oplus 2}$}};
\node at (27.5, 0.5) {\footnotesize{$\xi^{\oplus 2}$}};
\end{tikzpicture}
\caption{$E_2^{p,q}$ and $E^{p,q}_{\infty}$ of $Q$ implementing $n[N^{(4)}]^2=0$ with $\mathbb{Z}_2$ coefficients and $p+q \leq 12$ ($\xi = \mathbb{Z}_{(n,2)}$).}
\label{fig:n4sqnZ2coef}
\end{figure}

\begin{figure}
\centering
\begin{tikzpicture}[scale=0.53]
    % Axes
    \draw[->] (0,0) -- (13.2,0) node[right] {$p$};
    \draw[->] (0,0) -- (0,13.2) node[above] {$q$};
    % Grid (optional)
    \draw[step=1cm,gray!20] (0,0) grid (13,13);
    % Labels
    \foreach \x in {0,1,...,12}
        \node[below] at (\x+0.5, 0) {\small $\x$};
    \foreach \y in {0,1,..., 12}
        \node[left] at (0,\y+0.5) {\small $\y$};
\node(00) at (0.5, 0.5) {$\mathbb{Z}_3$};
\node(07) at (0.5, 7.5) {$\mathbb{Z}_3$};
\node(011) at (0.5, 11.5) {$\mathbb{Z}_3$};
\node(012) at (0.5, 12.5) {$\mathbb{Z}_3$};
\node(40) at (4.5, 0.5) {$\mathbb{Z}_3$};
\node(47) at (4.5, 7.5) {$\mathbb{Z}_3$};
\node(80) at (8.5, 0.5) {$\mathbb{Z}_3^{\oplus 2}$};
\node(90) at (9.5, 0.5) {$\mathbb{Z}_3$};
\node(120) at (12.5, 0.5) {$\mathbb{Z}_3^2$};
\node(130) at (13.5, 0.5){};
\draw[->] (07)-- (80);
\draw[->] (47)-- (120);
\draw[->, dashed] (011)--(120);
\draw[ dashed] (012)--node[above]{$2n$} (12.8, 1.2);
%
%
    % Axes
    \draw[->] (15,0) -- (28.2,0) node[right] {$p$};
    \draw[->] (15,0) -- (15,13.2) node[above] {$q$};
    % Grid (optional)
    \draw[step=1cm,gray!20] (15,0) grid (28,13);
    % Labels
    \foreach \x in {0,1,...,12}
        \node[below] at (\x+15.5, 0) {\small $\x$};
    \foreach \y in {0,1,..., 12}
        \node[left] at (15,\y+0.5) {\small $\y$};
\node at (15.5, 0.5) {$\mathbb{Z}_3$};
\node at (15.5, 7.5) {$\zeta$};
\node at (15.5, 11.5) {$ \zeta$};
\node at (15.5, 12.5) {$\zeta$};
\node at (19.5, 0.5) {$\mathbb{Z}_3$};
\node at (19.5, 7.5) {$\zeta$};
\node at (23.4, 0.5) {$\tilde{\zeta}$};
\node at (24.6, 0.5) {$\mathbb{Z}_3$};
\node at (27.5, 0.5) {$\zeta^{\oplus 2}$};
\end{tikzpicture}
\caption{$E_2^{p,q}$ and $E^{p,q}_{\infty}$ of $Q$ implementing $n[N^{(4)}]^2=0$ with $\mathbb{Z}_3$ coefficients and $p+q \leq 12$ ($\zeta = \mathbb{Z}_{(n,3)}$ and $\tilde{\zeta} = \mathbb{Z}_3 \oplus \mathbb{Z}_{(n,3)}$).}
\label{fig:n4sqnZ3coef}
\end{figure}
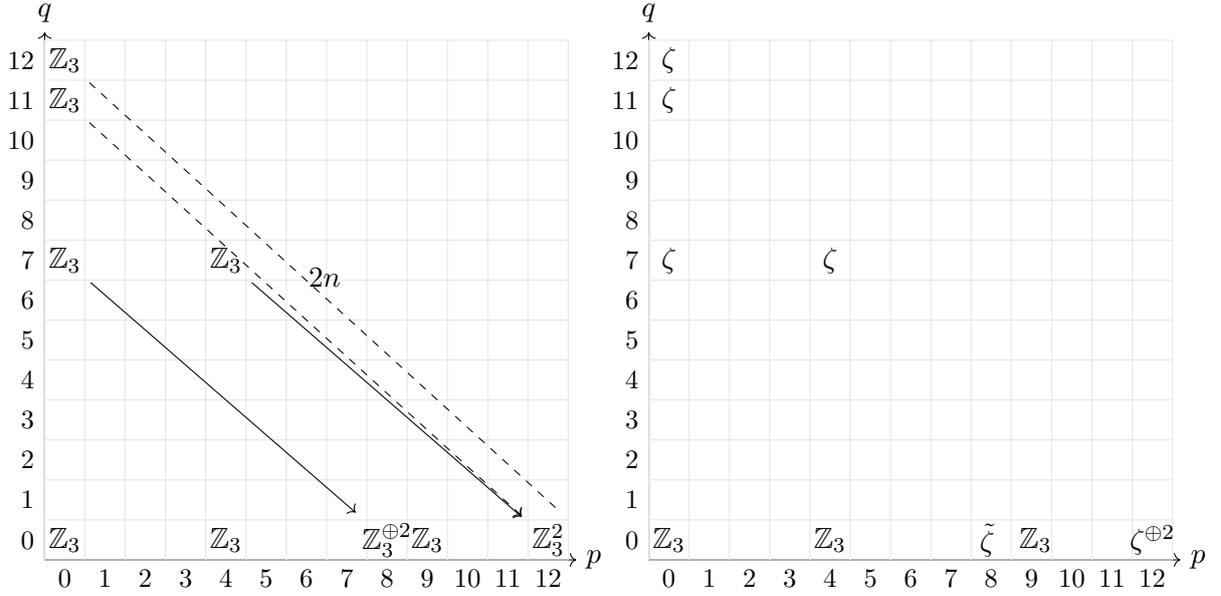

\noindent First, let us discuss the times $n$ constraint, with homotopy fiber $Q$ described by
\begin{equation}
    K(\mathbb{Z},7) \longrightarrow Q \longrightarrow K(\mathbb{Z},4) \,.
\end{equation}
We provide the cohomology groups of the fiber and base in Appendix \ref{app:EMachom}, which leads to the second page of the LSSS indicated in Figure \ref{fig:E2N4sqn}. 
The only low degree differentials act by multiplication by $n$ and we can immediately read off the integer cohomology groups of $Q$ up to degree 9. For degree 10, 11 and 12 we need to determine whether there are non-trivial differentials $d_{11}: E_{11}^{0,10} \rightarrow E_{11}^{11,0}$, or $d_{13}:E^{0, 12}_{13}\rightarrow E^{13, 0}_{13}$. Beyond that, there is a non-trivial extension problem in degree 12, where one has
\begin{equation}
    0 \longrightarrow \mathbb{Z}_n \longrightarrow H^{12}(Q;\mathbb{Z}) \longrightarrow \mathbb{Z}_2 \oplus \mathbb{Z}_3 \longrightarrow 0 \,.
\end{equation}
As above, this can be non-trivial only in cases when $n$ is divisible by $2$ or $3$. To resolve these issues we once more can use the LSSS with $\mathbb{Z}_2$ and $\mathbb{Z}_3$ coefficients, see Figure~\ref{fig:n4sqnZ2coef} and Figure~\ref{fig:n4sqnZ3coef}.

In Figure~\ref{fig:n4sqnZ2coef} we show the second and the $\infty$-page. The $\mathbb{Z}_2$ version of the differential implementing the constraint is given by
\begin{equation}
    d \tilde{v}_7 = \mathfrak{m}_2 \big( n v_4^2\big) = n \, \tilde{v}_4^2 \,,
\end{equation}
where $\tilde{v}_i$ generates $H^i \big( K(\mathbb{Z},i); \mathbb{Z}_2\big) \simeq \mathbb{Z}_2$ and $v_i$ generates $H^i \big( K(\mathbb{Z},i); \mathbb{Z}\big) \simeq \mathbb{Z}$. All other differentials follow from the Steenrod algebra structure. In particular, one has with transgression map $\tau$,
\begin{equation}
    \tau \big( Sq^4 \tilde{v}_7 \big) = Sq^4 \big( \tau (\tilde{v}_7) \big) = n \, Sq^4 \big( \tilde{v}_4^2 \big) = n \big( Sq^2 \tilde{v}_4 \big)^2 \,,
\end{equation}
which is non-trivial for $n$ odd. This leads to the non-trivial differential indicated by the dashed arrow in Figure~\ref{fig:n4sqnZ2coef}. This allows us to obtain
\begin{equation}
    \begin{array}{c | c c c c c c c c c}
    k & 4 & 5 & 6 & 7 & 8 & 9 & 10 & 11 & 12 \\ \hline
    H^k(Q;\mathbb{Z}_2) & \mathbb{Z}_2 & 0 & \mathbb{Z}_2 & \mathbb{Z}_2 \oplus \mathbb{Z}_{(n,2)} & \mathbb{Z}_{(n,2)} & \mathbb{Z}_2 & \mathbb{Z}_2^{\oplus 3} & \mathbb{Z}_2^{\oplus 2} \oplus \mathbb{Z}_{(n,2)}^{\oplus 2} & \mathbb{Z}_2 \oplus \mathbb{Z}_{(n,2)}^{\oplus 2}
    \end{array}
\end{equation}
from the $\infty$-page with $\mathbb{Z}_2$ coefficients.

Finally, to resolve the extensions for the prime $p=3$ parts, we use the spectral sequence with $\mathbb{Z}_3$ coefficients, Figure~\ref{fig:n4sqnZ3coef}, together with the mod $3$ Steenrod algebra of the Eilenberg-MacLane space (remember that $P^1$ increases the degree by $4$), where we denote the generators of $H^i \big( K(\mathbb{Z},i); \mathbb{Z}_3 \big)$ by $v_i'$,
\begin{equation}
\begin{split}
    H^\bullet \big(K(\mathbb{Z},4); \mathbb{Z}_3\big) &= \mathbb{Z}_3[v'_4, P^1 v'_4, \beta P^1 v'_4, \dots ] \,, \\
    H^\bullet \big(K(\mathbb{Z}, 7); \mathbb{Z}_3 \big) &= \mathbb{Z}_3[v'_7, P^1 v'_7, \beta P^1 v'_7,\dots] \,.
\end{split}
\end{equation}
Further using $d_8 v'_7 = \tau(v_7') = \mathfrak{m}_3 (n v_4^2) = n (v_4')^2$ we find that all differentials are trivial for $n$ divisible by $3$ and otherwise use that Steenrod operations commute with transgression. So for example one has
\begin{equation}
\begin{split}
    d_{11} (P^1 v_7') &= \tau(P^1 v_7') = P^1 \big(\tau (v_7') \big) = n \, P^1 (v_4')^2 = 2n \, v_4' P^1 v_4'  \,, \\
    d_{12} (\beta P^1 v_7') &= 2n \, v_4' \beta P^1 v_4' \,,
\end{split}
\end{equation}
which are both non-trivial for $n$ not divisible by 3, and are indicated by the dashed lines in Figure~\ref{fig:n4sqnZ3coef}. With this we find
\begin{equation}
    \begin{array}{c | c c c c c c c c c}
    k & 4 & 5 & 6 & 7 & 8 & 9 & 10 & 11 & 12 \\ \hline
    H^k(Q;\mathbb{Z}_3) & \mathbb{Z}_3 & 0 & 0 & \mathbb{Z}_{(n,3)} & \mathbb{Z}_3 \oplus \mathbb{Z}_{(n,3)} & \mathbb{Z}_3 & 0 & \mathbb{Z}_{(n,3)}^{\oplus 2} & \mathbb{Z}_{(n,3)}^{\oplus 3}
    \end{array}
\end{equation}
from the $\infty$-page.

With this knowledge we can return to cohomology with integer coefficients, fixing extension problems with the application of the universal coefficient theorem. We obtain
\begin{equation}
    \begin{array}{c | c c c c c c c c c}
    k & 4 & 5 & 6 & 7 & 8 & 9 & 10 & 11 & 12 \\ \hline
    H^k(Q;\mathbb{Z}) & \mathbb{Z} & 0 & 0 & \mathbb{Z}_2 & \mathbb{Z}_n & \mathbb{Z}_3 & \mathbb{Z}_2 & \mathbb{Z}_2^{\oplus 2} & \mathbb{Z}_n \oplus \mathbb{Z}_{(n,2)} \oplus \mathbb{Z}_{(n,3)}
    \end{array}
\end{equation}
as the final result.

We see that the cohomology groups are sensitive to the prime factors of $n$, since some differentials become trivial in case $n$ is divisible by 2 or 3. However, this only happens in degree 12, and below we find a straightforward identification of characteristic classes of $Q$ and in terms of $K(\mathbb{Z},4)$ and $K(\mathbb{Z},7)$. Especially, we note that for $k \geq 2$, $v_4^k$ pulls back to the generator of the $\mathbb{Z}_n$ factor of $H^{4k}(Q;\mathbb{Z})$ under the projection, which is very similar what we saw for the modification of $c_1^2$ above.

\vspace{0.5cm}

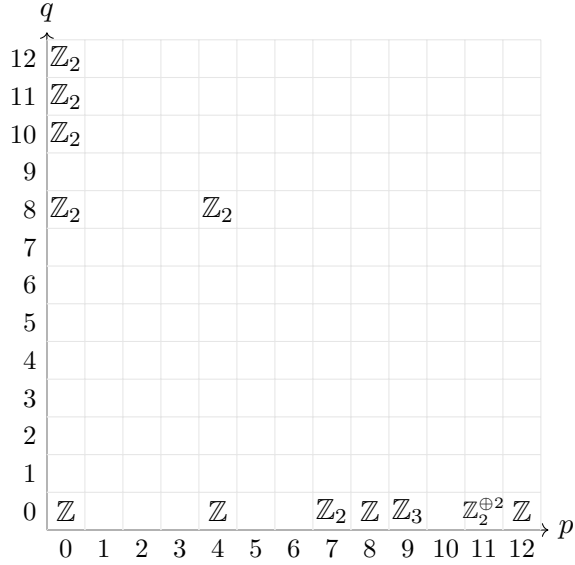
\begin{figure}
    \centering
    \begin{tikzpicture}[scale=0.5]
    % Axes
    \draw[->] (0,0) -- (13.2,0) node[right] {$p$};
    \draw[->] (0,0) -- (0,13.2) node[above] {$q$};
    % Grid (optional)
    \draw[step=1cm,gray!20] (0,0) grid (13,13);
    % Labels
    \foreach \x in {0,1,...,12}
        \node[below] at (\x+0.5, 0) {\small $\x$};
    \foreach \y in {0,1,..., 12}
        \node[left] at (0,\y+0.5) {\small $\y$};
\node(00) at (0.5, 0.5) {$\mathbb{Z}$};
\node(40) at (4.5, 0.5) {$\mathbb{Z}$};
\node(70) at (7.5, 0.5) {$\mathbb{Z}_2$};
\node(80) at (8.5, 0.5) {$\mathbb{Z}$};
\node(90) at (9.5,0.5) {$\mathbb{Z}_3$};
\node(110) at (11.5, 0.5) {\footnotesize{$\mathbb{Z}_2^{\oplus2}$}};
\node(120) at (12.5, 0.5) {$\mathbb{Z}$};
\node(08) at (0.5, 8.5) {$\mathbb{Z}_2$};
\node(48) at (4.5, 8.5) {$\mathbb{Z}_2$};
\node(010) at (0.5, 10.5) {$\mathbb{Z}_2$};
\node(011) at (0.5, 11.5) {$\mathbb{Z}_2$};
\node(012) at (0.5, 12.5) {$\mathbb{Z}_{2}$};
\end{tikzpicture}
    \caption{$E_2^{p,q}$ for $Q$ implementing $\mathfrak{m}_2([N^{(4)}]^2)=0$, for $p+q \leq 12$.}
    \label{fig:m2nsqfor3formfield}
\end{figure}

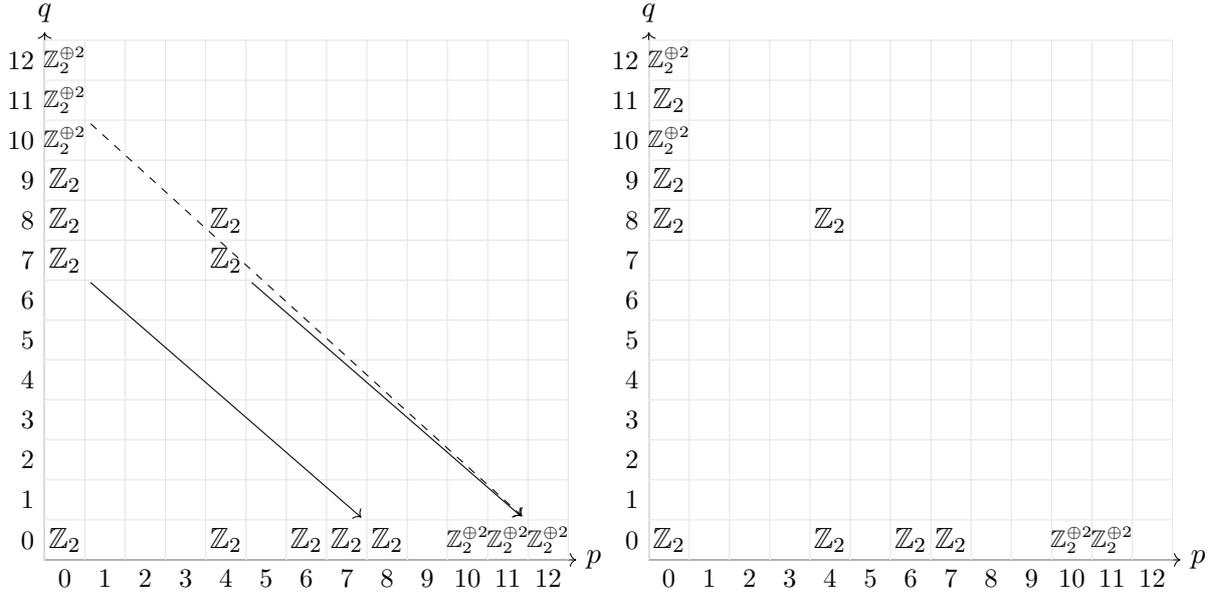
\begin{figure}
\centering
\begin{tikzpicture}[scale=0.53]
    % Axes
    \draw[->] (0,0) -- (13.2,0) node[right] {$p$};
    \draw[->] (0,0) -- (0,13.2) node[above] {$q$};
    % Grid (optional)
    \draw[step=1cm,gray!20] (0,0) grid (13,13);
    % Labels
    \foreach \x in {0,1,...,12}
        \node[below] at (\x+0.5, 0) {\small $\x$};
    \foreach \y in {0,1,..., 12}
        \node[left] at (0,\y+0.5) {\small $\y$};
\node(00) at (0.5, 0.5) {$\mathbb{Z}_2$};
\node(07) at (0.5, 7.5) {$\mathbb{Z}_2$};
\node(08) at (0.5, 8.5) {$\mathbb{Z}_2$};
\node(09) at (0.5, 9.5) {$\mathbb{Z}_2$};
\node(010) at (0.5, 10.5) {\footnotesize{$\mathbb{Z}_2^{\oplus 2}$}};
\node(011) at (0.5, 11.5) {\footnotesize{$\mathbb{Z}_2^{\oplus 2}$}};
\node(012) at (0.5, 12.5) {\footnotesize{$\mathbb{Z}_2^{\oplus 2}$}};
\node(40) at (4.5, 0.5) {$\mathbb{Z}_2$};
\node(47) at (4.5, 7.5) {$\mathbb{Z}_2$};
\node(48) at (4.5, 8.5) {$\mathbb{Z}_2$};
\node(60) at (6.5, 0.5) {$\mathbb{Z}_2$};
\node(70) at (7.5, 0.5) {$\mathbb{Z}_2$};
\node(80) at (8.5, 0.5) {$\mathbb{Z}_2$};
\node(100) at (10.5, 0.5) {\footnotesize{$\mathbb{Z}_2^{\oplus 2}$}};
\node(110) at (11.5, 0.5) {\footnotesize{$\mathbb{Z}_2^{\oplus 2}$}};
\node(120) at (12.5, 0.5) {\footnotesize{$\mathbb{Z}_2^{\oplus 2}$}};
\draw[->] (07)--(80);
\draw[->] (47)--(120);
\draw[->, dashed] (011)--(120);
%
%
    % Axes
    \draw[->] (15,0) -- (28.2,0) node[right] {$p$};
    \draw[->] (15,0) -- (15,13.2) node[above] {$q$};
    % Grid (optional)
    \draw[step=1cm,gray!20] (15,0) grid (28,13);
    % Labels
    \foreach \x in {0,1,...,12}
        \node[below] at (\x+15.5, 0) {\small $\x$};
    \foreach \y in {0,1,..., 12}
        \node[left] at (15,\y+0.5) {\small $\y$};
\node at (15.5, 0.5) {$\mathbb{Z}_2$};
\node at (15.5, 8.5) {$\mathbb{Z}_2$};
\node at (15.5, 9.5) {$\mathbb{Z}_2$};
\node at (15.5, 10.5) {\footnotesize{$\mathbb{Z}_2^{\oplus 2}$}};
\node at (15.5, 11.5) {$\mathbb{Z}_2$};
\node at (15.5, 12.5) {\footnotesize{$\mathbb{Z}_2^{\oplus 2}$}};
\node at (19.5, 0.5) {$\mathbb{Z}_2$};
\node at (19.5, 8.5) {$\mathbb{Z}_2$};
\node at (21.5, 0.5) {$\mathbb{Z}_2$};
\node at (22.5, 0.5) {$\mathbb{Z}_2$};
\node at (25.5, 0.5) {\footnotesize{$\mathbb{Z}_2^{\oplus 2}$}};
\node at (26.5, 0.5) {\footnotesize{$\mathbb{Z}_2^{\oplus 2}$}};
\end{tikzpicture}
\caption{$E_2^{p,q}$ and $E^{p,q}_{\infty}$ of $Q$ implementing $\mathfrak{m}_2([N^{(4)}]^2) = 0$ with $\mathbb{Z}_2$ coefficients and $p+q \leq 12$.}
\label{fig:n4sqmod2Z2coef}
\end{figure}

\noindent Now we consider the mod $2$ constraint of the 3-form gauge field.\footnote{The case of $n$ odd prime can be treated analogously using the mod-$p$ Steenrod algebra.} We have the homotopy fiber $Q$ described by
\begin{equation}
    Q \longrightarrow K(\mathbb{Z}, 4)\longrightarrow  K(\mathbb{Z}_2, 8) \,,
\end{equation}
where $u_8$, the generator of $H^8 \big( K(\mathbb{Z}_2,8); \mathbb{Z}\big)$, pulls back to $\mathfrak{m}_2 (v_4^2) \in H^8 \big( K(\mathbb{Z},4); \mathbb{Z}_2\big)$. Extending to the left, leads to the more useful fibration
\begin{equation}
   K(\mathbb{Z}_2, 7) \longrightarrow  Q \longrightarrow K(\mathbb{Z}, 4) \,.
\end{equation}
With the cohomology groups summarized in Appendix~\ref{app:EMachom}, this fibration leads to second page of the LSSS indicated in Figure~\ref{fig:m2nsqfor3formfield}.
To determine differentials and extensions we can use the same approach as described for the other cases above, in particular, the switch to $\mathbb{Z}_2$ coefficients, see Figure~\ref{fig:n4sqmod2Z2coef}. From this we can read off the cohomology groups with $\mathbb{Z}_2$ coefficients:
\begin{equation}
    \begin{array}{c | c c c c c c c c c }
    k & 4 & 5 & 6 & 7 & 8 & 9 & 10 & 11 & 12 \\ \hline
    H^k(Q;\mathbb{Z}_2) & \mathbb{Z}_2 & 0 & \mathbb{Z}_2 & \mathbb{Z}_2 & \mathbb{Z}_2 & \mathbb{Z}_2 & \mathbb{Z}_2^{\oplus 4} & \mathbb{Z}_2^{\oplus 3} & \mathbb{Z}_2^{\oplus 3}
    \end{array}
\end{equation}
This helps to determine the extensions and one finds:
\begin{equation}
    \begin{array}{c | c c c c c c c c c }
    k & 4 & 5 & 6 & 7 & 8 & 9 & 10 & 11 & 12 \\ \hline
    H^k(Q;\mathbb{Z}) & \mathbb{Z} & 0 & 0 & \mathbb{Z}_2 & \mathbb{Z} & \mathbb{Z}_3 & \mathbb{Z}_2 & \mathbb{Z}_2^{\oplus 3} & \mathbb{Z}
    \end{array}
\end{equation}
a rather rich structure for the cohomology groups with integer coefficients, which indicates a non-trivial extension for the $\mathbb{Z}$ factor in degree 8 and 12.

This non-trivial extension also helps for the physical interpretation of the results, by pulling back classes from the base $K(\mathbb{Z};4)$ to the homotopy fiber $Q$, using the associated projection map $p$. For example we have
\begin{equation}
p^*(v_4) = \alpha_4 \,, \quad p^* (v_4^2) = 2 \alpha_8 \,, 
\end{equation}
where $\alpha_{4i}$ are generators of the $\mathbb{Z}$ factor of $H^{4i} (Q;\mathbb{Z})$. The specific realization of $p^* (v_4^3)$ requires a full knowledge of the differentials from elements in $E_j^{13-r,r}$, which goes beyond the our scope here. From the physics interpretation we would expect $p^*(v_4^3) = 2 \alpha_{12}$, inherited from $p^*(v_4^2)$. If this is indeed correct it does require the existence of a non-trivial differential.

\subsection{Multiple constraints}
\label{subsec:multconst}

In this section we want to explore what happens if we implement more than one constraint. This can be done by consecutive constructions of homotopy fibers in the following sense. Let us start with $BG$ on which we want to restrict two specific topological classes of the gauge bundle via the homotopy fiber construction. Then we implement the first constraint via the fibration
\begin{equation}
    Q_1 \longrightarrow BG \longrightarrow Y_1 \,,
\end{equation}
after which we can implement the second constraint on $Q_1$:
\begin{equation}
    Q \longrightarrow Q_1 \longrightarrow Y_2 \,.
\end{equation}
Equivalently, we can switch the order of implementation with the two fibers given by
\begin{equation}
\begin{split}
    Q_2 & \longrightarrow BG \longrightarrow Y_2 \,, \\
    \widetilde{Q} & \longrightarrow Q_2 \longrightarrow Y_1 \,.
\end{split}
\end{equation}
That these two procedures lead to the same result, i.e., that $Q$ is homotopy equivalent to $\widetilde{Q}$, can be seen by combining this two-step process into a single fibration. Suppose we aim to trivialize the topological classes $\alpha \in H^n(BG; A)$ and $\beta \in  H^m(BG; B)$, represented by maps labeled by the same name
\begin{equation}
\begin{split}
    \alpha:& \quad BG \rightarrow K(A, n) = Y_1 \,, \\
    \beta:& \quad BG \rightarrow K(B, m) = Y_2 \,.
\end{split}
\end{equation}
In the first step we construct a fibration $Q_1 \xrightarrow {p_1} BG $ for trivializing the class $\alpha$ and then, in order to further trivialize the class $\beta$ in $BG$, we implement the homotopy fiber for the class represented by $\pi = \beta \circ p_1$ in $H^m(Q_1; B)$ that leads to the fibration $Q \xrightarrow {p} Q_1$ as follows
 \begin{equation}\label{eq:doublefiberdiag}
\begin{tikzcd}
K(B, m-1) \arrow[r] & Q \arrow[d, "p"]  \\
 K(A, n-1) \arrow[r, "i"] & Q_1 \arrow[r, "\pi"]\arrow[d, "p_1"] & K(B,m)    \\
 & BG \arrow[r, "\alpha"] \arrow [ur, "\beta"]& K(A, n) \,.
\end{tikzcd}
\end{equation}
We can combine this sequence of fibrations into a single fibration $Q \xrightarrow {p_1\circ p} BG $ with fiber $Z$ given by a fibration of $K(B, m-1)$ over $K(A, n-1)$ twisted by a class $i^*\pi\in H^m(K(A, n-1); B)$
\begin{equation}
\begin{tikzcd}
 K(B, m-1) \arrow[r] & Z \arrow[d]  \\
 & K(A, n-1) \arrow[r, "i^*\pi"] & K(B, m) \,.
\end{tikzcd}
\end{equation}
Since from the diagram \ref{eq:doublefiberdiag} one sees that $\pi\circ i\simeq \beta\circ p_1 \circ i$ is homotopy trivial since $p_1\circ i$ is homotopy trivial as maps in a fibration, the class that measures the twisting in the fibration above $i^* \pi = 0$ is trivial. This means that $Z$ is homotopy equivalent to a direct product leading to
\begin{equation}
\begin{tikzcd}
K(A, n-1)\times K(B, m-1) \arrow[r] & Q \arrow[d]  \\
 & BG \arrow[r, "\alpha\times \beta"] & K(A, n) \times K(B, m) \,.
\end{tikzcd}
\end{equation}
Analogously, by doing the procedure in the opposite order, one would arrive at the same fibration leading to the conclusion that the order does not matter and we can even do it simultaneously, with the fiber given by the product $\Omega Y_1 \times \Omega Y_2$.

Note that the product structure of the fiber is due to the fact that we impose both constraints in the same space $BG$. Alternatively, one could impose one constraint on $BG$ and another, independent one, on the homotopy fiber $Q_1$. This new constraint can now involve classes that did not exist in $BG$, which for example originate from the fiber of $Q_1$. If this is the case, the result would not be a product space and there might be non-trivial twists. In the following, we will focus on the situation where we impose constraints on the same space $BG$.

\vspace{0.5cm}

\noindent Let us illustrate this in a simple example, where we impose two different constraints in various ways. For that we consider gauge theory with a U$(1)$ 1-form and impose the two constraints
\begin{equation}
\begin{split}
    &m c_1 = 0 \in H^2(X;\mathbb{Z}) \,, \\
    &\mathfrak{m}_n (c_1) = 0 \in H^2(X;\mathbb{Z}_n) \,,
\end{split}
\end{equation}
i.e., we simultaneously impose a `times $m$' and `mod $n$' constraint.

If we first construct the homotopy fiber $Q_1$ to trivialize $\mathfrak{m}_n(v_2)$ (as usual $v_2$ is the generator of $H^2\big( B \mathrm{U}(1); \mathbb{Z} \big)$) we end up with the space discussed in \ref{eq:fibrationforc_1modn}, for which we already computed the cohomology groups $H^{\bullet} (Q_1; \mathbb{Z}) \cong \mathbb{Z} [x]$, with $x \in H^2(Q_1;\mathbb{Z})$ a generator in degree two. Under the pull-back with respect to $p_1: Q_1 \rightarrow B \mathrm{U}(1)$ one has
\begin{equation}
    p_1^{\ast} (v_2) = n x \,.
\end{equation}
Then, imposing the constraint $m v_2 = 0 \in H^2\big(B\mathrm{U}(1); \mathbb{Z}\big)$ translates to the constraint
\begin{equation}
    p_1^{\ast} (m v_2) = m \, p_1^\ast (v_2) = m n x = 0 \in H^2(Q_1;\mathbb{Z}) \,,
\end{equation}
on cohomology classes on $Q_1$, defining the second homotopy fiber to be given by
\begin{equation}
    Q \rightarrow Q_1 \xrightarrow{\cdot mn} K(\mathbb{Z},2) \,,
\end{equation}
where $\cdot nm$ is understood in the same way as in \eqref{eq:fibrationfornc_1}. Using the result of the analysis there, we find,
\begin{equation}
    H^{k} (Q; \mathbb{Z}) =\begin{cases} \mathbb{Z} \,, \quad k = 0 \,, \\ \mathbb{Z}_{mn} \,, \quad k>0 \,, \text{even} \,, \\ 0 \,, \quad \text{otherwise} \,.  \end{cases}
\end{equation}
From the full construction we can follow the fate of the generator $v_2 \in H^2 \big( B \mathrm{U}(1); \mathbb{Z} \big)$ and its powers, through the pull-backs using the projection maps
\begin{equation}
    p^{\ast} \big( p_1^{\ast} (v_2^k) \big) = p^* \big(p_1^* (v_2)^k\big) = p^* (n^k x^k) = n^ky^k
\end{equation}
where the generator $y = p^*(x) \in H^2(Q;\mathbb{Z})$ satisfies $nm y =0$, and equivalently for its powers. Note that even in the case $m = n$ one does not fully trivialize $c_1$ since one has
\begin{equation}
    (p_1 \circ p)^* v_2 = n y \,,
\end{equation}
which generically is non-zero since the triviality condition for $n = m$ reads $n^2 y = 0$.\footnote{To get an intuition for that, consider a manifold with $H^2(X;\mathbb{Z}) = \mathbb{Z}_{n^2}$, then there are elements that are zero when multiplied by $n$ but are also $n$ times an element. In particular these are given by the $\mathbb{Z}_n \hookrightarrow \mathbb{Z}_{n^2}$ subgroup.}

Now let us implement the two constraints in different order. First, we construct the fiber $Q_2$ imposing $m c_1 = 0$, which we have already considered in \ref{eq:fibrationfornc_1}, and leading to
\begin{equation}
    H^k(Q_2;\mathbb{Z}) = \begin{cases} \mathbb{Z} \,, \quad k=0 \,, \\ \mathbb{Z}_m \,, \quad k>0 \,, \text{even} \,, \\ 0 \quad \text{otherwise} \,.\end{cases}
\end{equation}
We denote the generator of $H^2(Q_2;\mathbb{Z}) \simeq \mathbb{Z}_m$ by $x_2$, whose powers also generate $H^{2i}(Q;\mathbb{Z})$ and satisfies
\begin{equation}
    p_2^* (v_2) = x_2 \,.
\end{equation}
The second constraint $\mathfrak{m}_n (v_2) = 0$ is lifted to
\begin{equation}
    \mathfrak{m}_n (x_2) = 0 \in H^2(Q_2;\mathbb{Z}) \,,
\end{equation}
in terms of cohomology classes of $Q_2$. As usual we implement that by the homotopy fiber construction obtaining
\begin{equation}
    K(\mathbb{Z}_n,1) \longrightarrow \widetilde{Q} \longrightarrow Q_2 \,,
\end{equation}
where $\widetilde{Q}$ implements both constraints. Let us determine the LSSS with $\mathbb{Z}_n$ coefficients, which leads to the second page 
\begin{equation}
    E_2^{p,q} = H^p \big( Q_2; H^q \big(K(\mathbb{Z}_n,1);\mathbb{Z}_n\big)\big)\implies H^{p+q}(\widetilde{Q};\mathbb{Z}_n) \,,
\end{equation}
in Figure~\ref{fig:E2andEinfcc1modnaftermc1}. Combining this with the spectral sequence with integer coefficients, Figure~\ref{fig:E2andEinfcc1modnaftermc1integral}, one deduces that
\begin{equation}
    H^k (\widetilde{Q};\mathbb{Z}_n) = \mathbb{Z}_n \,, \quad k \geq 0 \,,
\end{equation}
due to non-trivial extensions. This in turn determines the cohomology with integer coefficients
\begin{equation}
    H^k(\widetilde{Q};\mathbb{Z}) = \begin{cases} \mathbb{Z} \,, \quad k = 0 \,, \\ \mathbb{Z}_{mn} \,, \quad k>0 \,, \text{even} \,, \\ 0 \,, \quad \text{otherwise} \,, \end{cases}
\end{equation}
as necessary for $Q \simeq \widetilde{Q}$.
\begin{figure}
\centering
\begin{tikzpicture}[scale=0.65]
    % Axes
    \draw[->] (0,0) -- (6.2,0) node[right] {$p$};
    \draw[->] (0,0) -- (0,6.2) node[above] {$q$};
    % Grid (optional)
    \draw[step=1cm,gray!20] (0,0) grid (6,6);
    % Labels
    \foreach \x in {0,1,...,5}
        \node[below] at (\x+0.5, 0) {\small $\x$};
    \foreach \y in {0,1,..., 5}
        \node[left] at (0,\y+0.5) {\small $\y$};
\node(00) at (0.5, 0.5) {$\mathbb{Z}_n$};
\node(10) at (1.5, 0.5) {$\mathbb{Z}_d$};
\node(20) at (2.5, 0.5) {$\mathbb{Z}_d$};
\node(30) at (3.5, 0.5) {$\mathbb{Z}_d$};
\node(40) at (4.5, 0.5) {$\mathbb{Z}_d$};
\node(50) at (5.5, 0.5) {$\mathbb{Z}_d$};
\node(01) at (0.5, 1.5) {$\mathbb{Z}_n$};
\node(11) at (1.5, 1.5) {$\mathbb{Z}_d$};
\node(21) at (2.5, 1.5) {$\mathbb{Z}_d$};
\node(31) at (3.5, 1.5) {$\mathbb{Z}_d$};
\node(41) at (4.5, 1.5) {$\mathbb{Z}_d$};
\node(51) at (5.5, 1.5) {$\mathbb{Z}_d$};
\node(02) at (0.5, 2.5) {$\mathbb{Z}_n$};
\node(12) at (1.5, 2.5) {$\mathbb{Z}_d$};
\node(22) at (2.5, 2.5) {$\mathbb{Z}_d$};
\node(32) at (3.5, 2.5) {$\mathbb{Z}_d$};
\node(42) at (4.5, 2.5) {$\mathbb{Z}_d$};
\node(52) at (5.5, 2.5) {$\mathbb{Z}_d$};
\node(03) at (0.5, 3.5) {$\mathbb{Z}_n$};
\node(13) at (1.5, 3.5) {$\mathbb{Z}_d$};
\node(23) at (2.5, 3.5) {$\mathbb{Z}_d$};
\node(33) at (3.5, 3.5) {$\mathbb{Z}_d$};
\node(43) at (4.5, 3.5) {$\mathbb{Z}_d$};
\node(53) at (5.5, 3.5) {$\mathbb{Z}_d$};
\node(04) at (0.5, 4.5) {$\mathbb{Z}_n$};
\node(14) at (1.5, 4.5) {$\mathbb{Z}_d$};
\node(24) at (2.5, 4.5) {$\mathbb{Z}_d$};
\node(34) at (3.5, 4.5) {$\mathbb{Z}_d$};
\node(44) at (4.5, 4.5) {$\mathbb{Z}_d$};
\node(54) at (5.5, 4.5) {$\mathbb{Z}_d$};
\node(05) at (0.5, 5.5) {$\mathbb{Z}_n$};
\node(15) at (1.5, 5.5) {$\mathbb{Z}_d$};
\node(25) at (2.5, 5.5) {$\mathbb{Z}_d$};
\node(35) at (3.5, 5.5) {$\mathbb{Z}_d$};
\node(45) at (4.5, 5.5) {$\mathbb{Z}_d$};
\node(55) at (5.5, 5.5) {$\mathbb{Z}_d$};
\node(60) at (6.5, 0.5) {};
\node(62) at (6.5, 2.5) {};
\node(64) at (6.5, 4.5) {};
\draw[->](01)-- node{} (20);
\draw[->](11)-- node{} (30);
\draw[->](21)-- node{} (40);
\draw[->](31)-- node{} (50);
\draw[->](03)-- node{} (22);
\draw[->](13)-- node{} (32);
\draw[->](23)-- node{} (42);
\draw[->](33)-- node{} (52);
\draw[->](05)-- node{} (24);
\draw[->](15)-- node{} (34);
\draw[->](25)-- node{} (44);
\draw[->](35)-- node{} (54);
\draw[->] (8,0) -- (14.2,0) node[right] {$p$};
\draw[->] (8,0) -- (8,6.2) node[above] {$q$};2
    % Grid (optional)
    \draw[step=1cm,gray!20] (8,0) grid (14,6);
    % Labels
    \foreach \x in {0,1,...,5}
        \node[below] at (\x+8.5, 0) {\small $\x$};
    \foreach \y in {0,1,..., 5}
        \node[left] at (0,\y+0.5) {\small $\y$};
\node(80) at (8.5, 0.5) {$\mathbb{Z}_{n}$};
\node(82) at (8.5, 1.5) {$\mathbb{Z}_{\frac{n}{d}}$};
\node(82) at (8.5, 2.5) {$\mathbb{Z}_{n}$};
\node(82) at (8.5, 3.5) {$\mathbb{Z}_{\frac{n}{d}}$};
\node(84) at (8.5, 4.5) {$\mathbb{Z}_{n}$};
\node(82) at (8.5, 5.5) {$\mathbb{Z}_{\frac{n}{d}}$};
\node(80) at (9.5, 0.5) {$\mathbb{Z}_d$};
\node(82) at (9.5, 2.5) {$\mathbb{Z}_d$};
\node(84) at (9.5, 4.5) {$\mathbb{Z}_d$};
\end{tikzpicture}
\caption{$E^{p,q}_2$ and $E^{p,q}_{\infty}$ of $\widetilde{Q}$ implementing $\mathfrak{m}_n(c_1) = 0$ after $m c_1 = 0$ with $\mathbb{Z}_n$ coefficients ($d=(n,m)$).}
\label{fig:E2andEinfcc1modnaftermc1}
\end{figure}
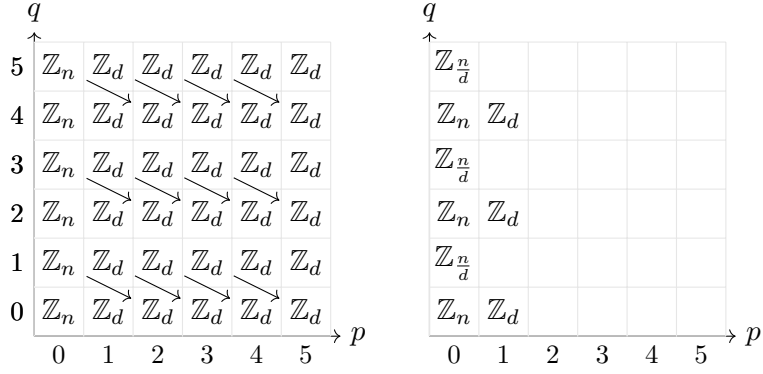
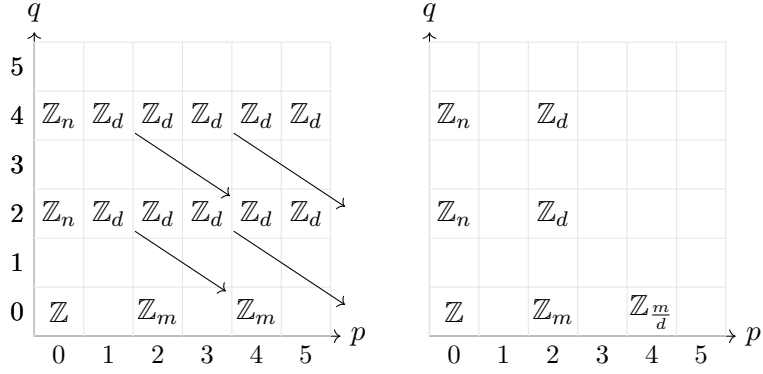
\begin{figure}
\centering
\begin{tikzpicture}[scale=0.65]
    % Axes
    \draw[->] (0,0) -- (6.2,0) node[right] {$p$};
    \draw[->] (0,0) -- (0,6.2) node[above] {$q$};
    % Grid (optional)
    \draw[step=1cm,gray!20] (0,0) grid (6,6);
    % Labels
    \foreach \x in {0,1,...,5}
        \node[below] at (\x+0.5, 0) {\small $\x$};
    \foreach \y in {0,1,..., 5}
        \node[left] at (0,\y+0.5) {\small $\y$};
\node(00) at (0.5, 0.5) {$\mathbb{Z}$};
\node(20) at (2.5, 0.5) {$\mathbb{Z}_m$};
\node(40) at (4.5, 0.5) {$\mathbb{Z}_m$};
\node(02) at (0.5, 2.5) {$\mathbb{Z}_n$};
\node(12) at (1.5, 2.5) {$\mathbb{Z}_d$};
\node(22) at (2.5, 2.5) {$\mathbb{Z}_d$};
\node(32) at (3.5, 2.5) {$\mathbb{Z}_d$};
\node(42) at (4.5, 2.5) {$\mathbb{Z}_d$};
\node(52) at (5.5, 2.5) {$\mathbb{Z}_d$};
\node(04) at (0.5, 4.5) {$\mathbb{Z}_n$};
\node(14) at (1.5, 4.5) {$\mathbb{Z}_d$};
\node(24) at (2.5, 4.5) {$\mathbb{Z}_d$};
\node(34) at (3.5, 4.5) {$\mathbb{Z}_d$};
\node(44) at (4.5, 4.5) {$\mathbb{Z}_d$};
\node(54) at (5.5, 4.5) {$\mathbb{Z}_d$};
\node(60) at (6.5, 0.5) {};
\node(62) at (6.5, 2.5) {};
\node(64) at (6.5, 4.5) {};
\draw[->](12)-- node{} (40);
\draw[->](32)-- node{} (60);
\draw[->](14)-- node{} (42);
\draw[->](34)-- node{} (62);
\draw[->] (8,0) -- (14.2,0) node[right] {$p$};
\draw[->] (8,0) -- (8,6.2) node[above] {$q$};2
    % Grid (optional)
    \draw[step=1cm,gray!20] (8,0) grid (14,6);
    % Labels
    \foreach \x in {0,1,...,5}
        \node[below] at (\x+8.5, 0) {\small $\x$};
    \foreach \y in {0,1,..., 5}
        \node[left] at (0,\y+0.5) {\small $\y$};
\node(80) at (8.5, 0.5) {$\mathbb{Z}$};
\node(82) at (8.5, 2.5) {$\mathbb{Z}_{n}$};
\node(84) at (8.5, 4.5) {$\mathbb{Z}_{n}$};
\node(100) at (10.5, 0.5) {$\mathbb{Z}_m$};
\node(102) at (10.5, 2.5) {$\mathbb{Z}_d$};
\node(102) at (10.5, 4.5) {$\mathbb{Z}_d$};
\node(102) at (12.5, 0.5) {$\mathbb{Z}_{\frac{m}{d}}$};
\end{tikzpicture}
\caption{$E^{p,q}_2$ (left) and $E^{p,q}_{\infty}$ (right) page of $Q$ implementing $\mathfrak{m}_n(c_1) = 0$ after $mc_1=0$ with $\mathbb{Z}$ coefficients ($d = (n,m)$).}
\label{fig:E2andEinfcc1modnaftermc1integral}
\end{figure}

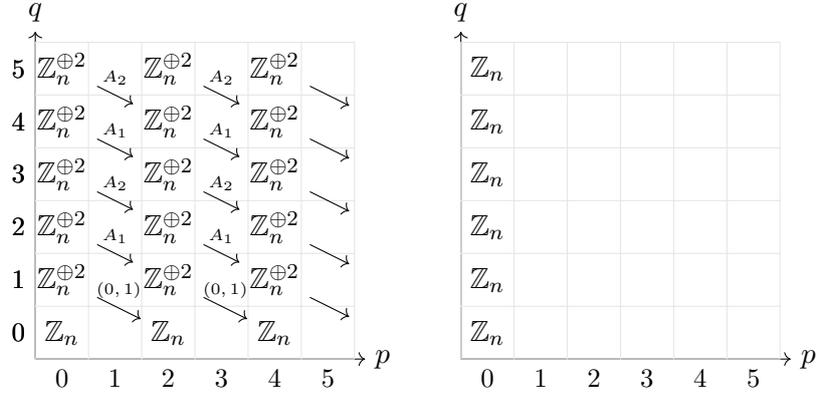
\begin{figure}
\centering
\begin{tikzpicture}[scale=0.7]
    % Axes
    \draw[->] (0,0) -- (6.2,0) node[right] {$p$};
    \draw[->] (0,0) -- (0,6.2) node[above] {$q$};
    % Grid (optional)
    \draw[step=1cm,gray!20] (0,0) grid (6,6);
    % Labels
    \foreach \x in {0,1,...,5}
        \node[below] at (\x+0.5, 0) {\small $\x$};
    \foreach \y in {0,1,..., 5}
        \node[left] at (0,\y+0.5) {\small $\y$};
\node(00) at (0.5, 0.5) {$\mathbb{Z}_n$};
\node(20) at (2.5, 0.5) {$\mathbb{Z}_n$};
\node(40) at (4.5, 0.5) {$\mathbb{Z}_n$};
\node(01) at (0.5, 1.5) {$\mathbb{Z}_n^{\oplus 2}$};
\node(21) at (2.5, 1.5) {$\mathbb{Z}_n^{\oplus 2}$};
\node(41) at (4.5, 1.5) {$\mathbb{Z}_n^{\oplus 2}$};
\node(02) at (0.5, 2.5) {$\mathbb{Z}_n^{\oplus 2}$};
\node(22) at (2.5, 2.5) {$\mathbb{Z}_n^{\oplus 2}$};
\node(42) at (4.5, 2.5) {$\mathbb{Z}_n^{\oplus 2}$};
\node(03) at (0.5, 3.5) {$\mathbb{Z}_n^{\oplus 2}$};
\node(23) at (2.5, 3.5) {$\mathbb{Z}_n^{\oplus 2}$};
\node(43) at (4.5, 3.5) {$\mathbb{Z}_n^{\oplus 2}$};
\node(04) at (0.5, 4.5) {$\mathbb{Z}_n^{\oplus 2}$};
\node(24) at (2.5, 4.5) {$\mathbb{Z}_n^{\oplus 2}$};
\node(44) at (4.5, 4.5) {$\mathbb{Z}_n^{\oplus 2}$};
\node(05) at (0.5, 5.5) {$\mathbb{Z}_n^{\oplus 2}$};
\node(25) at (2.5, 5.5) {$\mathbb{Z}_n^{\oplus 2}$};
\node(45) at (4.5, 5.5) {$\mathbb{Z}_n^{\oplus 2}$};
\node(60) at (6.5, 0.5) {};
\node(61) at (6.5, 1.5) {};
\node(62) at (6.5, 2.5) {};
\node(63) at (6.5, 3.5) {};
\node(64) at (6.5, 4.5) {};
\draw[->](01)-- node[above]{\tiny$(0,1)$} (20);
\draw[->](21)-- node[above]{\tiny$(0,1)$} (40);
\draw[->](41)-- (5.9,0.8);
\draw[->](02)-- node[above]{\tiny$A_1$} (21);
\draw[->](22)-- node[above]{\tiny$A_1$} (41);
\draw[->](42)-- (5.9,1.8);
\draw[->](03)-- node[above]{\tiny$A_2$} (22);
\draw[->](23)-- node[above]{\tiny$A_2$} (42);
\draw[->](43)-- (5.9,2.8);
\draw[->](04)-- node[above]{\tiny$A_1$} (23);
\draw[->](24)-- node[above]{\tiny$A_1$} (43);
\draw[->](44)-- (5.9,3.8);
\draw[->](05)-- node[above]{\tiny$A_2$} (24);
\draw[->](25)-- node[above]{\tiny$A_2$} (44);
\draw[->](45)-- (5.9,4.8);
\draw[->] (8,0) -- (14.2,0) node[right] {$p$};
\draw[->] (8,0) -- (8,6.2) node[above] {$q$};2
    % Grid (optional)
    \draw[step=1cm,gray!20] (8,0) grid (14,6);
    % Labels
    \foreach \x in {0,1,...,5}
        \node[below] at (\x+8.5, 0) {\small $\x$};
    \foreach \y in {0,1,..., 5}
        \node[left] at (0,\y+0.5) {\small $\y$};
\node(80) at (8.5, 0.5) {$\mathbb{Z}_{n}$};
\node(81) at (8.5, 1.5) {$\mathbb{Z}_{n}$};
\node(82) at (8.5, 2.5) {$\mathbb{Z}_{n}$};
\node(83) at (8.5, 3.5) {$\mathbb{Z}_{n}$};
\node(84) at (8.5, 4.5) {$\mathbb{Z}_{n}$};
\node(85) at (8.5, 5.5) {$\mathbb{Z}_{n}$};
\end{tikzpicture}
\caption{$E^{p,q}_2$ and $E^{p,q}_{\infty}$ of $Q$ implementing $\mathfrak{m}_n(c_1) = 0$ and $nc_1 = 0$ simultaneously, with $\mathbb{Z}_n$ coefficients.}
\label{fig:E2andEinfsimultwithZncoef}
\end{figure}

Finally, we implement both constraints simultaneously using
\begin{equation}
    Q \longrightarrow B \mathrm{U}(1) \xrightarrow{((m v_2 \,, \mathfrak{m}_n(v_2))} K(\mathbb{Z},2) \times K(\mathbb{Z}_n,2) \,,
\end{equation}
leading to a fiber given by a product of Eilenberg-MacLane spaces
\begin{equation}
    K(\mathbb{Z},1) \times K(\mathbb{Z}_n,1) \longrightarrow Q \overset{P}{\longrightarrow} B \mathrm{U}(1) \,.
\end{equation}
For the sake of simplicity we focus on the case $m = n$ and use K\"unneth theorem to obtain the cohomology groups of the fiber as well as its ring structure, from the two individual factors (e.g. \cite{BottTu1982})
\begin{equation}
    H^{\bullet} \big( K(\mathbb{Z},1) \times K(\mathbb{Z}_n, 1) ; A\big) = H^{\bullet} \big( K(\mathbb{Z},1); A\big) \otimes H^{\bullet} \big( K(\mathbb{Z}_n,1); A \big) \,,
\end{equation}
for a field $A$, e.g., one can choose $A = \mathbb{Z}_p$ with $p$ prime. Here, because cohomology of $K(\mathbb{Z},1) \simeq S^1$ is free, we can even use this for $\mathbb{Z}$ and $\mathbb{Z}_n$, since there are no extra terms.\footnote{In general one has, \cite{hatcher2002},
\begin{equation}
    0 \rightarrow \oplus_i \big( H_i (X;R) \otimes_R H_{k-i} (Y;R) \big) \rightarrow H_k (X \times Y;R) \rightarrow \oplus_i \text{Tor}_R \big( H_i (X;R) ; H_{k - i -1} (Y;R) \big) \rightarrow 0 \,,
\end{equation}
for a principle ideal domain $R$, which splits, and similar for cohomology.}
Starting with $\mathbb{Z}_n$ coefficients we find Figure~\ref{fig:E2andEinfsimultwithZncoef}, where the differentials are given in terms of the $(2 \times 2)$-matrices $A_i \in M_2 (\mathbb{Z}_n)$. Choosing the basis 
\begin{equation}
\{ \tilde{v}_1 \tilde{u}_j \,, \tilde{u}_{j+1} \} \in H^{j+1} \big( K(\mathbb{Z},1) \times K(\mathbb{Z}_n,1); \mathbb{Z}_n \big) \,,
\end{equation}
where $\tilde{v}_1$ generates $H^1\big( K(\mathbb{Z},1); \mathbb{Z}_n \big) \simeq \mathbb{Z}_n$ and $\tilde{u}_j$ generates $H^j \big( K(\mathbb{Z}_n,1); \mathbb{Z}_n\big)$, see also \eqref{eq:BZnringstrodd} and \eqref{eq:BZnringstreven}, these are given by\footnote{The elements for $p > 0$ are given by multiplication with the appropriate power of $\mathfrak{m}_n (v_2^k)$.}
\begin{equation}
A_1 = \begin{pmatrix} 1 & 0 \\ 0 & 0 \end{pmatrix}  \,, \quad A_2 = \begin{pmatrix} 0 & 0 \\ 0 & 1\end{pmatrix}  \,,
\end{equation}
encoding the differential $d_2 \tilde{u}_1 = \mathfrak{m}_n (v_2)$
From that it is apparent that
\begin{equation}
    H^k (Q; \mathbb{Z}) = \mathbb{Z}_n \,, \quad  k \geq 0 \,.
\end{equation}
Combining this with the LSSS with integer coefficients in Figure~\ref{fig:E2andEinfsimultwithZcoef}, one finds that there are non-trivial differentials $d_2$ as well as $d_4$. This leads to the final result
\begin{equation}
     H^k(Q;\mathbb{Z}) = \begin{cases} \mathbb{Z} \,, \quad k = 0 \,, \\ \mathbb{Z}_{n^2} \,, \quad k>0 \,, \text{even} \,, \\ 0 \,, \quad \text{otherwise} \,, \end{cases}
\end{equation}
precisely as expected in the case $n = m$. Moreover, pulling back under the projection map for the product fiber $P$, we can track the fate of the of the characteristic classes of $B\mathrm{U}(1)$
\begin{equation}
\begin{split}
    P^*(v_2) &= n x \in H^2(Q;\mathbb{Z}) \simeq \mathbb{Z}_{n^2} \,, \\
    P^*(v_2^k) &= n^k x^k = 0 \in H^{2k}(Q;\mathbb{Z}) \simeq \mathbb{Z}_{n^2}
\end{split}
\end{equation}
with $x$ generating $H^2(Q;\mathbb{Z})$. Again, this matches our expectations from the step-wise implementation for the special case $m = n$.
\begin{figure}
\centering
\begin{tikzpicture}[scale=0.65]
    % Axes
    \draw[->] (0,0) -- (6.2,0) node[right] {$p$};
    \draw[->] (0,0) -- (0,6.2) node[above] {$q$};
    % Grid (optional)
    \draw[step=1cm,gray!20] (0,0) grid (6,6);
    % Labels
    \foreach \x in {0,1,...,5}
        \node[below] at (\x+0.5, 0) {\small $\x$};
    \foreach \y in {0,1,..., 5}
        \node[left] at (0,\y+0.5) {\small $\y$};
\node(00) at (0.5, 0.5) {$\mathbb{Z}$};
\node(20) at (2.5, 0.5) {$\mathbb{Z}$};
\node(40) at (4.5, 0.5) {$\mathbb{Z}$};
\node(01) at (0.5, 1.5) {$\mathbb{Z}$};
\node(21) at (2.5, 1.5) {$\mathbb{Z}$};
\node(41) at (4.5, 1.5) {$\mathbb{Z}$};
\node(02) at (0.5, 2.5) {$\mathbb{Z}_n$};
\node(22) at (2.5, 2.5) {$\mathbb{Z}_n$};
\node(42) at (4.5, 2.5) {$\mathbb{Z}_n$};
\node(03) at (0.5, 3.5) {$\mathbb{Z}_n$};
\node(23) at (2.5, 3.5) {$\mathbb{Z}_n$};
\node(43) at (4.5, 3.5) {$\mathbb{Z}_n$};
\node(04) at (0.5, 4.5) {$\mathbb{Z}_n$};
\node(24) at (2.5, 4.5) {$\mathbb{Z}_n$};
\node(44) at (4.5, 4.5) {$\mathbb{Z}_n$};
\node(05) at (0.5, 5.5) {$\mathbb{Z}_n$};
\node(25) at (2.5, 5.5) {$\mathbb{Z}_n$};
\node(45) at (4.5, 5.5) {$\mathbb{Z}_n$};
\node(60) at (6.5, 0.5) {};
\node(61) at (6.5, 1.5) {};
\node(62) at (6.5, 2.5) {};
\node(63) at (6.5, 3.5) {};
\node(64) at (6.5, 4.5) {};
\draw[->](01)--(20);
\draw[->](21)--(40);
\draw(41)--(6,0.7);
\draw[->,dashed](03)-- node{} (40);
\draw[dashed](23)-- node{} (6,0.8);
\draw[->,dashed](05)-- node{} (42);
\draw[dashed](25)-- node{} (6,2.8);
\draw[->] (8,0) -- (14.2,0) node[right] {$p$};
\draw[->] (8,0) -- (8,6.2) node[above] {$q$};2
    % Grid (optional)
    \draw[step=1cm,gray!20] (8,0) grid (14,6);
    % Labels
    \foreach \x in {0,1,...,5}
        \node[below] at (\x+8.5, 0) {\small $\x$};
    \foreach \y in {0,1,..., 5}
        \node[left] at (0,\y+0.5) {\small $\y$};
\node(80) at (8.5, 0.5) {$\mathbb{Z}$};
\node(82) at (8.5, 2.5) {$\mathbb{Z}_{n}$};
\node(84) at (8.5, 4.5) {$\mathbb{Z}_{n}$};
\node(100) at (10.5, 0.5) {$\mathbb{Z}_n$};
\node(102) at (10.5, 2.5) {$\mathbb{Z}_{n}$};
\node(104) at (10.5, 4.5) {$\mathbb{Z}_{n}$};
\end{tikzpicture}
\caption{$E^{p,q}_2$ and $E^{p,q}_{\infty}$ of $Q$ implementing $\mathfrak{m}_n(c_1) = 0$ and $n c_1=0$ simultaneously with $\mathbb{Z}$ coefficients.}
\label{fig:E2andEinfsimultwithZcoef}
\end{figure}

\vspace{0.5cm}

\noindent We close this section with another illustrative and important example for the implementation of two constraints, now in different degrees. Assume, we want to modify a 1-form U$(1)$ gauge theory by
\begin{equation}
    n c_1 = 0 \quad \text{and} \quad n c_1^2 = 0 \,.
\end{equation}
Since the first condition implies the second, one would naively expect to obtain the same modified classifying space $Q_1$ as in \eqref{eq:fibrationfornc_1}, i.e., that only of the constraint $nc_1 = 0$. However, since we still introduce the fiber for the second constraint 
\begin{equation}
    K(\mathbb{Z},3) \longrightarrow Q \longrightarrow Q_1 \,,
\end{equation}
which is now trivially fibered, one finds the product space
\begin{equation}
    Q \simeq Q_1 \times K(\mathbb{Z},3) \,.
\end{equation}
Thus, the characteristic classes are enhanced to include those of an additional U$(1)$ 2-form gauge field, that does not appear in the Bianchi identities of the modified gauge theory. Using K\"unneth theorem, now with potential torsion \cite{hatcher2002}, we find
\begin{equation}
    \begin{array}{c | c c c c c c c c }
    k & 0 & 1 & 2 & 3 & 4 & 5 & 6 & 7 \\ \hline
    H^k(Q;\mathbb{Z}) & \mathbb{Z} & 0 & \mathbb{Z}_n & \mathbb{Z} & \mathbb{Z}_n & \mathbb{Z}_n & \mathbb{Z}_n \oplus \mathbb{Z}_2 & \mathbb{Z}_n \oplus \mathbb{Z}_{(n,2)}
    \end{array}
\end{equation}
as the cohomology groups with integer coefficient.

\section{Bordism groups and anomalies}
\label{sec:bordism}

The modification of the gauge theory can also have interesting and important consequences for the possible anomalies of the theory. In fact, excluding gravitational anomalies, the group cohomology of $Q$, that we determined above, is precisely what captures such an anomaly for bosonic theories, see, e.g., \cite{Kapustin:2014zva}. However, once we include diffeomorphisms as a symmetry\footnote{This is not yet necessarily gravity, for which we would also have to make the spacetime manifold dynamical and sum over different configurations in the path integral.} of the system anomalies are described by the Anderson dual of the associated bordism group, \cite{Freed:2016rqq, Yonekura:2018ufj} . The Anderson dual, $(I_{\mathbb{Z}} \Omega^{\xi})^{D+2} (Q)$, is given by the short exact sequence \cite{Hopkins:2002rd, Tachikawa:2021mby}
\begin{equation}
    0 \longrightarrow \mathrm{Ext} \big( \Omega^{\xi}_{D+1} (Q); \mathbb{Z} \big) \longrightarrow (I_{\mathbb{Z}} \Omega^{\xi})^{D+2} (Q) \longrightarrow \mathrm{Hom} \big( \Omega^{\xi}_{D+2} (Q); \mathbb{Z} \big) \longrightarrow 0 \,,
\end{equation}
where $D$ denotes the spacetime dimension of the theory of interest and $\xi$ its tangential structure, i.e., a restriction of the allowed spacetime manifolds. For a bosonic theory for example one can choose to work on oriented spacetime manifolds, $\xi = \mathrm{SO}$, in the presence of fermions one needs to include a Spin structure, $\xi = \text{Spin}$, or generalizations thereof. In the following we will focus on these two possibilities. We see that the relevant bordism groups are $\Omega^{\text{SO/Spin}}_{D+1}(Q)$ and $\Omega^{\text{SO/Spin}}_{D+2} (Q)$, which roughly capture the potential non-perturbative and perturbative anomalies, respectively. 

Since the bordism groups, as deformation classes of smooth manifolds with classifying maps into $Q$, depend on the modification of the gauge theory, also anomalies generically change. This becomes obvious when considering the Atiyah-Hirzebruch spectral sequence (AHSS) for oriented and Spin bordism, whose second page is given by
\begin{equation}
    E^2_{p,q} = H_p \big( Q; \Omega_q^{\text{SO/Spin}} (\text{pt})\big) \Rightarrow \Omega_{p+q}^{\text{SO/Spin}} (Q) \,.
\end{equation}
Since $Q$ has different (co)homology groups when compared to the original classifying space $BG$, also the bordism groups and hence the associated potential anomalies can be different. From the spectral sequence we further see that if $H_{\bullet}(Q;A)$ changes only in degree $k$ and above, the bordism groups can change earliest in degree $(k-1)$.

In the oriented case, there is a significant simplification due to the statement that  all differentials in the spectral sequence for oriented bordism groups have an image given by odd torsion groups and also extensions only affect these odd torsion pieces \cite{94db963e-8a07-3135-b0ec-dcdc8d13c27b, ConnerFloyd1964}, implying that:
\begin{equation}
    \Omega_{k}^{\text{SO}}(Q) \cong \oplus_p \, H_{p} \big(Q;\Omega_{k-p}^{\text{SO}}(\text{pt})\big) \quad \text{ modulo odd torsion} \,,
    \label{eq:Walltheorem}
\end{equation}
as a sum of the homology group of $Q$. This once more underlines the importance of the (co)homology groups in the evaluation of anomalies.

Before we perform some explicit calculations of the oriented and Spin bordism groups of some of the theories above, let us provide a very simple example: Consider U$(1)$ gauge theory with fermions in $D = 2$, which has a pure gauge anomaly captured by
\begin{equation}
    \tfrac{1}{2} c_1^2 \in \mathrm{Hom}\big( \Omega^{\text{Spin}}_4 \big(B\mathrm{U}(1)\big); \mathbb{Z} \big) \,.
\end{equation}
In fact this presents a generator, in that the minimal value (e.g., for $S^2 \times S^2$ with single unit of flux on each sphere) is given by $1$. Now let us implement the `mod $n$' constraint on $c_1$ with new classifying space $Q$. While it is still true that $\tfrac{1}{2} c_1^2$ is an element of $\mathrm{Hom}\big( \Omega^{\text{Spin}}_4 (Q); \mathbb{Z} \big)$ it is not a generator any more since we can divide it by $n^2$ and still obtain an integer, changing the anomaly.\footnote{A fermionic spectrum with anomaly for $B\mathrm{U}(1)$ would still be anomalous for $Q$, but due to the additional $n^2$ prefactor might be easier to cancel.} This easy example also suggests that it is interesting to compare the anomalies of the modified theory with the original one in order to be sensitive to such relative factors.

Now let us move to some more interesting examples.

\subsection{Bordism groups for modified \texorpdfstring{$c_1^2$}{square of c1}}

Let us focus on the bordism groups for Maxwell theory with a modification of $c_1^2$. 

\vspace{0.5cm}

\begin{figure}
    \centering
    \begin{tikzpicture}[scale=0.7]
    % Axes
    \draw[->] (0,0) -- (8.2,0) node[right] {$p$};
    \draw[->] (0,0) -- (0,8.2) node[above] {$q$};
    % Grid (optional)
    \draw[step=1cm,gray!20] (0,0) grid (8,8);
    % Labels
    \foreach \x in {0,1,...,7}
        \node[below] at (\x+0.5, 0) {\small $\x$};
    \foreach \y in {0,1,...,7}
        \node[left] at (0,\y+0.5) {\small $\y$};
\node at (0.5, 0.5) {$\mathbb{Z}$};
\node at (0.5, 4.5) {$\mathbb{Z}$};
\node at (0.5, 5.5) {$\mathbb{Z}_2$};
\node at (2.5, 0.5) {$\mathbb{Z}$};
\node at (2.5, 4.5) {$\mathbb{Z}$};
\node at (2.5, 5.5) {$\mathbb{Z}_2$};
\node at (3.5, 0.5) {$\mathbb{Z}_n$};
\node at (3.5, 4.5) {$\mathbb{Z}_n$};
\node at (5.5, 0.5) {$\kappa$};
\node at (7.5, 0.5) {\footnotesize{$\kappa \oplus \mathbb{Z}_3$}};
\end{tikzpicture}
    \caption{$E_2^{p,q}$ page for oriented bordism of $Q$ implementing $n c_1^2 = 0$ up to order $p+q\leq7$ ($\kappa$ denotes $\mathbb{Z}_n \oplus \mathbb{Z}_2$).}
    \label{fig:E2AHSSSOnc12}
\end{figure}
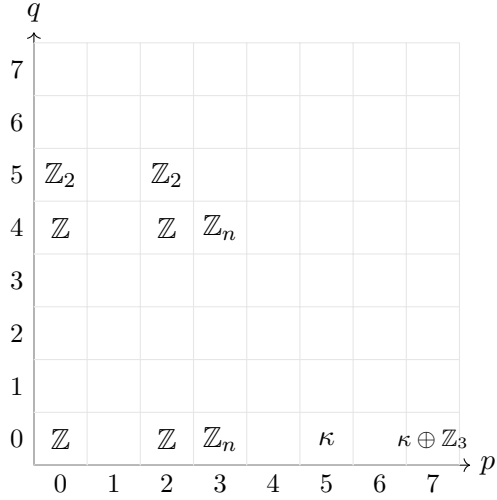

\noindent As above we start the discussion with the `times $n$' constraint, $n c_1^2 = 0$. With our knowledge of the integer homology groups of $Q$, and the homology groups with $\mathbb{Z}_2$ coefficients after application of the universal coefficient theorem (e.g., \cite{BottTu1982}), given by:
\begingroup
\renewcommand*{\arraystretch}{1.2}
\begin{equation}
    \begin{array}{c | c c }
    k & H_k(Q;\mathbb{Z}) & H_k (Q;\mathbb{Z}_2) \\ \hline
    0 & \mathbb{Z} & \mathbb{Z}_2 \\
    1 & 0 & 0 \\
    2 & \mathbb{Z} & \mathbb{Z}_2 \\
    3 & \mathbb{Z}_n & \mathbb{Z}_{(n,2)} \\ 
    4 & 0 & \mathbb{Z}_{(n,2)} \\
    5 & \mathbb{Z}_n \oplus \mathbb{Z}_2 & \mathbb{Z}_{(n,2)} \oplus \mathbb{Z}_2 \\
    6 & 0 & \mathbb{Z}_{(n,2)} \oplus \mathbb{Z}_2 \\
    7 & \mathbb{Z}_n \oplus \mathbb{Z}_3 \oplus \mathbb{Z}_2 & \mathbb{Z}_{(n,2)} \oplus \mathbb{Z}_2 \\
    8 & \mathbb{Z}_{(n,2)} & \mathbb{Z}_{(n,2)}^{\oplus 2} \oplus \mathbb{Z}_2 \\
    9 & \mathbb{Z}_n \oplus \mathbb{Z}_3 \oplus \mathbb{Z}_2 \oplus \mathbb{Z}_{(n,2)} & \mathbb{Z}_{(n,2)}^{\oplus 3} \oplus \mathbb{Z}_2
    \end{array}
\end{equation}
\endgroup
we can apply the Atiyah-Hirzebruch spectral sequence (see \cite{Garcia-Etxebarria:2018ajm} for an introduction) in order to obtain information about $\Omega^{\text{SO/Spin}}_{\bullet} (Q)$. To determine the second page we also need to provide the groups $\Omega^{\text{SO/Spin}}_{\bullet} (\text{pt})$, which up to degree 8 are given by
\begingroup
\renewcommand*{\arraystretch}{1.2}
\begin{equation}
    \begin{array}{c | c c c c c c c c c }
    k & 0 & 1 & 2 & 3 & 4 & 5 & 6 & 7 & 8 \\ \hline 
    \Omega^{\text{SO}}_k (\text{pt}) & \mathbb{Z} & 0 & 0 & 0 & \mathbb{Z} & \mathbb{Z}_2 & 0 & 0 & \mathbb{Z} \oplus \mathbb{Z} \\
    \Omega^{\text{Spin}}_k(\text{pt}) & \mathbb{Z} & \mathbb{Z}_2 & \mathbb{Z}_2 & 0 & \mathbb{Z} & 0 & 0 & 0 & \mathbb{Z} \oplus \mathbb{Z} 
    \end{array}
    \label{eq:bordismgroupspt}
\end{equation}
\endgroup
With the knowledge of the bordism groups of a point we can easily find the second page of the AHSS for $\Omega^{\mathrm{SO}}_\bullet (Q)$, see Figure~\ref{fig:E2AHSSSOnc12}. 
Below degree 7 we do not need to worry about differentials or extensions and find \footnote{One can argue that differentials in the higher degrees are also trivial since the image has to be odd, see \eqref{eq:Walltheorem}, but there could possibly be non-trivial extensions. For instance, in degree 7 in case $n$ is odd there is an extension problem of the form $0\rightarrow \mathbb{Z}_n\rightarrow E \subset \Omega^{\mathrm{SO}}_7 (Q) \rightarrow\mathbb{Z}_n\oplus\mathbb{Z}_3\rightarrow 0$.}
\begingroup
\renewcommand*{\arraystretch}{1.2}
\begin{equation}
    \begin{array}{c | c c c c c c c }
    k & 0 & 1 & 2 & 3 & 4 & 5 & 6 \\ \hline
    \Omega^{\mathrm{SO}}_k (Q) & \mathbb{Z} & 0 & \mathbb{Z} & \mathbb{Z}_n & \mathbb{Z} & \mathbb{Z}_n \oplus \mathbb{Z}_2 \oplus \mathbb{Z}_2 & \mathbb{Z} \\ 
    \Omega^{\mathrm{SO}}_k \big( B \mathrm{U}(1) \big) & \mathbb{Z} & 0 & \mathbb{Z} & 0 & \mathbb{Z} \oplus \mathbb{Z} & \mathbb{Z}_2 & \mathbb{Z} \oplus \mathbb{Z}
    \end{array}
\end{equation}
\endgroup
which clearly differs from $\Omega^{\mathrm{SO}}_{\bullet} \big( B \mathrm{U}(1) \big)$, which we also indicate. The topological classes that detect the $\mathbb{Z}_n$ factors in degree 3 and 5 are $\mathfrak{m}_n (c_1^2)$ and $\mathfrak{m}_n(c_1^3)$, the mod $n$ reduction of powers of the Chern class in cohomology \footnote{Note the usual shift in degree between torsion in homology and cohomology.}. The two factors of $\mathbb{Z}_2$ in degree 5 are detected by $w_2 \cup N^{(3)}$   and $w_2 \cup w_3$, where $w_i$ denotes the Stiefel-Whitney classes of the tangent bundle of $X$ and $N^{(3)}$ the pull-back of the fiber class $v_3 \in H^3\big( K(\mathbb{Z},3); \mathbb{Z}\big)$ of $Q$ to $X$. The $\mathbb{Z}$ factors in degree 4 and 6 are inherited from the unmodified U$(1)$ theory.

For example one generator of $\Omega^{\text{SO}}_6 \big( B \mathrm{U}(1) \big)$ in dimension 6 given by $\mathbb{CP}^3$, with $c_1$ given by the generator of $H^2(\mathbb{CP}^3; \mathbb{Z})$, does not survive the modification, since now one needs to satisfy
\begin{equation}
    n c_1^3 = c_1 (n c_1^2) = 0 \,,
\end{equation}
which would not be satisfied for $\mathbb{CP}^3$ with the indicated flux. Instead one only finds a remnant $\mathbb{Z}_n$ factor of the pure gauge anomaly that now shifts into $\Omega^{\text{SO}}_5 (Q)$, indicating that it becomes a non-perturbative anomaly. From its origin in the spectral sequence we see that it is associated to the element $H_5(Q;\mathbb{Z})$, which, since it is torsion, appears as part of $H^6(Q;\mathbb{Z})$, which can be identified with $\mathfrak{m}_n (c_1^3)$. From this logic we find, that a new generator can be chosen to be $L^5_n$, a five-dimensional lens space with a flat U$(1)$ gauge bundle classified by $c_1 \in H^2(L^5_n; \mathbb{Z}) \simeq \mathbb{Z}_n$. This version of an anomaly interplay, \cite{Davighi:2020bvi, Davighi:2020uab}, resembles that of a $\mathbb{Z}_n$ gauge anomaly\footnote{See \cite{Ibanez:1991hv, Garcia-Etxebarria:2018ajm, Hsieh:2018ifc, Monnier:2018nfs, Dierigl:2022zll, Dierigl:2025rfn, Cheng:2025ikd} for some of the works on $\mathbb{Z}_n$ anomalies and their cancellation.} with the difference that the mixed gauge gravitational piece is still $\mathbb{Z}$-valued and perturbative. Thus it seems that, at least when it comes to the description of topological sectors, the resulting theory has features of both a U$(1)$ and $\mathbb{Z}_n$ gauge theory. 

Another way to understand this modification is as an incomplete anomaly cancellation mechanism, in which not the full U$(1)$ pure gauge anomaly is canceled, but it is reduced to a remnant $\mathbb{Z}_n$. The mixed gauge gravitational anomaly on the other hand is unmodified, since it would require the homotopy fiber to be sensitive to gravitational characteristic classes, e.g., $p_1 \in H^4 (X;\mathbb{Z})$, and would rather be described by a twisted string structure (see, e.g., \cite{wang2008geometric, Sati:2008kz, Sati:2009ic, Basile:2023knk} and references therein).

Now let us move to the evaluation of $\Omega^{\text{Spin}}_{\bullet} (Q)$ for which the second page of the AHSS is depicted in Figure~\ref{fig:E2AHSSSpinnc12}.
\begin{figure}
    \centering
    \begin{tikzpicture}[scale=0.7]
    % Axes
    \draw[->] (0,0) -- (8.2,0) node[right] {$p$};
    \draw[->] (0,0) -- (0,8.2) node[above] {$q$};
    % Grid (optional)
    \draw[step=1cm,gray!20] (0,0) grid (8,8);
    % Labels
    \foreach \x in {0,1,...,7}
        \node[below] at (\x+0.5, 0) {\small $\x$};
    \foreach \y in {0,1,...,7}
        \node[left] at (0,\y+0.5) {\small $\y$};
\node at (0.5, 0.5) {$\mathbb{Z}$};
\node at (0.5, 1.5) {$\mathbb{Z}_2$};
\node at (0.5, 2.5) {$\mathbb{Z}_2$};
\node at (0.5, 4.5) {$\mathbb{Z}$};
%\node at (0.5, 8.5) {\footnotesize $\mathbb{Z}^{\oplus 2}$};
%\node at (0.5, 9.5) {\footnotesize $\mathbb{Z}_2^{\oplus 2}$};
\node at (2.5, 0.5) {$\mathbb{Z}$};
\node at (2.5, 1.5) {$\mathbb{Z}_2$};
\node(22) at (2.5, 2.5) {$\mathbb{Z}_2$};
\node at (2.5, 4.5) {$\mathbb{Z}$};
%\node at (2.5, 8.5) {\footnotesize $\mathbb{Z}^{\oplus 2}$};
\node at (3.5, 0.5) {$\mathbb{Z}_n$};
\node at (3.5, 1.5) {$\xi$};
\node at (3.5, 2.5) {$\xi$};
\node at (3.5, 4.5) {$\mathbb{Z}_n$};
\node(41) at (4.5, 1.5) {$\xi$};
\node at (4.5, 2.5) {$\xi$};
\node at (5.5, 0.5) {$\kappa$};
\node at (5.5, 1.5) {$\tilde{\kappa}$};
\node at (5.5, 2.5) {$\tilde{\kappa}$};
\node at (5.5, 4.5) {$\kappa$};
\node at (6.5, 1.5) {$\tilde{\kappa}$};
\node at (6.5, 2.5) {$\tilde{\kappa}$};
\node at (7.5, 0.5) {\footnotesize{$\kappa \oplus \mathbb{Z}_3$}};
%\draw[->] (41) -- (22);
\end{tikzpicture}
    \caption{$E_2^{p,q}$ for AHSS for Spin bordism of $Q$ implementing $n c_1^2 = 0$ up to order $p+q\leq7$ ($\xi$ denotes $\mathbb{Z}_{(n,2)}$, $\kappa$ denotes $\mathbb{Z}_n \oplus \mathbb{Z}_2$, and $\tilde{\kappa}$ denotes $\mathbb{Z}_{(n,2)} \oplus \mathbb{Z}_2$).}
    \label{fig:E2AHSSSpinnc12}
\end{figure}
Again, this looks very different from that of Maxwell theory see, e.g., \cite{Garcia-Etxebarria:2018ajm}. Let us focus on the anomalies for theories in dimension $D \leq 4$, for which we are interested in $\Omega^{\text{Spin}}_k (Q)$ with $k \leq 6$.

Since $H_{\bullet} (Q;A)$, for Abelian $A$, is modified earliest in degree three and there are no non-trivial differentials from the diagonal $p+q = 3$ the bordism groups with $k \in \{ 0 \,, 1\,, 2\}$ are unchanged and isomorphic to that of $B\mathrm{U}(1)$. 

The first change happens for $k =3$, where still the differentials are trivial but one encounters the extension problem for even $n$
\begin{equation}
    0 \longrightarrow \mathbb{Z}_2\longrightarrow \Omega^{\text{Spin}}_3 (Q) \longrightarrow \mathbb{Z}_n \longrightarrow 0 \,.
\end{equation}
In higher degrees differentials can also be non-trivial and we encounter the typical problem of deducing their action. Luckily, since $Q$ is the total space of a fibration, we can employ another approach, see also \cite{Saito:2025idl}.

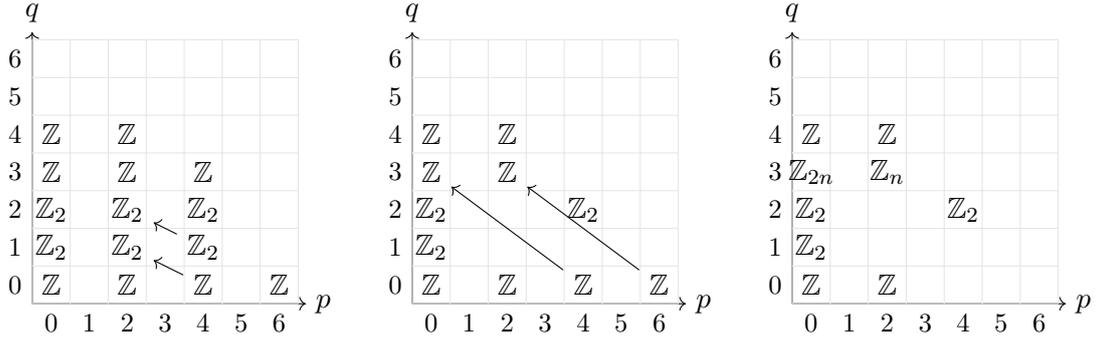
\begin{figure}
    \centering
    \begin{tikzpicture}[scale=0.5]
    % Axes
    \draw[->] (0,0) -- (7.2,0) node[right] {$p$};
    \draw[->] (0,0) -- (0,7.2) node[above] {$q$};
    % Grid (optional)
    \draw[step=1cm,gray!20] (0,0) grid (7,7);
    % Labels
    \foreach \x in {0,1,...,6}
        \node[below] at (\x+0.5, 0) {\small $\x$};
    \foreach \y in {0,1,...,6}
        \node[left] at (0,\y+0.5) {\small $\y$};
\node at (0.5, 0.5) {$\mathbb{Z}$};
\node at (0.5, 1.5) {$\mathbb{Z}_2$};
\node at (0.5, 2.5) {$\mathbb{Z}_2$};
\node at (0.5, 3.5) {$\mathbb{Z}$};
\node at (0.5, 4.5) {$\mathbb{Z}$};
\node at (2.5, 0.5) {$\mathbb{Z}$};
\node(21) at (2.5, 1.5) {$\mathbb{Z}_2$};
\node(22) at (2.5, 2.5) {$\mathbb{Z}_2$};
\node at (2.5, 3.5) {$\mathbb{Z}$};
\node at (2.5, 4.5) {$\mathbb{Z}$};
\node(40) at (4.5, 0.5) {$\mathbb{Z}$};
\node(41) at (4.5, 1.5) {$\mathbb{Z}_2$};
\node at (4.5, 2.5) {$\mathbb{Z}_2$};
\node at (4.5, 3.5) {$\mathbb{Z}$};
\node at (6.5, 0.5) {$\mathbb{Z}$};
\draw[->] (40) -- (21);
\draw[->] (41) -- (22);
%
%
    % Axes
    \draw[->] (10,0) -- (17.2,0) node[right] {$p$};
    \draw[->] (10,0) -- (10,7.2) node[above] {$q$};
    % Grid (optional)
    \draw[step=1cm,gray!20] (10,0) grid (17,7);
    % Labels
    \foreach \x in {0,1,...,6}
        \node[below] at (\x+10.5, 0) {\small $\x$};
    \foreach \y in {0,1,...,6}
        \node[left] at (10,\y+0.5) {\small $\y$};
\node at (10.5, 0.5) {$\mathbb{Z}$};
\node at (10.5, 1.5) {$\mathbb{Z}_2$};
\node at (10.5, 2.5) {$\mathbb{Z}_2$};
\node(b) at (10.5, 3.5) {$\mathbb{Z}$};
\node at (10.5, 4.5) {$\mathbb{Z}$};
\node at (12.5, 0.5) {$\mathbb{Z}$};
\node(d) at (12.5, 3.5) {$\mathbb{Z}$};
\node at (12.5, 4.5) {$\mathbb{Z}$};
\node(a) at (14.5, 0.5) {$\mathbb{Z}$};
\node at (14.5, 2.5) {$\mathbb{Z}_2$};
\node(c) at (16.5, 0.5) {$\mathbb{Z}$};
\draw[->] (a) -- (b);
\draw[->] (c) -- (d);
%
%
    % Axes
    \draw[->] (20,0) -- (27.2,0) node[right] {$p$};
    \draw[->] (20,0) -- (20,7.2) node[above] {$q$};
    % Grid (optional)
    \draw[step=1cm,gray!20] (20,0) grid (27,7);
    % Labels
    \foreach \x in {0,1,...,6}
        \node[below] at (\x+20.5, 0) {\small $\x$};
    \foreach \y in {0,1,...,6}
        \node[left] at (20,\y+0.5) {\small $\y$};
\node at (20.5, 0.5) {$\mathbb{Z}$};
\node at (20.5, 1.5) {$\mathbb{Z}_2$};
\node at (20.5, 2.5) {$\mathbb{Z}_2$};
\node at (20.5, 3.5) {$\mathbb{Z}_{2n}$};
\node at (20.5, 4.5) {$\mathbb{Z}$};
\node at (22.5, 0.5) {$\mathbb{Z}$};
\node at (22.5, 3.5) {$\mathbb{Z}_n$};
\node at (22.5, 4.5) {$\mathbb{Z}$};
\node at (24.5, 2.5) {$\mathbb{Z}_2$};
\node at (18.5, 0.5) {\phantom{$\mathbb{Z}$}};
\end{tikzpicture}
    \caption{$E_2^{p,q}$ of the AHSS for the Spin bordism of $Q$ implementing $n c_1^2 = 0$, where we utilize the fibration structure of $Q$ (left), as well as $E^4_{p,q}$ (middle) and $E^{\infty}_{4,0}$ (right) of the same construction, for $p+q \leq 6$. }
    \label{fig:E2AHSSSpinnc12mod}
\end{figure}

The alternative version of the AHSS is given by, see \cite{Saito:2025idl},
\begin{equation}
    E^2_{p,q} = H_p \big( B\mathrm{U}(1); \Omega^{\text{Spin}}_q \big( K(\mathbb{Z},3)\big)\big) \Rightarrow \Omega^{\text{Spin}}_{p+q} (Q) \,.
\end{equation}
We calculate the Spin bordism of $K(\mathbb{Z},3)$ in Appendix~\ref{app:bordism}, with result up to degree $7$;
\begin{equation}
    \begin{array}{c | c c c c c c c c}
        k & 0 & 1 & 2 & 3 & 4 & 5 & 6 & 7 \\ \hline
        \Omega^{\text{Spin}}_k \big( K(\mathbb{Z},3)\big) & \mathbb{Z} & \mathbb{Z}_2 & \mathbb{Z}_2 & \mathbb{Z} & \mathbb{Z} & 0 & 0 & \mathbb{Z}
    \end{array}
\end{equation}
With that we can determine the second page, which we summarize in Figure~\ref{fig:E2AHSSSpinnc12mod}.

Since the cohomology of $K(\mathbb{Z},3)$ is concentrated in degree 3 and above, one has the isomorphism
\begin{equation}
    \Omega^{\text{Spin}}_k (\text{pt}) \simeq \Omega_k^{\text{Spin}} \big( K(\mathbb{Z},3)\big) \,, \quad \text{for} \enspace k \in \{0 \,, 1 \,, 2\} \,.
\end{equation}
This also means that the differentials in the lowest two rows are given by the dual of the Steenrod square operation, e.g., \cite{Teichner:1993aij, Zhubr1999}, which we know how to evaluate. The fibration of $K(\mathbb{Z},3)$ over $B\mathrm{U}(1)$, on the other hand is, as inherited from the LSSS above, encoded in a differential on the fourth page $E^4_{p,q}$. More explicitly, since $\Omega^{\text{Spin}}_3 \big( K(\mathbb{Z},3) \big) \simeq H_3 \big( K(\mathbb{Z},3) \big)$, we can use the homological version of the LSSS to show that the differential there is essentially multiplication by $n$
\begin{equation}
    H_{4} \big(B\mathrm{U}(1);\mathbb{Z} \big) \simeq \mathbb{Z} \xrightarrow{\cdot n} \mathbb{Z} \simeq H_3 \big( K (\mathbb{Z},3)\big) \,,
\end{equation}
and analogously for $H_6 \big( B \mathrm{U}(1); \mathbb{Z}\big)$. This induces the action of the differential on $E^4_{4,0} = \text{ker} (d_2)$, given by the even elements in $H_4 \big( B \mathrm{U}(1); \mathbb{Z}\big)$ and $E^4_{6,0} \simeq H_6 \big( B \mathrm{U}(1); \mathbb{Z} \big)$. The result is shown in Figure~\ref{fig:E2AHSSSpinnc12mod}, from which we can read off the Spin bordism groups in degree 5 and below
\begin{equation}
    \begin{array}{c | c c c c c c}
    k & 0 & 1 & 2 & 3 & 4 & 5 \\ \hline
    \Omega^{\text{Spin}}_k (Q) & \mathbb{Z} & \mathbb{Z}_2 & \mathbb{Z} \oplus \mathbb{Z}_2 & \mathbb{Z}_{2n} & \mathbb{Z} & \mathbb{Z}_n
    \end{array}
\end{equation}
While we find the same $\mathbb{Z}_n$ as for oriented bordism above, the $\mathbb{Z}_n$ in degree is extended to $\mathbb{Z}_{2n}$. This is a remnant of the fact that $c_1^2$ is divisible by $2$ on Spin 4-manifolds (the same does not hold for $c_1^3$). Similarly, the $\mathbb{Z}$ factor in degree 6 now is detected by $\tfrac{1}{48} p_1$ as opposed to the oriented case. For the two $\mathbb{Z}_2$ factors, coming from $\Omega^{\mathrm{Spin}}_{\bullet} (\text{pt})$, in degree 1 and 2 one further needs to include fermionic classes, like $\eta$-invariants.

This leaves us with an extension problem in degree 6, given by
\begin{equation}
    0 \rightarrow \mathbb{Z}\rightarrow \Omega_6^{\text{Spin}}(Q)\rightarrow \mathbb{Z}_2 \rightarrow 0 \,.
\end{equation}
While this does not change the fact, that there is a perturbative anomaly associated to a single $\mathbb{Z}$ factor it might change its normalization, i.e., the normalization of $c_1 p_1$ in the associated anomaly polynomial.

Now let us assume we are in a four dimensional theory, with U$(1)$ charged chiral fermions. From the structure of the spectral sequences we see that while an anomaly free theory still requires a cancellation of a perturbative mixed anomaly, the absence of pure gauge anomalies is no longer $\mathbb{Z}$-valued. This means that the typical requirement 
\begin{equation}
    \sum_q n_q \, q^3 = 0 \in \mathbb{Z} \,,
\end{equation}
is replaced by some discrete remnant in the modified Abelian gauge theory. Indeed this can be understood as a form of incomplete anomaly cancellation \cite{Saito:2025idl}, as described above.

\vspace{0.5cm}

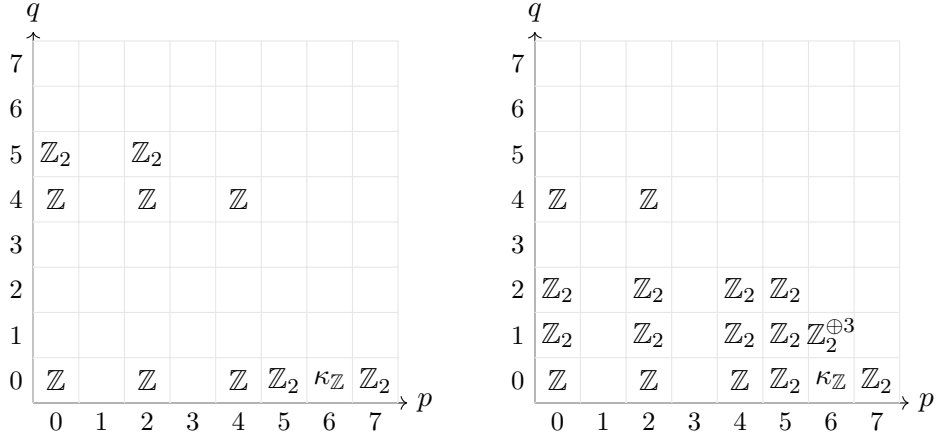
\begin{figure}
    \centering
    \begin{tikzpicture}[scale=0.6]
    % Axes
    \draw[->] (0,0) -- (8.2,0) node[right] {$p$};
    \draw[->] (0,0) -- (0,8.2) node[above] {$q$};
    % Grid (optional)
    \draw[step=1cm,gray!20] (0,0) grid (8,8);
    % Labels
    \foreach \x in {0,1,...,7}
        \node[below] at (\x+0.5, 0) {\small $\x$};
    \foreach \y in {0,1,...,7}
        \node[left] at (0,\y+0.5) {\small $\y$};
\node at (0.5, 0.5) {$\mathbb{Z}$};
\node at (0.5, 4.5) {$\mathbb{Z}$};
\node at (0.5, 5.5) {$\mathbb{Z}_2$};
%\node at (0.5, 8.5) {$\mathbb{Z}^{\oplus 2}$};

\node at (2.5, 0.5) {$\mathbb{Z}$};
\node at (2.5, 4.5) {$\mathbb{Z}$};
\node at (2.5, 5.5) {$\mathbb{Z}_2$};
\node at (4.5, 0.5) {$\mathbb{Z}$};
\node at (4.5, 4.5) {$\mathbb{Z}$};
\node at (5.5, 0.5) {$\mathbb{Z}_2$};
\node at (6.5, 0.5) {$\kappa_{\mathbb{Z}}$};
\node at (7.5, 0.5) {$\mathbb{Z}_2$};
%\node at (8.5, 0.5) {$\kappa_{\mathbb{Z}}$};
%
% Axes
    \draw[->] (11,0) -- (19.2,0) node[right] {$p$};
    \draw[->] (11,0) -- (11,8.2) node[above] {$q$};
    % Grid (optional)
    \draw[step=1cm,gray!20] (11,0) grid (19,8);
    % Labels
    \foreach \x in {0,1,...,7}
        \node[below] at (\x+11.5, 0) {\small $\x$};
    \foreach \y in {0,1,...,7}
        \node[left] at (11,\y+0.5) {\small $\y$};
\node at (11.5, 0.5) {$\mathbb{Z}$};
\node at (11.5, 1.5) {$\mathbb{Z}_2$};
\node at (11.5, 2.5) {$\mathbb{Z}_2$};
\node at (11.5, 4.5) {$\mathbb{Z}$};
\node at (13.5, 0.5) {$\mathbb{Z}$};
\node at (13.5, 1.5) {$\mathbb{Z}_2$};
\node at (13.5, 2.5) {$\mathbb{Z}_2$};
\node at (13.5, 4.5) {$\mathbb{Z}$};
\node at (15.5, 0.5) {$\mathbb{Z}$};
\node at (15.5, 1.5) {$\mathbb{Z}_2$};
\node at (15.5, 2.5) {$\mathbb{Z}_2$};
\node at (16.5, 0.5) {$\mathbb{Z}_2$};
\node at (16.5, 1.5) {$\mathbb{Z}_2$};
\node at (16.5, 2.5) {$\mathbb{Z}_2$};
\node at (17.5, 0.5) {$\kappa_{\mathbb{Z}}$};
\node at (17.5, 1.5) {$\mathbb{Z}_2^{\oplus 3}$};
\node at (18.5, 0.5) {$\mathbb{Z}_2$};
\end{tikzpicture}
    \caption{$E^2_{p,q}$ for the AHSS for oriented (left) and Spin (right) bordism of $Q$ implementing $\mathfrak{m}_2 (c_1^2) = 0$ up to order $p+q \leq 7$. ($\kappa_{\mathbb{Z}}$ denotes $\mathbb{Z} \oplus \mathbb{Z}_2$)}
    \label{fig:E2AHSSSOSpinmodnc12}
\end{figure}

\noindent Next, let us move to the `mod $n$' constraint (again for $n =2$) for which the relevant homology groups with $\mathbb{Z}$ and $\mathbb{Z}_2$ coefficients can be determined from \eqref{eq:Qmodn_intcohom}, and are given by
\begin{equation}
    \begin{array}{c | c c }
    k & H_{k}(Q;\mathbb{Z}) & H_k(Q;\mathbb{Z}_2) \\ \hline 
    0 & \mathbb{Z} & \mathbb{Z}_2 \\ 
    1 & 0 & 0 \\
    2 & \mathbb{Z} & \mathbb{Z}_2 \\
    3 & 0 & 0 \\
    4 & \mathbb{Z} & \mathbb{Z}_2 \\
    5 & \mathbb{Z}_2 & \mathbb{Z}_2 \\
    6 & \mathbb{Z} \oplus \mathbb{Z}_2 & \mathbb{Z}_2^{\oplus 3} \\ 
    7 & \mathbb{Z}_2 & \mathbb{Z}_2^{\oplus 2} \\
    8 & \mathbb{Z} \oplus \mathbb{Z}_2 & \mathbb{Z}_2^{\oplus 3}
    \end{array}
\end{equation}
With that it is straightforward to obtain the second page of the AHSS for oriented and Spin bordism which we summarize in Figure~\ref{fig:E2AHSSSOSpinmodnc12}.
Since in this case neither $H_{\bullet}(Q;\mathbb{Z})$ nor $\Omega^{\mathrm{SO}}_{\bullet} (\text{pt})$ contain odd torsion factors the AHSS collapses on the second page \eqref{eq:Walltheorem} and we can simply read off the oriented bordism groups:
\begin{equation}
    \begin{array}{c | c c c c c c c }
    k & 0 & 1 & 2 & 3 & 4 & 5 & 6  \\ \hline
    \Omega^{\text{SO}}_k (Q) & \mathbb{Z} & 0 & \mathbb{Z} & 0 & \mathbb{Z} \oplus \mathbb{Z} & \mathbb{Z}_2 \oplus \mathbb{Z}_2 & \mathbb{Z} \oplus \mathbb{Z} \oplus \mathbb{Z}_2
    \end{array}
\end{equation}
Again this differs, from the groups of $B\mathrm{U}(1)$ in several ways. For instance it contains extra $\mathbb{Z}_2$ factors, which capture anomalous contributions of the fiber (e.g., one of the $\mathbb{Z}_2$ in degree 5 is of this kind and can be detected for example by $w_2 \cup f^*_Q (\tilde{u}_3)$, with $\tilde{u}_3 \in H^3\big( K (\mathbb{Z}_2; 3); \mathbb{Z}_2 \big)$). But also the $\mathbb{Z}$ factors differ in their normalization, as discussed already at the beginning of this Section~\ref{sec:bordism}. For example the constraint $\mathfrak{m}_2 (c_1^2) = 0$ implies, that one can divide $c_1^2$ by $2$ in the modified gauge theory.\footnote{This is typically not the case for oriented manifolds; take for example $\mathbb{CP}^2$, with $c_1$ given by the generator of $H^{2}(\mathbb{CP}^2;\mathbb{Z}) \simeq \mathbb{Z}$.}

\begin{figure}
    \centering
    \begin{tikzpicture}[scale=0.6]
    % Axes
    \draw[->] (0,0) -- (7.2,0) node[right] {$p$};
    \draw[->] (0,0) -- (0,7.2) node[above] {$q$};
    % Grid (optional)
    \draw[step=1cm,gray!20] (0,0) grid (7,7);
    % Labels
    \foreach \x in {0,1,...,6}
        \node[below] at (\x+0.5, 0) {\small $\x$};
    \foreach \y in {0,1,...,6}
        \node[left] at (0,\y+0.5) {\small $\y$};
\node at (0.5, 0.5) {$\mathbb{Z}$};
\node at (0.5, 1.5) {$\mathbb{Z}_2$};
\node at (0.5, 2.5) {$\mathbb{Z}_2$};
\node at (0.5, 3.5) {$\mathbb{Z}_2$};
\node at (0.5, 4.5) {$\mathbb{Z}$};
\node at (2.5, 0.5) {$\mathbb{Z}$};
\node(b) at (2.5, 1.5) {$\mathbb{Z}_2$};
\node(d) at (2.5, 2.5) {$\mathbb{Z}_2$};
\node(f) at (2.5, 3.5) {$\mathbb{Z}_2$};
\node at (2.5, 4.5) {$\mathbb{Z}$};
\node(a) at (4.5, 0.5) {$\mathbb{Z}$};
\node(c) at (4.5, 1.5) {$\mathbb{Z}_2$};
\node at (4.5, 2.5) {$\mathbb{Z}_2$};
\node(e) at (6.5, 0.5) {$\mathbb{Z}$};
\draw[->] (a) -- (b);
\draw[->] (c) -- (d);
\draw[->, dashed] (e) -- (f);
%
%
% Axes
    \draw[->] (10,0) -- (17.2,0) node[right] {$p$};
    \draw[->] (10,0) -- (10,7.2) node[above] {$q$};
    % Grid (optional)
    \draw[step=1cm,gray!20] (10,0) grid (17,7);
    % Labels
    \foreach \x in {0,1,...,6}
        \node[below] at (\x+10.5, 0) {\small $\x$};
    \foreach \y in {0,1,...,6}
        \node[left] at (10,\y+0.5) {\small $\y$};
\node at (10.5, 0.5) {$\mathbb{Z}$};
\node at (10.5, 1.5) {$\mathbb{Z}_2$};
\node at (10.5, 2.5) {$\mathbb{Z}_2$};
\node at (10.5, 3.5) {$\mathbb{Z}_2$};
\node at (10.5, 4.5) {$\mathbb{Z}$};
\node at (12.5, 0.5) {$\mathbb{Z}$};
\node at (12.5, 4.5) {$\mathbb{Z}$};
\node at (14.5, 0.5) {$\mathbb{Z}$};
\node at (14.5, 2.5) {$\mathbb{Z}_2$};
\node at (16.5, 0.5) {$\mathbb{Z}$};
\end{tikzpicture}
    \caption{$E^2_{p,q}$ and $E^{\infty}_{p,q}$ of the AHSS for Spin bordism of $Q$ implementing $\mathfrak{m}_2 (c_1^2) = 0$ using its fibration structure, with $p+q\leq 6$. }
    \label{fig:E2andEinfAHSSSOforc_12viafibration}
\end{figure}
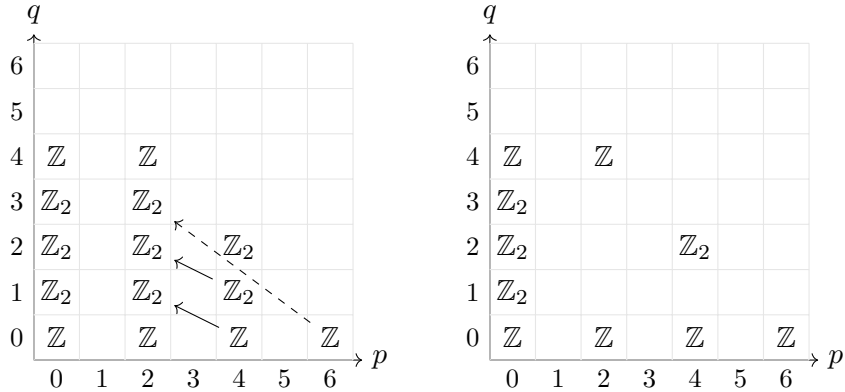

For Spin bordism the story becomes more complicated, since now the lower rows in the AHSS are non-trivial as can be seen in Figure~\ref{fig:E2AHSSSOSpinmodnc12}. 
Once more one way to proceed is to consider a different AHSS for the fibration \ref{eq:fibrationforc1sqmod2} with the second page:
\begin{equation}
    E^2_{p,q} = H_p \big( B\mathrm{U}(1) ; \Omega_q^{\text{Spin}} \big( K(\mathbb{Z}_2, 3) \big) \big) \,,
\end{equation}
where we use the bordism groups computed in the Appendix~\ref{app:bordism} to obtain the second page depicted in Figure~\ref{fig:E2andEinfAHSSSOforc_12viafibration}. The differentials in the two lowest rows are one more inherited from the Spin bordism of $B\mathrm{U}(1)$. The action of a differential on the fourth page, analogously to the discussion above, is fixed by LSSS for cohomology. In the range of interest in the example this gives the action
\begin{equation}
    d_4: \quad E^4_{6,0} \simeq H_6\big( B\mathrm{U}(1);\mathbb{Z}\big) \simeq \mathbb{Z} \xrightarrow{\mathfrak{m}_2} \mathbb{Z}_2 \simeq E^4_{2,3} \,. 
\end{equation}
This allows us to determine the $\infty$-page for Spin bordism, see Figure~\ref{fig:E2andEinfAHSSSOforc_12viafibration}, up to $p+q \leq 6$ and leads to the Spin bordism groups
\begin{equation}
     \begin{array}{c | c c c c c c}
        k & 0 & 1 & 2 & 3 & 4 & 5 \\ \hline
        \Omega^{\text{Spin}}_k \big( Q\big) & \mathbb{Z} & \mathbb{Z}_2 & \mathbb{Z}\oplus\mathbb{Z}_2 & \mathbb{Z}_2 & \mathbb{Z}\oplus \mathbb{Z} & 0
    \end{array}
\end{equation}
The most interesting entry is the factor $\mathbb{Z}_2$ in degree $3$. This element does not exist for the unmodified U$(1)$ gauge theory, because of a non-trivial differential dual to the  Steenrod square $Sq^2$ from $H^2\big( B \mathrm{U}(1); \mathbb{Z}_2\big)$ to $H^4 \big( B \mathrm{U}(1);\mathbb{Z}_2 \big)$. However, for $Q$ this differential is trivial and the associated class survives. From Figure~\ref{fig:E2andEinfAHSSSOforc_12viafibration} we also see that it is associated to the non-trivial $\Omega^{\text{Spin}}_3 \big( K(\mathbb{Z}_2, 3)\big) \simeq \mathbb{Z}_2$ related to the mod $2$ class $u_3 \in H^3\big( K(\mathbb{Z}_2,3); \mathbb{Z} \big)$, see Appendix~\ref{app:bordism}. It is therefore suggestive that this non-trivial class is detected by a non-trivial $\mathbb{Z}_2$-valued integral of the pull-back of this fiber class to spacetime $X$.

The first extension problem appears in degree $6$, in which the $\mathbb{Z}_2$ factor could be absorbed by the $\mathbb{Z} \simeq E^{\infty}_{2,4}$, precisely as above. Again, this would only change the normalization, but not the fact that there is perturbative mixed gravitational anomaly in four dimensions. As opposed to the `times $n$' constraint now we also have a perturbative pure gauge anomaly, which however has a different quantization, since $c_1^3$ can be divided by $2$. We also find additional bordism groups due to the fiber, for example $\Omega^{\text{Spin}}_3 (Q)\cong \mathbb{Z}_2$.

\vspace{0.5cm}

We see that even in these rather minimal examples the modification of the topological sectors of Abelian gauge theories does have a profound impact on both the characteristic classes, as well as the anomalies classified by bordism groups. While it indeed can be used to reduce the conserved charges and anomalies, the homotopy fiber construction often introduces new topological sectors, which also can contribute to anomalies and hence need to be considered

\section{Conclusions and outlook}
\label{sec:concl}

We have demonstrated that the modification of the characteristic classes of an Abelian gauge theory at the level of the classifying space has several important consequences. These include the proliferation of constraints to higher-dimensional characteristic classes as well as the appearance of additional topological gauge field configurations coming from the modified classifying space as a fibration. These modifications at the level of (co)homology will also impact the deformation classes of the associated modified Abelian gauge theories described by bordism groups. Since these classify the potential anomalies of the theory the implementation of the constraints further affects the anomalies of the modified theory. With the general techniques of the homotopy fiber construction and the corresponding phenomena, we now want to point out several avenues for a continuation of this work.

We already mentioned the close connection between the homotopy fiber construction and higher groups. However, at least when the higher group contains a 0-form symmetry there is one additional piece of data that specifies its higher-group structure and is given roughly by the action of the 0-form symmetry on the higher-form pieces \cite{Kapustin:2013uxa, Benini:2018reh, Bhardwaj:2022scy}. To implement such an action one would likely have to pass to cohomology with local coefficients, whose local nature encode the non-trivial 0-form symmetry action as an automorphism on the higher-symmetry pieces. This will also change the realization of characteristic classes as well as anomalies (see, e.g., \cite{Benini:2018reh, Wan:2018bns, Davighi:2023luh, Debray:2023rlx}) and might provide further applications in the context of anomaly cancellation. Similarly, one should include the characteristic classes of the tangent bundle and its mixing with the gauge symmetries to allow for a treatment of twisted tangential structures, which often appear in type II string theories and their compactification \cite{Tachikawa:2018njr, Fiorenza:2019ckz, Debray:2023yrs, Cheng:2023owv, Chakrabhavi:2025bfi}, and are essential for the heterotic string (e.g., \cite{Sati:2008kz, Sati:2009ic, Fiorenza:2010mh}, for a mathematical description), see also \cite{BoyleSmith:2025duo}.

Similar techniques as discussed here, can also be used for non-Abelian 0-form symmetries, e.g., for non-Abelian gauge theories described by a Lie group. Beyond the implications for anomalies an their cancellation, see \cite{Saito:2025idl}, this can have clear phenomenological applications. In particular, a restriction on the sum of topological sectors or a coupling to a topological field theory sector, see, e.g., \cite{Pantev:2005rh, Pantev:2005wj, Pantev:2005zs, Seiberg:2010qd, Kapustin:2014gua, Freed:2017rlk, Tanizaki:2019rbk, Brennan:2023kpw, Najjar:2025htp}, can have important consequences for fields that couple to the topological charge density, such as axions, e.g., \cite{Choi:2023pdp, Reece:2023iqn, Agrawal:2023sbp, Cordova:2023her, Dierigl:2024cxm} and \cite{Reece:2023czb} for a review. Since the group cohomology classes $H^D \big(Q;\mathrm{U}(1) \big)$ or bordism groups $\Omega^{\xi}_D (Q)$ classify such topological couplings they will generically be affected by the modification. It would further be interesting to connect the modification of a characteristic class described by a top-form when pulled back to spacetime with the decomposition of the theory in several individual universes, see, e.g., \cite{Hellerman:2006zs}. Moreover, restriction of the type discussed in this work can be implemented on the lattice (e.g., \cite{Abe:2022nfq, Kan:2023yhz} for modifications of $c_1$) which should allow for a treatment of challenging strong coupling dynamics.

The transition from group (co)homology to bordism groups is in general very interesting, since it allows one to understand the effect of gravity in the modification of (generalized) symmetries and the connected non-conservation of some of the charges \cite{Yonekura:2020ino}. This is closely related to the Swampland Cobordism Conjecture \cite{McNamara:2019rup}. In particular, the homotopy fiber analysis encodes a gauging of these global symmetries and verifies the expectation that such a gauging can introduce new conserved charges which need to be violated in the full theory of quantum gravity.

Finally, in this work we mainly focused on the topological classes, which for U$(1)$ gauge theories are concentrated in even degrees and thus most apparent for theories in even dimensions. These modifications would however also modify the Chern-Simons terms in odd dimensions and can allow for fractional levels, even in the absence of a Spin structure, see \cite{Seiberg:2016rsg} and references therein for the importance of Spin structure. The generality of the homotopy fiber approach which is viable also for higher-form gauge fields might therefore be a unifying description of these and related phenomena, such as the anyonic statistics of extended objects described in \cite{Kobayashi:2024dqj, Feng:2025mdg}.

\section*{Acknowledgments}

We thank Joe Davighi, Arun Debray, Ling Lin, Miguel Montero, and Ethan Torres for useful discussions and comments. We are especially thankful to Iñaki Garc\'ia-Etxebarria and Yuji Tachikawa for valuable comments on the draft. We are further grateful to the CERN short-term visitor program, which allowed in-person interaction during the later stages of this project. The work of MD is funded by the European Union (ERC, SymQuaG, 101163591).

\begin{appendix}

\section{Homotopy fibers and Bianchi identities}
\label{app:homBianchi}

We use this appendix to make the connection between the homotopy fiber construction and Bianchi identities more explicit and point out some subtleties associated to the realization of the higher-form fields. We will concentrate on the `times $n$' constraint for $c_1^2$, which involves the appearance of a U$(1)$ higher-form field.

If we write out the action of the differential in the top left of Figure~\ref{fig:E2c1sqn}, we obtain the equation
\begin{equation}
    d v_3 = n \, v_2^2 \,,
    \label{eq:expdiffer}
\end{equation}
where $v_3$ is the generator of $H^3\big( K(\mathbb{Z},3); \mathbb{Z} \big)$ and $v_2$ the generator of $H^2\big(B \mathrm{U}(1); \mathbb{Z}\big)$. Writing this in a more physics-oriented way, by pulling back to spacetime $X$ and using Chern-Weil representatives this becomes
\begin{equation}
    d H^{(3)} = \frac{n}{4 \pi^2} F^{(2)} \wedge F^{(2)} \,.
\label{eq:expBianchi}
\end{equation}
This precisely corresponds to a usual Bianchi identity. However, there are two important subtleties:
\begin{itemize}
    \item{The equation \eqref{eq:expBianchi} does not include all the information of \eqref{eq:expdiffer}, because it is not sensitive to torsion classes.}
    \item{The equation \eqref{eq:expBianchi} contains way too much information, since it is a local identity of differential forms, i.e., valid at each point in $X$, whereas \eqref{eq:expdiffer} only equates the cohomology classes, independent of their representatives.}
\end{itemize}
Instead of imposing the Bianchi identity on top of the equations of motion, we can also include it as enforced by an equation of motion of an additional Lagrange multiplier field. In order to do so we however need to fix the number of spacetime dimensions, $D$, and use a specific Lagrangian description. To do this consistently we move to the description in terms of differential characters, \cite{Cheeger1985DifferentialCA, Hopkins:2002rd, Hsieh:2020jpj, Moore:2025tmt}, see \cite{Sati:2009ic, 2012arXiv1201.2919R} for a mathematical description of Bianchi identities on the level of differential cohomology. In this framework, the equation \eqref{eq:expdiffer} is contained in the topological class of the composite differential character $\check{A} \ast \check{A}$ and can be written as
\begin{equation}
n \big[N_{A \ast A}^{(4)} \big] = n \big[ N_A^{(2)} \big] \cup \big[ N_A^{(2)} \big] = 0 \in H^4 (X;\mathbb{Z}) \,.
\end{equation}
This triviality of the topological class is enforced in the Bianchi identity \eqref{eq:expBianchi} by providing an explicit trivialization. This choice of trivialization in \eqref{eq:expBianchi} is precisely what contains the additional local information and beyond restriction of the topological class, also determines the field strength $F^{(4)}_{A \ast A} = F^{(2)} \wedge F^{(2)}$.

Let us now describe, how this can be encoded in the properties of a Lagrange multiplier $\check{\lambda}$ and we include a term
\begin{equation}
    \Delta_{\check{\lambda}} S = 2 \pi i \int_X \big(\check{\lambda}, \check{q}(\check{A})\big) 
    \label{eq:Lagrangemult}
\end{equation}
in the Euclidean action. Here, $\check{q}(\check{A}) \in \check{H}^4(X)$ denotes a differential character that depends on $\check{A}$, and $(\enspace, \enspace)$ is the differential cohomology pairing (see also \cite{GarciaEtxebarria:2024fuk}).

\subsection{Flat gauging}

The first realization is also known as `flat gauging' in the literature \cite{Antinucci:2024zjp} and is used to implement the topological constraint \eqref{eq:expdiffer}. For that we want  $\check{\lambda} \in \check{H}^{D-3}(X)$ to be a flat differential character, i.e., using the short exact sequence 
\begin{equation}
    0 \longrightarrow H^{k-1} \big( X; \mathrm{U}(1) \big) \longrightarrow \check{H}^k (X) \longrightarrow \Omega^{k}_{\mathbb{Z}} (X) \longrightarrow 0 \,,
\end{equation}
we see that $\check{\lambda}$ is fully characterized as an element in $H^{D-4} \big( X; \mathrm{U}(1) \big)$. Further choosing $\check{q}(\check{A}) = n \check{A} \ast \check{A}$, \eqref{eq:Lagrangemult} becomes
\begin{equation}\label{eq:topologicallagrangemultiplier}
    \Delta_{\check{\lambda}} S = 2 \pi i n \int_X \lambda \cup [N^{(2)}_A \cup N_A^{(2)}]
\end{equation}
which realizes the perfect pairing
\begin{equation}
    H^{D-k} \big(X;\mathrm{U}(1) \big) \times H^k (X;\mathbb{Z}) \rightarrow \mathrm{U}(1)
\end{equation}
described in \cite{Freed:2006yc}. The path integral over $\lambda \in H^{D-4} \big( X; \mathrm{U}(1) \big)$ then implements \eqref{eq:expdiffer} without additional local constraints.

But then one might ask, whether the flatness condition on $\check{\lambda}$ should also be implemented at the level of the equations of motion. This can be done using the other perfect pairing described in \cite{Freed:2006yc}, given by
\begin{equation}
    \Omega^{D-k}_{\mathbb{Z}} (X) \times \frac{\Omega^{k} (X)}{\Omega^k_{\mathbb{Z}}(X)} \rightarrow \mathrm{U}(1) \,.
    \label{eq:pairing2}
\end{equation}
For that we introduce another Lagrange multiplier $\check{\alpha}\in \check{H}^4(X)$ which is a topologically trivial differential character described by a globally-defined differential form $\alpha^{(3)}\in\frac{\Omega^{3} (X)}{\Omega^3_{\mathbb{Z}}(X)}$ defined up to large gauge transformations\footnote{One is inclined to call a topologically trivial gauge field an $\mathbb{R}$ gauge field, but there is a subtlety that $\mathbb{R}$ gauge field transformations are only `small' in the sense that gauge transformations are always exact forms, and for the purpose of using \ref{eq:pairing2}, it is important to include `large' gauge transformations.}. Now we modify the action \ref{eq:Lagrangemult} by adding:
\begin{equation}\label{eq:topologicallytrivialterm}
    \Delta_{\check{\alpha}} S=2\pi i \int_X \big(\check{\lambda},\check{\alpha}   \big)=2\pi i\int_X F_{\lambda}^{(D-3)}\wedge \alpha^{(3)} \,,
\end{equation}
where the second equality follows from the simplification of the differential cohomology pairing in case one of the differential characters is topologically trivial. Integrating out $\check{\alpha}$ in \ref{eq:topologicallytrivialterm} imposes the flatness condition $F_{\lambda} = 0$ on the field $\check{\lambda}$ reducing \ref{eq:Lagrangemult} to \ref{eq:topologicallagrangemultiplier}. 

\subsection{Full gauging}

If instead we take the Lagrange multiplier $\check{\lambda} \in \check{H}^{D-3}(X)$ to be an unconstrained differential character, Equation~\ref{eq:Lagrangemult} implements a stronger constraint. In particular, the path integral over $\check{\lambda}$ now includes the integration over topologically trivial characters described by elements in $\Omega^{D-4}(X) / \Omega^{D-4}_{\mathbb{Z}} (X)$, which due to the pairing \eqref{eq:pairing2}, enforces
\begin{equation}
    F^{(4)}_{q(A)} = 0 \in \Omega^4_{\mathbb{Z}}(X) \,.
\end{equation}
For the same choice as above, i.e., $\check{q} = n (\check{A} \ast \check{A})$, this yields
\begin{equation}
    \frac{n}{4 \pi^2} \big( F^{(2)} \wedge F^{(2)} \big) = 0 \in \Omega^4_{\mathbb{Z}} (X) \,,
\end{equation}
which is the Bianchi identity \eqref{eq:expBianchi} for trivial $H^{(3)}$ (or the `homogeneous solution'). To modify this to obtain \ref{eq:expdiffer} and \ref{eq:expBianchi} one needs to introduce a new field $\check{H}\in\check{C}^3(X)$ which is a differential cochain, defined for example in \cite{Hopkins:2002rd}, i.e., it is not necessarily closed. Including this field we consider the action:
\begin{equation}\label{eq:Lagrangemultiplierfull}
    \Delta S=2\pi i \int_X \big(\check{\lambda}, \check{q}(\check{A})-d \check{H}    \big) \,,
\end{equation}
where $d$ is a differential defined on differential cochains.\footnote{If a differential cochain $\check{c}\in \check{C}^p(X)$ is given by the triplet $(c,  h_c, F_c)\in C^p(X;\mathbb{Z})\times C^{p-1}(X;\mathbb{R})\times \Omega^p(X)$, the differential $d:\check{C}^p(X)\rightarrow \check{C}^{p+1}(X)$ is given by $d(c,  h_c, F_c)=(dc, F_c-dh_c-c,dF_c)$ where the differentials on the individual entries represent differentials for singular cochains and differential forms, respectively.} Since $d^2=0$ we see that $d\check{H}$ is closed, i.e., the Lagrangian contains a well-defined differential character. Integrating out the Lagrange multiplier field $\check{\lambda}$ in \ref{eq:Lagrangemultiplierfull} one obtains the constraint, see also \cite{Monnier:2018nfs},
\begin{equation}
    d\check{H} = \check{q}(\check{A}) \,,
\end{equation}
which written in components, $\check{H} = (h, h_H, H^{(3)})$, for the case  $\check{q}(\check{A}) = n(\check{A}*\check{A})$ contains
\begin{equation}
    d H^{(3)}=\frac{n}{4\pi^2}F^{(2)}\wedge F^{(2)} \,,
\end{equation}
identical to \eqref{eq:expBianchi}, as desired.

\section{(Co)homology of Eilenberg-MacLane spaces}
\label{app:EMachom}

In this appendix we summarize the (co)homology of Eilenberg-MacLane spaces of the form $K(\mathbb{Z},m)$ and $K(\mathbb{Z}_n,m)$ with a focus on $\mathbb{Z}_n = \mathbb{Z}_2$. These spaces are defined by their homotopy groups, see, e.g., \cite{hatcher2002, hatcherspectral}, 
\begin{equation}
\pi_k \big( K(A,m) \big) = \begin{cases} A \,, \quad k = m \,, \\
0 \,, \quad k>0 \,, \text{ otherwise} \,, \end{cases}
\end{equation}
where in our case $A$ is Abelian.

\subsection{\texorpdfstring{$K(\mathbb{Z},m)$}{K(Z,m)}}

These Eilenberg-MacLane spaces can be interpreted as the classifying spaces of U$(1)$ $(m-1)$-from fields. In low degree they are given by
\begin{equation}
    K(\mathbb{Z},1) = \mathrm{U}(1) = S^1 \,, \quad K(\mathbb{Z},2) = B\mathrm{U}(1) = \mathbb{CP}^{\infty} \,.
\end{equation}
From that we can easily extract their cohomology groups with integer coefficients:
\begin{equation}
    H^k \big( K(\mathbb{Z},1); \mathbb{Z}\big) = \begin{cases} \mathbb{Z} \,, \quad k \in \{ 0,1\} \,, \\ 0 \,, \quad \text{otherwise} \,, \end{cases} \quad H^k \big( K(\mathbb{Z},2); \mathbb{Z}\big) = \begin{cases} \mathbb{Z} \,, \quad k \geq 0 \,, k \enspace \text{even} \,, \\ 0 \,, \quad \text{otherwise}  \,. \end{cases}
\end{equation}
Moreover, the cohomology of $K(\mathbb{Z},2)$ is built from a single generator $v_2$ in degree $2$, defining the (polynomial) ring structure
\begin{equation}
    H^{\bullet} \big( K(\mathbb{Z},2) ; \mathbb{Z} \big) = \mathbb{Z} [v_2] \,.
\end{equation}
In higher degrees $m \geq 3$, the cohomology groups become more complicated quickly and can be extracted from the homology groups with integer coefficients in \cite{2013arXiv1312.5676B}, give by

{\renewcommand{\arraystretch}{1.3} \footnotesize
\begin{equation}
    \begin{array}{c | c c c c c c c c c c }
    k & 3 & 4 & 5 & 6 & 7 & 8 & 9 & 10 & 11 & 12 \\ \hline
    H_k\big( K(\mathbb{Z},3); \mathbb{Z}\big) & \mathbb{Z} & 0 & \mathbb{Z}_2 & 0 & \mathbb{Z}_3 & \mathbb{Z}_2 & \mathbb{Z}_2 & \mathbb{Z}_3 & \mathbb{Z}_2 \oplus \mathbb{Z}_5 & \mathbb{Z}_2 \\
    H_k\big( K(\mathbb{Z},4); \mathbb{Z}\big) & 0 & \mathbb{Z} & 0 & \mathbb{Z}_2 & 0 & \mathbb{Z} \oplus \mathbb{Z}_3 & 0 & \mathbb{Z}_2 \oplus \mathbb{Z}_2 & 0 & \mathbb{Z} \oplus \mathbb{Z}_5 \oplus \mathbb{Z}_3 \oplus \mathbb{Z}_4 \\
%    H^k\big( K(\mathbb{Z},5); \mathbb{Z}\big) & 0 & 0 & \mathbb{Z} & 0 & 0 & \mathbb{Z}_2 & 0 & \mathbb{Z}_2 \oplus \mathbb{Z}_3 & 0 & \mathbb{Z}_2 \\
    H_k\big( K(\mathbb{Z},7); \mathbb{Z}\big) & 0 & 0 & 0 & 0 & \mathbb{Z} & 0 & \mathbb{Z}_2 & 0 & \mathbb{Z}_2 \oplus \mathbb{Z}_3 & 0 
    \end{array}
\end{equation}}
which translates to: 
{\renewcommand{\arraystretch}{1.3} \footnotesize
\begin{equation}\label{eq:tableforintecoh}
    \begin{array}{c | c c c c c c c c c c }
    k & 3 & 4 & 5 & 6 & 7 & 8 & 9 & 10 & 11 & 12 \\ \hline
    H^k\big( K(\mathbb{Z},3); \mathbb{Z}\big) & \mathbb{Z} & 0 & 0 & \mathbb{Z}_2 & 0 & \mathbb{Z}_3 & \mathbb{Z}_2 & \mathbb{Z}_2 & \mathbb{Z}_3 & \mathbb{Z}_2 \oplus \mathbb{Z}_5\\
    H^k\big( K(\mathbb{Z},4); \mathbb{Z}\big) & 0 & \mathbb{Z} & 0 & 0 & \mathbb{Z}_2 & \mathbb{Z} & \mathbb{Z}_3 & 0 & \mathbb{Z}_2 \oplus \mathbb{Z}_2 & \mathbb{Z} \\
%    H^k\big( K(\mathbb{Z},5); \mathbb{Z}\big) & 0 & 0 & \mathbb{Z} & 0 & 0 & \mathbb{Z}_2 & 0 & \mathbb{Z}_2 \oplus \mathbb{Z}_3 & 0 & \mathbb{Z}_2 \\
    H^k\big( K(\mathbb{Z},7); \mathbb{Z}\big) & 0 & 0 & 0 & 0 & \mathbb{Z} & 0 & 0 & \mathbb{Z}_2 & 0 & \mathbb{Z}_2 \oplus \mathbb{Z}_3
    \end{array}
\end{equation}}
where we always have that $H^0\big( K(\mathbb{Z},m);\mathbb{Z}\big) = \mathbb{Z}$ since these spaces are connected as well as $H^k\big( K(\mathbb{Z},m);\mathbb{Z}\big) = 0$ for $0<k<m$. One further has the generic generator of $H^m\big( K(\mathbb{Z},m);\mathbb{Z}\big) = \mathbb{Z}$ in degree $m$, which we will denote by $v_m$.

\begin{figure}
\centering
\begin{tikzpicture}[x=0.5cm,y=0.5cm]
\draw[fill] (1,1) circle (2pt) node[anchor=west] {\footnotesize{$\tilde{v}_7$}};
\draw[fill] (1,3) circle (2pt);
\draw[fill] (1,4) circle (2pt);
\draw[fill] (1,5) circle (2pt);
\draw[fill] (1,7) circle (2pt);
\draw[fill] (1,8) circle (2pt);
\draw[fill] (1,6) circle (2pt);
\draw[thick] (1,3) -- (1,4);
\draw[thick] (1,5) -- (1,6);
\draw[thick] (1,7) -- (1,8);
\draw[thick] (1,1) .. controls (0.3,2) .. (1,3);
\draw[thick] (1,5) .. controls (1.7,6) .. (1,7);
\draw[thick] (1,4) .. controls (0.3,5) .. (1,6);
\end{tikzpicture}
\caption{The $\mathcal{A}(1)$ structure of $K(\mathbb{Z},7)$}
\label{fig:SqKZ7}
\end{figure}
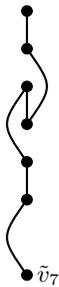

Using the universal coefficient theorem we can also extract the cohomology with $\mathbb{Z}_2$ coefficients, that will be useful for the Adams spectral sequence below. In particular, one has:

{\renewcommand{\arraystretch}{1.3} \footnotesize
\begin{equation}
    \begin{array}{c | c c c c c c c c c c }
    k & 3 & 4 & 5 & 6 & 7 & 8 & 9 & 10 & 11 & 12 \\ \hline
    H^k\big( K(\mathbb{Z},3); \mathbb{Z}_2\big) & \mathbb{Z}_2 & 0 & \mathbb{Z}_2 & \mathbb{Z}_2 & 0 & \mathbb{Z}_2 & \mathbb{Z}_2^{\oplus 2} & \mathbb{Z}_2 & \mathbb{Z}_2 & \mathbb{Z}_2^{\oplus 2} \\
    H^k\big( K(\mathbb{Z},4); \mathbb{Z}_2\big) & 0 & \mathbb{Z}_2 & 0 & \mathbb{Z}_2 & \mathbb{Z}_2 & \mathbb{Z}_2 & 0 & \mathbb{Z}_2^{\oplus 2} & \mathbb{Z}_2^{\oplus 2} & \mathbb{Z}_2^{\oplus 2} \\
    H^k\big( K(\mathbb{Z},7); \mathbb{Z}_2\big) & 0 & 0 & 0 & 0 & \mathbb{Z}_2 & 0 & \mathbb{Z}_2 & \mathbb{Z}_2 & \mathbb{Z}_2 & \mathbb{Z}_2
    \end{array}
    \label{eq:KZmZ2cohom}
\end{equation}}
The non-trivial generators can be identified with the action of all admissible actions of the Steenrod operations $Sq^n$, without the appearance of $Sq^1$, see, e.g., \cite{hatcherspectral}, on the generator $\tilde{v}_m \in H^m\big(K(\mathbb{Z},m); \mathbb{Z}_2\big) \simeq \mathbb{Z}_2$. For example for $K(\mathbb{Z},7)$ one has
\begin{equation}
    H^{\bullet}\big( K(\mathbb{Z},7) ;\mathbb{Z}_2 \big) = \mathbb{Z}_2 [\tilde{v}_7 \,, Sq^2 \tilde{v}_7 \,, Sq^3 \tilde{v}_7 \,, Sq^4 \tilde{v}_7 \,, Sq^5 \tilde{v}_7 \,, \dots ] \,, 
\end{equation}
where we indicated all elements in degree $\leq 12$. Its $\mathcal{A}(1)$ structure is captured by Figure~\ref{fig:SqKZ7}. Indeed just considering the different elements we find a match with \eqref{eq:KZmZ2cohom}, derived from the universal coefficient theorem as needed.

\subsection{\texorpdfstring{$K(\mathbb{Z}_n,m)$}{K(Zn,m)}}

As above, the low degree case is known to be determined by
\begin{equation}
K(\mathbb{Z}_n,1) = B\mathbb{Z}_n = L^{\infty}_n \,, 
\end{equation}
leading to the cohomology groups
\begin{equation}
H^k \big( K(\mathbb{Z}_n,1); \mathbb{Z} \big) = \begin{cases} \mathbb{Z} \,, \quad k =0 \,, \\ \mathbb{Z}_n \,, \quad k>0 \,, k \enspace \text{even} \,, \\ 0 \,, \quad \text{otherwise} \,.\end{cases}
\end{equation}
and with $\mathbb{Z}_n$ coefficients which will also be useful for us
\begin{equation}
    H^k \big( K(\mathbb{Z}_n,1); \mathbb{Z}_n \big) =\mathbb{Z}_n \,, \quad k \geq 0 \,.
\end{equation}
Again, there is a ring structure which in the case of $n$ odd is
\begin{equation}\label{eq:BZnringstrodd}
H^{\bullet} (B \mathbb{Z}_n;\mathbb{Z}_n) = \frac{\mathbb{Z}_n [x,y]}{(x^2)}\,,
\end{equation}
where $x$ and $y$ are or degree 1 and 2, respectively. For the case of $n$ even 
\begin{equation}\label{eq:BZnringstreven}
H^{\bullet} (B \mathbb{Z}_n;\mathbb{Z}_n) = \frac{\mathbb{Z}_n [x,y]}{(x^2-\frac{n}{2}y)} \,,
\end{equation}
with the action of a Bockstein map relating the two, see \cite{hatcher2002}.

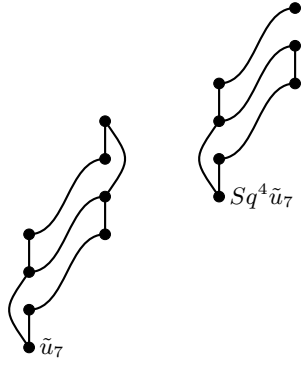
\begin{figure}
\centering
\begin{tikzpicture}[x=0.5cm,y=0.5cm]
\draw[fill] (1,1) circle (2pt) node[anchor=west] {\footnotesize{$\tilde{u}_7$}};
\draw[fill] (1,2) circle (2pt);
\draw[fill] (1,3) circle (2pt);
\draw[fill] (1,4) circle (2pt);
\draw[thick] (1,1) .. controls (0.3,2) .. (1,3);
\draw[thick] (1,3) -- (1,4);
\draw[thick] (1,1) -- (1,2);
\draw[fill] (3,4) circle (2pt);
\draw[fill] (3,5) circle (2pt);
\draw[fill] (3,6) circle (2pt);
\draw[fill] (3,7) circle (2pt);
\draw[thick] (3,4) -- (3,5);
\draw[thick] (3,6) -- (3,7);
\draw[thick] (3,5) .. controls (3.7,6) .. (3,7);
\draw[thick] (1,2) .. controls (2,2) and (2,4) .. (3,4);
\draw[thick] (1,3) .. controls (2,3) and (2,5) .. (3,5);
\draw[thick] (1,4) .. controls (2,4) and (2,6) .. (3,6);
\draw[fill] (6,5) circle (2pt) node[anchor=west] {\footnotesize{$Sq^4 \tilde{u}_7$}};
\draw[fill] (6,6) circle (2pt);
\draw[fill] (6,7) circle (2pt);
\draw[fill] (6,8) circle (2pt);
\draw[fill] (8,8) circle (2pt);
\draw[fill] (8,9) circle (2pt);
\draw[fill] (8,10) circle (2pt);
\draw[thick] (6,5) -- (6,6);
\draw[thick] (6,7) -- (6,8);
\draw[thick] (8,8) -- (8,9);
\draw[thick] (6,5) .. controls (5.3,6) .. (6,7);
\draw[thick] (6,6) .. controls (7,6) and (7,8) .. (8,8);
\draw[thick] (6,7) .. controls (7,7) and (7,9) .. (8,9);
\draw[thick] (6,8) .. controls (7,8) and (7,10) .. (8,10);
\end{tikzpicture}
\caption{The $\mathcal{A}(1)$ structure of $K(\mathbb{Z}_2,7)$}
\label{fig:SqKZ27}
\end{figure}

In higher degree the cohomology groups for general $n$ are calculated and we instead will focus on $n = 2$, for which some results can be found in \cite{Clementthesis}. In particular, the cohomology of $K(\mathbb{Z}_2,m)$ with $\mathbb{Z}_2$ coefficients can be extracted from the Steenrod action on a generator $\tilde{u}_m$ of $H^m \big( K(\mathbb{Z}_2,m); \mathbb{Z}_2 \big)$, \cite{MosherTangora1968, hatcherspectral}, in a way similar to the case $K(\mathbb{Z},m)$. However, this time one also allows for the appearance of $Sq^1$ in the enumeration of all admissible actions, \cite{hatcherspectral}. In the example of $K(\mathbb{Z}_2,7)$ one has
\begin{equation}\label{eq:kz27steenrdoalg}
\begin{split}
    H^{\bullet}\big( K(\mathbb{Z}_2,7); \mathbb{Z}_2\big) =& \mathbb{Z}_2 [\tilde{u}_7 \,, Sq^1 \tilde{u}_7 \,, Sq^2 \tilde{u}_7 \,, Sq^3 \tilde{u}_7 \,, Sq^2Sq^1 \tilde{u}_7 \,, Sq^4 \tilde{u}_7, \\ 
    & \phantom{\mathbb{Z}_2 [} \, Sq^3 Sq^1 \tilde{u}_7 \,, Sq^5 \tilde{u}_7 \,, Sq^4Sq^1 \tilde{u}_7 \,, \dots ] \,.
\end{split}
\end{equation}
Form this we can simply read off the cohomology groups
\begin{equation}
    \begin{array}{c | c c c c c c c c c c c c c }
    k & 0 & 1 & 2 & 3 & 4 & 5 & 6 & 7 & 8 & 9 & 10 & 11 & 12 \\ \hline
    H^{\bullet} \big(K(\mathbb{Z}_2,7); \mathbb{Z}_2\big) & \mathbb{Z}_2 & 0 & 0 & 0 & 0 & 0 & 0 & \mathbb{Z}_2 & \mathbb{Z}_2 & \mathbb{Z}_2 & \mathbb{Z}_2^{\oplus 2} & \mathbb{Z}_2^{\oplus 2} & \mathbb{Z}_2^{\oplus 2}
    \end{array}
\end{equation}
The $\mathcal{A}(1)$ structure in the relevant degree is given in Figure~\ref{fig:SqKZ27}. Forgetting about the $Sq^2$ data, we can also use this Adams spectral sequence for $H_{\bullet}\big( K(\mathbb{Z}_2,7); \mathbb{Z} \big)^{\vee}_2$, the 2-completion of its integer cohomology. Moreover, since we are interested in $K(\mathbb{Z}_2,7)$ and not $K(\mathbb{Z},7)$ the cohomology localizes in prime $2$, and the approach captures everything. The extensions now are taken with respect to $\mathcal{A}(0)$, generated only by $Sq^1$. With this we find the degrees relevant to us and convert to cohomology with the universal coefficient theorem\footnote{This can also be seen from an argument in \cite{hatcherspectral}, saying that if there are two elements $x \in H^k(X;\mathbb{Z}_2)$ and $y \in H^{k+1}(X;\mathbb{Z}_2)$ related via $y = Sq^1 x$, then there is a $\mathbb{Z}_2$ factor in $H_k(X;\mathbb{Z})$ where they both originate from.}, leading to 
{\renewcommand{\arraystretch}{1.3}
\begin{equation}\label{eq:cohofkz27integ}
    \begin{array}{c | c c c c c c c c c c c c c }
    k & 0 & 1 & 2 & 3 & 4 & 5 & 6 & 7 & 8 & 9 & 10 & 11 & 12 \\ \hline
    H_{\bullet} \big(K(\mathbb{Z}_2,7); \mathbb{Z}\big) &  \mathbb{Z} & 0 & 0 & 0 & 0 & 0 & 0 & \mathbb{Z}_2 & 0 & \mathbb{Z}_2 & \mathbb{Z}_2 & \mathbb{Z}_2 & \mathbb{Z}_2 \\
    H^{\bullet} \big(K(\mathbb{Z}_2,7); \mathbb{Z}\big) & \mathbb{Z} & 0 & 0 & 0 & 0 & 0 & 0 & 0 & \mathbb{Z}_2 & 0 & \mathbb{Z}_2 & \mathbb{Z}_2 & \mathbb{Z}_2
    \end{array}
\end{equation}}

With similar techniques one can derive the cohomology groups of $K(\mathbb{\mathbb{Z}}_2,3)$ and $K(\mathbb{Z}_2,2)$ with $\mathbb{Z}_2$ coefficients: 

{\renewcommand{\arraystretch}{1.5} \footnotesize
\begin{equation}\label{eq:tablewithz2coeffofz2em}
    \begin{array}{c | c c c c c c c c c c c }
    k & 2 & 3 & 4 & 5 & 6 & 7 & 8 & 9 & 10 & 11 & 12 \\ \hline
    H^k\big(K(\mathbb{Z}_2,2); \mathbb{Z}_2 \big) & \mathbb{Z}_2 & \mathbb{Z}_2 & \mathbb{Z}_2 & \mathbb{Z}_2^{\oplus 2} & \mathbb{Z}_2^{\oplus 2} & \mathbb{Z}_2^{\oplus 2} & \mathbb{Z}_2^{\oplus 3} & \mathbb{Z}_2^{\oplus 4} & \mathbb{Z}_2^{\oplus 4} & \mathbb{Z}_2^{\oplus 5} & \mathbb{Z}_2^{\oplus 6} \\
    H^k\big( K(\mathbb{Z}_2,3);\mathbb{Z}_2 \big) &  0 & \mathbb{Z}_2 & \mathbb{Z}_2 & \mathbb{Z}_2 & \mathbb{Z}_2^{\oplus 2} & \mathbb{Z}_2^{\oplus 2} & \mathbb{Z}_2^{\oplus 2} & \mathbb{Z}_2^{\oplus 4} & \mathbb{Z}_2^{\oplus 5} & \mathbb{Z}_2^{\oplus 5} & \mathbb{Z}_2^{\oplus 7}\\
    H^k\big( K(\mathbb{Z}_2,7);\mathbb{Z}_2 \big) & 0 & 0 & 0 & 0 & 0 & \mathbb{Z}_2 & \mathbb{Z}_2 & \mathbb{Z}_2 & \mathbb{Z}_2^{\oplus 2} & \mathbb{Z}_2^{\oplus 2} & \mathbb{Z}_2^{\oplus 2}
    \end{array}
\end{equation}}
which can also be verified by comparing with the integer cohomology groups calculated in \cite{Clementthesis}:

{\renewcommand{\arraystretch}{1.5} \footnotesize
\begin{equation}
    \begin{array}{c | c c c c c c c c c c }
    k & 3 & 4 & 5 & 6 & 7 & 8 & 9 & 10 & 11 & 12 \\ \hline
    H^k\big(K(\mathbb{Z}_2,2); \mathbb{Z} \big) & \mathbb{Z}_2 & 0 & \mathbb{Z}_4 & \mathbb{Z}_2 & \mathbb{Z}_2 & \mathbb{Z}_2 & \mathbb{Z}_2 \oplus \mathbb{Z}_8 & \mathbb{Z}_2^{\oplus 2} & \mathbb{Z}_2^{\oplus 2} & \mathbb{Z}_2^{\oplus 3} \\
    H^k\big( K(\mathbb{Z}_2,3);\mathbb{Z} \big) & 0 & \mathbb{Z}_2 & 0 & \mathbb{Z}_2 & \mathbb{Z}_2 & \mathbb{Z}_2 & \mathbb{Z}_2 & \mathbb{Z}_2^{\oplus 3} & \mathbb{Z}_2^{\oplus 2} & \mathbb{Z}_2^{\oplus 3}
    \end{array}
\end{equation}}
\vspace{0.5cm}

\noindent Using the Adams spectral sequence for $K(\mathbb{Z},7)$, one obtains the 2-completion
\begin{equation}
    \begin{array}{c | c c c c c c c c c c c c }
    k & 0 & 1 & 2&  3 & 4 & 5 & 6 & 7 & 8 & 9 & 10 & 11 \\ \hline
    H^{k} \big( K(\mathbb{Z},7); \mathbb{Z}\big)^{\vee}_2 & \mathbb{Z} & 0 & 0 & 0 & 0 & 0 & 0 & \mathbb{Z} & 0 & \mathbb{Z}_2 & 0 & \mathbb{Z}_2
    \end{array}
\end{equation}
but one needs to be careful regarding to the appearance of torsional elements at other primes, as we know appear in degree $12$, see \eqref{eq:tableforintecoh}.


\section{Bordism groups of Eilenberg-MacLane spaces}
\label{app:bordism}

Beyond the cohomology of the Eilenberg-MacLane spaces, we are also interested in their oriented and Spin bordism groups of the form
\begin{equation}
\Omega^{\text{SO/Spin}}_k \big( K(A,m) \big) \,.
\end{equation}
These can be constrained using the Atiyah-Hirzebruch and the Adams spectral sequence, see for example \cite{Stong1986SpinBordism, Lee:2022spd}.

\begin{figure}
\centering
\begin{tikzpicture}[scale=0.55]
    % Axes
    \draw[->] (0,0) -- (8.2,0) node[right] {$p$};
    \draw[->] (0,0) -- (0,8.2) node[above] {$q$};
    % Grid (optional)
    \draw[step=1cm,gray!20] (0,0) grid (8,8);
    % Labels
    \foreach \x in {0,1,...,7}
        \node[below] at (\x+0.5, 0) {\small $\x$};
    \foreach \y in {0,1,..., 7}
        \node[left] at (0,\y+0.5) {\small $\y$};
\node(00) at (0.5, 0.5) {$\mathbb{Z}$};
\node(04) at (0.5, 4.5) {$\mathbb{Z}$};
\node(05) at (0.5, 5.5) {$\mathbb{Z}_2$};
\node(30) at (3.5, 0.5) {$\mathbb{Z}_2$};
\node(34) at (3.5, 4.5) {$\mathbb{Z}_2$};
\node(50) at (5.5, 0.5) {$\mathbb{Z}_2$};
\node(60) at (6.5, 0.5) {$\mathbb{Z}_2$};
\node(70) at (7.5, 0.5) {$\mathbb{Z}_2$};
    % Axes
    \draw[->] (10,0) -- (18.2,0) node[right] {$p$};
    \draw[->] (10,0) -- (10,8.2) node[above] {$q$};
    % Grid (optional)
    \draw[step=1cm,gray!20] (10,0) grid (18,8);
    % Labels
    \foreach \x in {0,1,...,7}
        \node[below] at (\x+10.5, 0) {\small $\x$};
    \foreach \y in {0,1,..., 7}
        \node[left] at (10,\y+0.5) {\small $\y$};
\node at (10.5, 0.5) {$\mathbb{Z}$};
\node at (10.5, 4.5) {$\mathbb{Z}$};
\node at (10.5, 5.5) {$\mathbb{Z}_2$};
\node at (13.5, 0.5) {$\mathbb{Z}$};
\node at (13.5, 4.5) {$\mathbb{Z}$};
\node at (15.5, 0.5) {$\mathbb{Z}_2$};
\node at (17.5, 0.5) {$\mathbb{Z}_3$};
\end{tikzpicture}
\caption{The second page of the AHSS for the oriented bordism groups of $K(\mathbb{Z}_2,3)$ (left) and $K(\mathbb{Z},3)$ (right); for $p+q \leq 7$.}
\label{fig:AHSSorinetedKZ_23}
\end{figure}
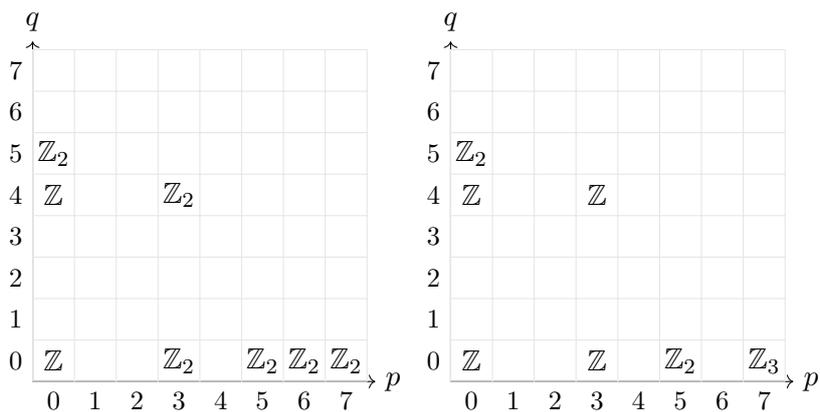

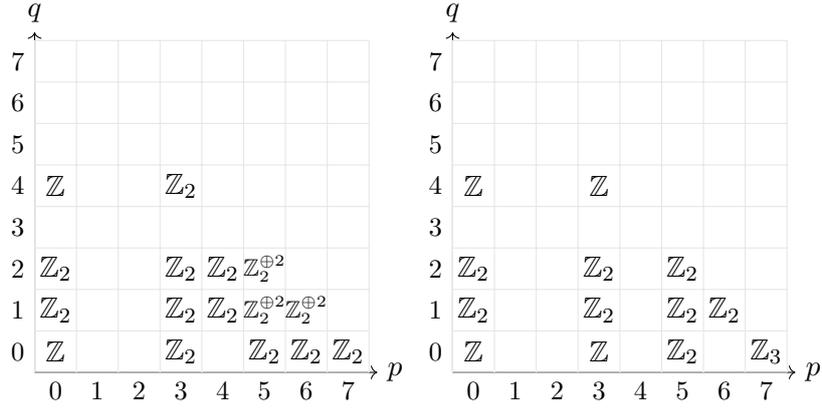
\begin{figure}
\centering
\begin{tikzpicture}[scale=0.55]
    % Axes
    \draw[->] (0,0) -- (8.2,0) node[right] {$p$};
    \draw[->] (0,0) -- (0,8.2) node[above] {$q$};
    % Grid (optional)
    \draw[step=1cm,gray!20] (0,0) grid (8,8);
    % Labels
    \foreach \x in {0,1,...,7}
        \node[below] at (\x+0.5, 0) {\small $\x$};
    \foreach \y in {0,1,..., 7}
        \node[left] at (0,\y+0.5) {\small $\y$};
\node at (0.5, 0.5) {$\mathbb{Z}$};
\node at (0.5, 1.5) {$\mathbb{Z}_2$};
\node at (0.5, 2.5) {$\mathbb{Z}_2$};
\node at (0.5, 4.5) {$\mathbb{Z}$};
\node at (3.5, 0.5) {$\mathbb{Z}_2$};
\node at (3.5, 4.5) {$\mathbb{Z}_2$};
\node at (3.5, 1.5) {$\mathbb{Z}_2$};
\node at (3.5, 2.5) {$\mathbb{Z}_2$};
\node at (4.5, 1.5) {$\mathbb{Z}_2$};
\node at (4.5, 2.5) {$\mathbb{Z}_2$};
\node at (5.5, 1.5) {\footnotesize{$\mathbb{Z}^{\oplus2}_2$}};
\node at (5.5, 2.5) {\footnotesize{$\mathbb{Z}^{\oplus2}_2$}};
\node at (5.5, 0.5) {$\mathbb{Z}_2$};
\node at (6.5, 1.5) {\footnotesize{$\mathbb{Z}^{\oplus2}_2$}};
\node at (6.5, 0.5) {$\mathbb{Z}_2$};
\node at (7.5, 0.5) {$\mathbb{Z}_2$};
    % Axes
    \draw[->] (10,0) -- (18.2,0) node[right] {$p$};
    \draw[->] (10,0) -- (10,8.2) node[above] {$q$};
    % Grid (optional)
    \draw[step=1cm,gray!20] (10,0) grid (18,8);
    % Labels
    \foreach \x in {0,1,...,7}
        \node[below] at (\x+10.5, 0) {\small $\x$};
    \foreach \y in {0,1,..., 7}
        \node[left] at (10,\y+0.5) {\small $\y$};
\node at (10.5, 0.5) {$\mathbb{Z}$};
\node at (10.5, 1.5) {$\mathbb{Z}_2$};
\node at (10.5, 2.5) {$\mathbb{Z}_2$};
\node at (10.5, 4.5) {$\mathbb{Z}$};
\node at (13.5, 0.5) {$\mathbb{Z}$};
\node at (13.5, 4.5) {$\mathbb{Z}$};
\node at (13.5, 1.5) {$\mathbb{Z}_2$};
\node at (13.5, 2.5) {$\mathbb{Z}_2$};
\node at (15.5, 0.5) {$\mathbb{Z}_2$};
\node at (15.5, 1.5) {$\mathbb{Z}_2$};
\node at (15.5, 2.5) {$\mathbb{Z}_2$};
\node at (16.5, 1.5) {$\mathbb{Z}_2$};
\node at (17.5, 0.5) {$\mathbb{Z}_3$};
\end{tikzpicture}
\caption{The second page of the AHSS for the Spin bordism groups of $K(\mathbb{Z}_2,3)$ (left) and $K(\mathbb{Z},3)$ (right); for $p+q \leq 7$.}
\label{fig:AHSSSpinKZ_23}
\end{figure}

\begin{figure}
\centering
\begin{tikzpicture}[x=0.5cm,y=0.5cm]
\draw[fill] (1,1) circle (2pt) node[anchor=west] {\footnotesize{$\tilde{u}_3$}};
\draw[fill] (1,2) circle (2pt);
\draw[fill] (1,3) circle (2pt);
\draw[fill] (1,4) circle (2pt);
\draw[fill] (1,5) circle (2pt);
\draw[fill] (1,6) circle (2pt);
\draw[fill] (1,7) circle (2pt);
\draw[fill] (1,8) circle (2pt);
\draw[fill] (1,10) circle (2pt);
\draw[fill] (3,4) circle (2pt);
\draw[fill] (3,5) circle (2pt);
\draw[thick] (1,1) -- (1,2);
\draw[thick] (1,3) -- (1,4);
\draw[thick] (1,5) -- (1,6);
\draw[thick] (1,7) -- (1,8);
\draw[thick] (3,4) -- (3,5);
\draw[thick] (1,1) .. controls (0.3,2) .. (1,3);
\draw[thick] (1,4) .. controls (0.3,5) .. (1,6);
\draw[thick] (1,8) .. controls (0.3,9) .. (1,10);
\draw[thick] (1,5) .. controls (1.7,6) .. (1,7);
\draw[thick] (1,2) .. controls (2,2) and (2,4) .. (3,4);
\draw[thick] (1,3) .. controls (2,3) and (2,5) .. (3,5);
\draw[fill] (6,6) circle (2pt) node[anchor=west] {\footnotesize{$\tilde{u}_3 Sq^2 \tilde{u}_3$}};
\draw[fill] (6,7) circle (2pt);
\draw[fill] (6,8) circle (2pt);
\draw[fill] (6,9) circle (2pt);
\draw[fill] (8,9) circle (2pt);
\draw[fill] (8,11) circle (2pt);
\draw[fill] (8,12) circle (2pt);
\draw[fill] (8,10) circle (2pt);
\draw[thick] (6,6) -- (6,7);
\draw[thick] (6,8) -- (6,9);
\draw[thick] (8,9) -- (8,10);
\draw[thick] (8,11) -- (8,12);
\draw[thick] (6,6) .. controls (5.3,7) .. (6,8);
\draw[thick] (8,10) .. controls (8.7,11) .. (8,12);
\draw[thick] (6,7) .. controls (7,7) and (7,9) .. (8,9);
\draw[thick] (6,8) .. controls (7,8) and (7,10) .. (8,10);
\draw[thick] (6,9) .. controls (7,9) and (7,11) .. (8,11);
\draw[fill] (11,7) circle (2pt) node[anchor=west] {\footnotesize{$\tilde{u}_3^3$}};
\draw[fill] (11,8) circle (2pt);
\draw[fill] (11,9) circle (2pt);
\draw[fill] (11,10) circle (2pt);
\draw[fill] (13,10) circle (2pt);
\draw[fill] (13,12) circle (2pt);
\draw[fill] (13,13) circle (2pt);
\draw[fill] (13,11) circle (2pt);
\draw[thick] (11,7) -- (11,8);
\draw[thick] (11,9) -- (11,10);
\draw[thick] (13,10) -- (13,11);
\draw[thick] (13,12) -- (13,13);
\draw[thick] (11,7) .. controls (10.3,8) .. (11,9);
\draw[thick] (13,11) .. controls (13.7,12) .. (13,13);
\draw[thick] (11,8) .. controls (12,8) and (12,10) .. (13,10);
\draw[thick] (11,9) .. controls (12,9) and (12,11) .. (13,11);
\draw[thick] (11,10) .. controls (12,10) and (12,12) .. (13,12);
\draw[fill] (16,7) circle (2pt) node[anchor=west] {\footnotesize{$Sq^4Sq^2 \tilde{u}_3$}};
\draw[fill] (16,8) circle (2pt);
\draw[fill] (16,10) circle (2pt);
\draw[fill] (16,9) circle (2pt);
\draw[fill] (16,11) circle (2pt) node[anchor=south] {\footnotesize{$\vdots$}};
\draw[thick] (16,7) -- (16,8);
\draw[thick] (16,9) -- (16,10);
\draw[thick] (16,8) .. controls (15.3,9) .. (16,10);
\draw[thick] (16,9) .. controls (16.7,10) .. (16,11);
\end{tikzpicture}
\caption{The $\mathcal{A}(1)$ structure of $K(\mathbb{Z}_2,3)$; $\tilde{u}_3$ is the generator of $H^3\big( K(\mathbb{Z}_2,3);\mathbb{Z}_2 \big)$.}
\label{fig:SqKZ23}
\end{figure}
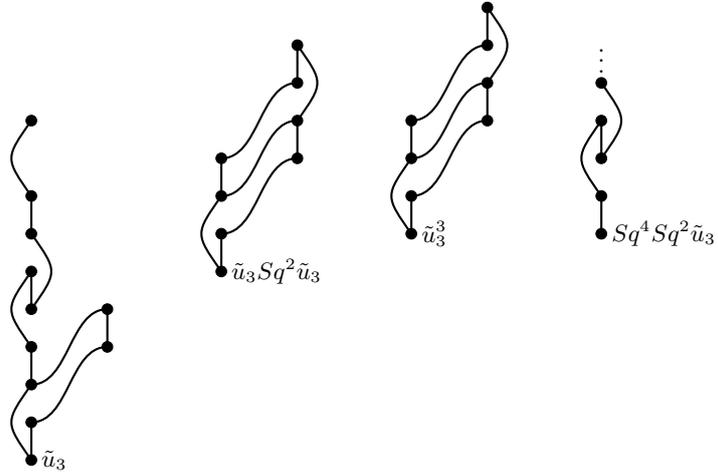

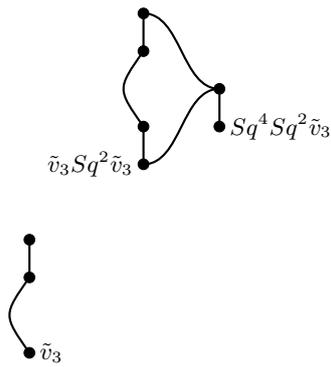
\begin{figure}
\centering
\begin{tikzpicture}[x=0.5cm,y=0.5cm]
\draw[fill] (1,1) circle (2pt) node[anchor=west] {\footnotesize{$\tilde{v}_3$}};
\draw[fill] (1,3) circle (2pt);
\draw[fill] (1,4) circle (2pt);
\draw[thick] (1,3) -- (1,4);
\draw[thick] (1,1) .. controls (0.3,2) .. (1,3);
\draw[fill] (4,6) circle (2pt) node[anchor=east] {\footnotesize{$\tilde{v}_3 Sq^2 \tilde{v}_3$}};
\draw[fill] (4,7) circle (2pt);
\draw[thick] (4,6) -- (4,7);
\draw[fill] (6,7) circle (2pt) node[anchor=west] {\footnotesize{$Sq^4Sq^2 \tilde{v}_3$}};
\draw[thick] (6,7) -- (6,8);
\draw[thick] (4,6) .. controls (5,6) and (5,8) .. (6,8);
\draw[fill] (6,8) circle (2pt);
\draw[fill] (4,9) circle (2pt);
\draw[fill] (4,10) circle (2pt);
\draw[thick] (4,9) -- (4,10);
\draw[thick] (6,8) .. controls (5,8) and (5,10) .. (4,10);
\draw[thick] (4,7) .. controls (3.3,8) .. (4,9);
\end{tikzpicture}
\caption{The $\mathcal{A}(1)$ structure of $K(\mathbb{Z},3)$; $\tilde{v}_3$ is the generator of $H^3\big( K(\mathbb{Z},3);\mathbb{Z}_2 \big)$.}
\label{fig:SqKZ3}
\end{figure}

The second page of the AHSS needs the integer homology groups $H_{\bullet}\big(K(G,m); \mathbb{Z}\big)$ as input. Let us illustrate this approach with $K(\mathbb{Z},3)$ and $K(\mathbb{Z}_2,3)$, which we also use in Section \ref{sec:bordism}. The second page of the AHSS is given by 
\begin{equation}
    E^2_{p,q} = H_p \big( K(A,m); \Omega^{\text{SO/Spin}}_q (\text{pt})\big) \,,
\end{equation}
where the oriented and Spin bordism groups of a point are given in \eqref{eq:bordismgroupspt}. We provide the second page of the AHSS for oriented bordism of $K(\mathbb{Z}_2,3)$ and $K(\mathbb{Z},3)$ in Figure \ref{fig:AHSSorinetedKZ_23}.
From there we see that in the range $p+q \leq 7$ there are no non-trivial differentials. The first extension problems only occur in degree 7 and above, and we find
{\renewcommand{\arraystretch}{1.3}
\begin{equation}
\begin{array}{c | c c c c c c c }
    k & 0 & 1 & 2 & 3 & 4 & 5 & 6 \\ \hline
    \Omega^{\text{SO}}_k \big( K(\mathbb{Z}_2,3)\big) & \mathbb{Z} & 0 & 0 & \mathbb{Z}_2 & \mathbb{Z} & \mathbb{Z}_2 \oplus \mathbb{Z}_2 & \mathbb{Z}_2 \\ 
    \Omega^{\text{SO}}_k \big( K(\mathbb{Z},3)\big) & \mathbb{Z} & 0 & 0 & \mathbb{Z} & \mathbb{Z} & \mathbb{Z}_2 \oplus \mathbb{Z}_2 & 0
\end{array}
\end{equation}}
For the Spin bordism group the second page of the AHSS looks a little more involved and is depicted in Figure~\ref{fig:AHSSSpinKZ_23}.
One way to make progress is to move to the Adams spectral sequence that will have information about the 2-completion and is sufficient to determine the Spin bordism groups in dimensions 6 and below. Its second page is given by:
\begin{equation}
    E_2^{s,t} = \text{Ext}^{s,t}_{\mathcal{A}(1)}\big((\mathbb{Z}_2\oplus\Sigma^8\mathbb{Z}_2\oplus \Sigma ^{10}J\oplus \dots )\otimes_{\mathbb{Z}_2}\widetilde{H}^{\bullet}(Y; \mathbb{Z}_2); \mathbb{Z}_2 \big)
\end{equation}
and converges to the 2-completion of the reduced Spin bordism group $\widetilde\Omega^{\text{spin}}_{*}(Y)^\vee_2$. Moreover, since we are only interested in degree $6$ and below this simplifies to
\begin{equation}
    E_2^{s,t} = \text{Ext}^{s,t}_{\mathcal{A}(1)} \big( \widetilde{H}^\bullet (Y;\mathbb{Z}_2); \mathbb{Z}_2 \big) \,,
\end{equation}
and we need the $\mathcal{A}(1)$ module structure of $Y = K(\mathbb{Z}_2,3)$ and $K(\mathbb{Z};3)$. This is depicted in Figure~\ref{fig:SqKZ23} and Figure~\ref{fig:SqKZ3}, respectively.
This can be converted to the Adams chart, see, e.g., \cite{beaudry2018guidecomputingstablehomotopy}, from which one can read off the bordism group. In dimension $6$ and below the Adams chart for $K(\mathbb{Z}_2,3)$ has a single $\mathbb{Z}_2$ in degree $3$.
For $K(\mathbb{Z},3)$, the first $\mathcal{A}(1)$ module, the `upside-down question mark', can be found in \cite{beaudry2018guidecomputingstablehomotopy}
From this we find\footnote{Using the exact sequence of $\mathcal{A}(1)$ modules we can further resolve the extension question in \cite{joyce2025bordismcategoriesorientationsmoduli}, confirming that $\Omega^{\text{Spin}}_7 \big( K(\mathbb{Z}_2, 3) \big) = \mathbb{Z}_4$.}
{\renewcommand{\arraystretch}{1.3}
\begin{equation}
    \begin{array}{c | c c c c c c c }
    k & 0 & 1 & 2 & 3 & 4 & 5 & 6 \\ \hline
    \widetilde{\Omega}^{\text{Spin}}_k \big( K(\mathbb{Z}_2, 3)\big)^{\vee}_2 & 0 & 0 & 0 & \mathbb{Z}_2 & 0 & 0 & 0 \\
    \widetilde{\Omega}^{\text{Spin}}_k \big( K(\mathbb{Z}, 3)\big)^{\vee}_2 & 0 & 0 & 0 & \mathbb{Z} & 0 & 0 & 0
    \end{array}
\end{equation}}
which combined with the results of the AHSS and $\Omega^{\text{Spin}}_{\bullet} (\text{pt})$ yields
{\renewcommand{\arraystretch}{1.3}
\begin{equation}
    \begin{array}{c | c c c c c c c }
    k & 0 & 1 & 2 & 3 & 4 & 5 & 6 \\ \hline
    \Omega^{\text{Spin}}_k \big( K(\mathbb{Z}_2, 3)\big) & \mathbb{Z} & \mathbb{Z}_2 & \mathbb{Z}_2 & \mathbb{Z}_2 & \mathbb{Z} & 0 & 0  \\
    \Omega^{\text{Spin}}_k \big( K(\mathbb{Z}, 3)\big) & \mathbb{Z} & \mathbb{Z}_2 & \mathbb{Z}_2 & \mathbb{Z} & \mathbb{Z} & 0 & 0
    \end{array}
\end{equation}}
which agrees with \cite{joyce2025bordismcategoriesorientationsmoduli} and we use as input for the AHSS
\begin{equation}
    E^2_{p,q} = H_p \big( B\mathrm{U}(1); \Omega_q^{\text{Spin}}\big( K(\mathbb{Z}_2, 3) \big) \big) \,,
\end{equation}
in Section \ref{sec:bordism}.

\end{appendix}

\bibliography{papers}{}

@ARTICLE{2013arXiv1312.5676B,
       author = {{Breen}, Lawrence and {Mikhailov}, Roman and {Touz{\'e}}, Antoine},
        title = "{Derived functors of the divided power functors}",
      journal = {arXiv e-prints},
         year = 2013,
        month = dec,
          eid = {arXiv:1312.5676},
        pages = {arXiv:1312.5676},
          doi = {10.48550/arXiv.1312.5676},
archivePrefix = {arXiv},
       eprint = {1312.5676},
 primaryClass = {math.AT},
       adsurl = {https://ui.adsabs.harvard.edu/abs/2013arXiv1312.5676B}
}

@ARTICLE{2012arXiv1201.2919R,
       author = {{Redden}, Corbett},
        title = "{Trivializations of differential cocycles}",
    journal      = {Journal of Homotopy and Related Structures},
  volume       = {10},
  pages        = {303--331},
  year         = {2015},
  publisher    = {Springer},
  doi          = {10.1007/s40062-013-0061-4},
archivePrefix = {arXiv},
       eprint = {1201.2919},
 primaryClass = {math.AT}
}

@book{MosherTangora1968,
  author    = {Robert E. Mosher and Martin C. Tangora},
  title     = {Cohomology Operations and Applications in Homotopy Theory},
  publisher = {Harper \& Row},
  address   = {New York},
  year      = {1968},
  series    = {Harper's Series in Modern Mathematics}
}

@book{BottTu1982,
  author    = {Raoul Bott and Loring W. Tu},
  title     = {Differential Forms in Algebraic Topology},
  series    = {Graduate Texts in Mathematics},
  volume    = {82},
  publisher = {Springer},
  address   = {New York, NY, USA},
  year      = {1982},
  isbn      = {978-0-387-90518-1},
  url       = {https://doi.org/10.1007/978-1-4757-3951-0}
}

@article{Brennan:2023mmt,
    author = "Brennan, T. Daniel and Hong, Sungwoo",
    title = "{Introduction to Generalized Global Symmetries in QFT and Particle Physics}",
    eprint = "2306.00912",
    archivePrefix = "arXiv",
    primaryClass = "hep-ph",
    month = "6",
    year = "2023"
}

@article{Schafer-Nameki:2023jdn,
    author = "Schafer-Nameki, Sakura",
    title = "{ICTP lectures on (non-)invertible generalized symmetries}",
    eprint = "2305.18296",
    archivePrefix = "arXiv",
    primaryClass = "hep-th",
    doi = "10.1016/j.physrep.2024.01.007",
    journal = "Phys. Rept.",
    volume = "1063",
    pages = "1--55",
    year = "2024"
}

@article{Heidenreich:2020pkc,
    author = "Heidenreich, Ben and McNamara, Jacob and Montero, Miguel and Reece, Matthew and Rudelius, Tom and Valenzuela, Irene",
    title = "{Chern-Weil global symmetries and how quantum gravity avoids them}",
    eprint = "2012.00009",
    archivePrefix = "arXiv",
    primaryClass = "hep-th",
    reportNumber = "ACFI-T20-16",
    doi = "10.1007/JHEP11(2021)053",
    journal = "JHEP",
    volume = "11",
    pages = "053",
    year = "2021"
}

@article{Bhardwaj:2023kri,
    author = "Bhardwaj, Lakshya and Bottini, Lea E. and Fraser-Taliente, Ludovic and Gladden, Liam and Gould, Dewi S. W. and Platschorre, Arthur and Tillim, Hannah",
    title = "{Lectures on generalized symmetries}",
    eprint = "2307.07547",
    archivePrefix = "arXiv",
    primaryClass = "hep-th",
    doi = "10.1016/j.physrep.2023.11.002",
    journal = "Phys. Rept.",
    volume = "1051",
    pages = "1--87",
    year = "2024"
}

@article{Witten:1999eg,
    author = "Witten, Edward",
    title = "{World-sheet corrections via $D$-instantons.}",
    eprint = "hep-th/9907041",
    archivePrefix = "arXiv",
    reportNumber = "IASSNS-HEP-99-56",
    doi = "10.1088/1126-6708/2000/02/030",
    journal = "JHEP",
    volume = "02",
    pages = "030",
    year = "2000"
}

@article{Chakrabhavi:2025bfi,
    author = "Chakrabhavi, Vivek and Debray, Arun and Dierigl, Markus and Heckman, Jonathan J.",
    title = "{Exploring Pintopia: Reflection Branes, Bordisms, and U-Dualities}",
    eprint = "2509.03573",
    archivePrefix = "arXiv",
    primaryClass = "hep-th",
    reportNumber = "CERN-TH-2025-180",
    month = "9",
    year = "2025"
}

@article{wang2008geometric,
  title={Geometric cycles, index theory and twisted K-homology},
  author={Wang, Bai-Ling},
  journal={Journal of Noncommutative Geometry},
  volume={2},
  number={4},
  pages={497--552},
  year={2008}
}

@article{94db963e-8a07-3135-b0ec-dcdc8d13c27b,
 ISSN = {0003486X, 19398980},
 URL = {http://www.jstor.org/stable/1970136},
 author = {C. T. C. Wall},
 journal = {Annals of Mathematics},
 number = {2},
 pages = {292--311},
 publisher = {[Annals of Mathematics, Trustees of Princeton University on Behalf of the Annals of Mathematics, Mathematics Department, Princeton University]},
 title = {Determination of the Cobordism Ring},
 urldate = {2026-02-11},
 volume = {72},
 year = {1960}
}

@article{Reece:2023iqn,
    author = "Reece, Matthew",
    title = "{Axion-gauge coupling quantization with a twist}",
    eprint = "2309.03939",
    archivePrefix = "arXiv",
    primaryClass = "hep-ph",
    doi = "10.1007/JHEP10(2023)116",
    journal = "JHEP",
    volume = "10",
    pages = "116",
    year = "2023"
}

@article{GarciaEtxebarria:2024fuk,
    author = "Garc{\'\i}a Etxebarria, I{\~n}aki and Hosseini, Saghar S.",
    title = "{Some aspects of symmetry descent}",
    eprint = "2404.16028",
    archivePrefix = "arXiv",
    primaryClass = "hep-th",
    doi = "10.1007/JHEP12(2024)223",
    journal = "JHEP",
    volume = "12",
    pages = "223",
    year = "2025"
}

@article{Freed:2006yc,
    author = "Freed, Daniel S. and Moore, Gregory W. and Segal, Graeme",
    title = "{Heisenberg Groups and Noncommutative Fluxes}",
    eprint = "hep-th/0605200",
    archivePrefix = "arXiv",
    reportNumber = "NSF-KITP-05-118",
    doi = "10.1016/j.aop.2006.07.014",
    journal = "Annals Phys.",
    volume = "322",
    pages = "236--285",
    year = "2007"
}

@article{Brennan:2024fgj,
    author = "Brennan, T. Daniel and Sun, Zhengdi",
    title = "{A SymTFT for continuous symmetries}",
    eprint = "2401.06128",
    archivePrefix = "arXiv",
    primaryClass = "hep-th",
    doi = "10.1007/JHEP12(2024)100",
    journal = "JHEP",
    volume = "12",
    pages = "100",
    year = "2024"
}

@book{HatcherKtheory,
  author = {Hatcher, A.},
  note = {http://www.math.cornell.edu/\~{}hatcher},
  title = {Vector Bundles and K-Theory},
  year = 2003
}

@article{Kobayashi:2024dqj,
    author = "Kobayashi, Ryohei and Li, Yuyang and Xue, Hanyu and Hsin, Po-Shen and Chen, Yu-An",
    title = "{Generalized Statistics on Lattices}",
    eprint = "2412.01886",
    archivePrefix = "arXiv",
    primaryClass = "quant-ph",
    doi = "10.1103/6k88-w52n",
    journal = "Phys. Rev. X",
    volume = "16",
    number = "1",
    pages = "011010",
    year = "2026"
}

@article{Feng:2025mdg,
    author = "Feng, Yitao and Xue, Hanyu and Li, Yuyang and Cheng, Meng and Kobayashi, Ryohei and Hsin, Po-Shen and Chen, Yu-An",
    title = "{Anyonic membranes and Pontryagin statistics}",
    eprint = "2509.14314",
    archivePrefix = "arXiv",
    primaryClass = "quant-ph",
    doi = "10.1103/4jww-6b6t",
    month = "9",
    year = "2025"
}

@article{Seiberg:2016rsg,
    author = "Seiberg, Nathan and Witten, Edward",
    title = "{Gapped Boundary Phases of Topological Insulators via Weak Coupling}",
    eprint = "1602.04251",
    archivePrefix = "arXiv",
    primaryClass = "cond-mat.str-el",
    doi = "10.1093/ptep/ptw083",
    journal = "PTEP",
    volume = "2016",
    number = "12",
    pages = "12C101",
    year = "2016"
}

@article{Reece:2023czb,
    author = "Reece, Matthew",
    title = "{TASI Lectures: (No) Global Symmetries to Axion Physics}",
    eprint = "2304.08512",
    archivePrefix = "arXiv",
    primaryClass = "hep-ph",
    doi = "10.22323/1.439.0008",
    journal = "PoS",
    volume = "TASI2022",
    pages = "008",
    year = "2024"
}

@article{Dierigl:2024cxm,
    author = "Dierigl, Markus and Novi{\v{c}}i{\'c}, Du{\v{s}}an",
    title = "{The axion is going dark}",
    eprint = "2409.02180",
    archivePrefix = "arXiv",
    primaryClass = "hep-th",
    reportNumber = "LMU-ASC 14/24",
    doi = "10.1007/JHEP12(2024)104",
    journal = "JHEP",
    volume = "12",
    pages = "104",
    year = "2024"
}

@article{Cordova:2023her,
    author = "Cordova, Clay and Hong, Sungwoo and Wang, Lian-Tao",
    title = "{Axion domain walls, small instantons, and non-invertible symmetry breaking}",
    eprint = "2309.05636",
    archivePrefix = "arXiv",
    primaryClass = "hep-ph",
    doi = "10.1007/JHEP05(2024)325",
    journal = "JHEP",
    volume = "05",
    pages = "325",
    year = "2024"
}

@article{Agrawal:2023sbp,
    author = "Agrawal, Prateek and Platschorre, Arthur",
    title = "{The monodromic axion-photon coupling}",
    eprint = "2309.03934",
    archivePrefix = "arXiv",
    primaryClass = "hep-th",
    doi = "10.1007/JHEP01(2024)169",
    journal = "JHEP",
    volume = "01",
    pages = "169",
    year = "2024"
}

@article{Choi:2023pdp,
    author = "Choi, Yichul and Forslund, Matthew and Lam, Ho Tat and Shao, Shu-Heng",
    title = "{Quantization of Axion-Gauge Couplings and Noninvertible Higher Symmetries}",
    eprint = "2309.03937",
    archivePrefix = "arXiv",
    primaryClass = "hep-ph",
    reportNumber = "YITP-SB-2023-27, MIT-CTP/5606",
    doi = "10.1103/PhysRevLett.132.121601",
    journal = "Phys. Rev. Lett.",
    volume = "132",
    number = "12",
    pages = "121601",
    year = "2024"
}

@article{Fiorenza:2010mh,
    author = "Fiorenza, Domenico and Schreiber, Urs and Stasheff, Jim",
    title = "{{\v{C}}ech cocycles for differential characteristic classes: an $\infty$-Lie theoretic construction}",
    eprint = "1011.4735",
    archivePrefix = "arXiv",
    primaryClass = "math.AT",
    doi = "10.4310/ATMP.2012.v16.n1.a5",
    journal = "Adv. Theor. Math. Phys.",
    volume = "16",
    number = "1",
    pages = "149--250",
    year = "2012"
}

@article{Christensen:2025ktc,
    author = "Christensen, Felix B. and Garc{\'\i}a Etxebarria, I{\~n}aki and Leung, Enoch",
    title = "{Anomaly-induced vanishing of brane partition functions}",
    eprint = "2510.19935",
    archivePrefix = "arXiv",
    primaryClass = "hep-th",
    month = "10",
    year = "2025"
}

@article{Sati:2009ic,
    author = "Sati, Hisham and Schreiber, Urs and Stasheff, Jim",
    title = "{Differential twisted String and Fivebrane structures}",
    eprint = "0910.4001",
    archivePrefix = "arXiv",
    primaryClass = "math.AT",
    doi = "10.1007/s00220-012-1510-3",
    journal = "Commun. Math. Phys.",
    volume = "315",
    pages = "169--213",
    year = "2012"
}

@article{Cheng:2023owv,
    author = "Cheng, Peng and Melnikov, Ilarion V. and Minasian, Ruben",
    title = "{Flat F-theory and friends}",
    eprint = "2306.00865",
    archivePrefix = "arXiv",
    primaryClass = "hep-th",
    doi = "10.1007/JHEP01(2024)027",
    journal = "JHEP",
    volume = "01",
    pages = "027",
    year = "2024"
}

@article{Debray:2023yrs,
    author = "Debray, Arun and Dierigl, Markus and Heckman, Jonathan J. and Montero, Miguel",
    title = "{The Chronicles of IIBordia: Dualities, Bordisms, and the Swampland}",
    eprint = "2302.00007",
    archivePrefix = "arXiv",
    primaryClass = "hep-th",
    reportNumber = "LMU-ASC 06/23, IFT-UAM/CSIC-23-7",
    year = "2024",
    journal = "ATMP",
    volume = "28",
    pages = "805",
    doi = "10.4310/ATMP.241028224804"
}

@article{Tachikawa:2018njr,
    author = "Tachikawa, Yuji and Yonekura, Kazuya",
    title = "{Why are fractional charges of orientifolds compatible with Dirac quantization?}",
    eprint = "1805.02772",
    archivePrefix = "arXiv",
    primaryClass = "hep-th",
    reportNumber = "IPMU-18-0067",
    doi = "10.21468/SciPostPhys.7.5.058",
    journal = "SciPost Phys.",
    volume = "7",
    number = "5",
    pages = "058",
    year = "2019"
}

@article{McNamara:2019rup,
    author = "McNamara, Jacob and Vafa, Cumrun",
    title = "{Cobordism Classes and the Swampland}",
    eprint = "1909.10355",
    archivePrefix = "arXiv",
    primaryClass = "hep-th",
    month = "9",
    year = "2019"
}

@article{Wan:2018bns,
    author = "Wan, Zheyan and Wang, Juven",
    title = "{Higher anomalies, higher symmetries, and cobordisms I: classification of higher-symmetry-protected topological states and their boundary fermionic/bosonic anomalies via a generalized cobordism theory}",
    eprint = "1812.11967",
    archivePrefix = "arXiv",
    primaryClass = "hep-th",
    doi = "10.4310/AMSA.2019.v4.n2.a2",
    journal = "Ann. Math. Sci. Appl.",
    volume = "4",
    number = "2",
    pages = "107--311",
    year = "2019"
}

@article{Debray:2023rlx,
    author = "Debray, Arun",
    title = "{Bordism for the 2-group symmetries of the heterotic and CHL strings}",
    eprint = "2304.14764",
    archivePrefix = "arXiv",
    primaryClass = "math.AT",
    doi = "10.1090/conm/802/16079",
    journal = "Contemp. Math.",
    volume = "802",
    pages = "227--98",
    year = "2024"
}

@article{Bhardwaj:2022scy,
    author = "Bhardwaj, Lakshya and Gould, Dewi S. W.",
    title = "{Disconnected 0-form and 2-group symmetries}",
    eprint = "2206.01287",
    archivePrefix = "arXiv",
    primaryClass = "hep-th",
    doi = "10.1007/JHEP07(2023)098",
    journal = "JHEP",
    volume = "07",
    pages = "098",
    year = "2023"
}

@article{Davighi:2023luh,
    author = "Davighi, Joe and Lohitsiri, Nakarin and Debray, Arun",
    title = "{Toric 2-group anomalies via cobordism}",
    eprint = "2302.12853",
    archivePrefix = "arXiv",
    primaryClass = "hep-th",
    doi = "10.1007/JHEP07(2023)019",
    journal = "JHEP",
    volume = "07",
    pages = "019",
    year = "2023"
}

@article{BoyleSmith:2025duo,
    author = "Boyle Smith, Philip and Davighi, Joe",
    title = "{Bosonisation Cohomology: Spin Structure Summation in Every Dimension}",
    eprint = "2511.13718",
    archivePrefix = "arXiv",
    primaryClass = "hep-th",
    month = "11",
    year = "2025"
}

@article{Antinucci:2024zjp,
    author = "Antinucci, Andrea and Benini, Francesco",
    title = "{Anomalies and gauging of U(1) symmetries}",
    eprint = "2401.10165",
    archivePrefix = "arXiv",
    primaryClass = "hep-th",
    reportNumber = "SISSA 01/2024/FISI",
    doi = "10.1103/PhysRevB.111.024110",
    journal = "Phys. Rev. B",
    volume = "111",
    number = "2",
    pages = "024110",
    year = "2025"
}

@incollection{Stong1986SpinBordism,
  author    = {Robert E. Stong},
  title     = {{Calculation of $\Omega^{\mathrm{spin}}_{11}(K(\mathbb{Z},4))$}},
  booktitle = {Unified String Theories},
  editor    = {Michael Green and David Gross},
  pages     = {430--437},
  publisher = {World Scientific},
  year      = {1986},
  address   = {Singapore}
}

@article{Abe:2022nfq,
    author = "Abe, Motokazu and Morikawa, Okuto and Suzuki, Hiroshi",
    title = "{Fractional topological charge in lattice Abelian gauge theory}",
    eprint = "2210.12967",
    archivePrefix = "arXiv",
    primaryClass = "hep-th",
    reportNumber = "KYUSHU-HET-248, OU-HET-1155",
    doi = "10.1093/ptep/ptad009",
    journal = "PTEP",
    volume = "2023",
    number = "2",
    pages = "023B03",
    year = "2023"
}

@article{Najjar:2025htp,
    author = "Najjar, Marwan",
    title = "{Modified instanton sum and 4-group structure in 4d $\mathcal{N}=1$$SU(M)$ SYM from holography}",
    eprint = "2503.17108",
    archivePrefix = "arXiv",
    primaryClass = "hep-th",
    month = "3",
    year = "2025"
}

@article{Kan:2023yhz,
    author = "Kan, Naoto and Morikawa, Okuto and Nagoya, Yuta and Wada, Hiroki",
    title = "{Higher-group structure in lattice Abelian gauge theory under instanton-sum modification}",
    eprint = "2302.13466",
    archivePrefix = "arXiv",
    primaryClass = "hep-th",
    reportNumber = "OU-HET-1173",
    doi = "10.1140/epjc/s10052-023-11616-6",
    journal = "Eur. Phys. J. C",
    volume = "83",
    number = "6",
    pages = "481",
    year = "2023",
    note = "[Erratum: Eur.Phys.J.C 84, 22 (2024)]"
}

@article{Brennan:2023kpw,
    author = "Brennan, T. Daniel and Hong, Sungwoo and Wang, Lian-Tao",
    title = "{Coupling a Cosmic String to a TQFT}",
    eprint = "2302.00777",
    archivePrefix = "arXiv",
    primaryClass = "hep-ph",
    doi = "10.1007/JHEP03(2024)145",
    journal = "JHEP",
    volume = "03",
    pages = "145",
    year = "2024"
}

@article{Tanizaki:2019rbk,
    author = {Tanizaki, Yuya and {\"U}nsal, Mithat},
    title = "{Modified instanton sum in QCD and higher-groups}",
    eprint = "1912.01033",
    archivePrefix = "arXiv",
    primaryClass = "hep-th",
    doi = "10.1007/JHEP03(2020)123",
    journal = "JHEP",
    volume = "03",
    pages = "123",
    year = "2020"
}

@article{Freed:2017rlk,
    author = "Freed, Daniel S. and Komargodski, Zohar and Seiberg, Nathan",
    title = "{The Sum Over Topological Sectors and $\theta$ in the 2+1-Dimensional $\mathbb{C}\mathbb{P}^1$ $\sigma$-Model}",
    eprint = "1707.05448",
    archivePrefix = "arXiv",
    primaryClass = "cond-mat.str-el",
    doi = "10.1007/s00220-018-3093-0",
    journal = "Commun. Math. Phys.",
    volume = "362",
    number = "1",
    pages = "167--183",
    year = "2018"
}

@article{Kapustin:2014gua,
    author = "Kapustin, Anton and Seiberg, Nathan",
    title = "{Coupling a QFT to a TQFT and Duality}",
    eprint = "1401.0740",
    archivePrefix = "arXiv",
    primaryClass = "hep-th",
    doi = "10.1007/JHEP04(2014)001",
    journal = "JHEP",
    volume = "04",
    pages = "001",
    year = "2014"
}

@article{Seiberg:2010qd,
    author = "Seiberg, Nathan",
    title = "{Modifying the Sum Over Topological Sectors and Constraints on Supergravity}",
    eprint = "1005.0002",
    archivePrefix = "arXiv",
    primaryClass = "hep-th",
    doi = "10.1007/JHEP07(2010)070",
    journal = "JHEP",
    volume = "07",
    pages = "070",
    year = "2010"
}

@book{milnor_stasheff_1974,
  title        = {Characteristic Classes},
  author       = {Milnor, John W. and Stasheff, James D.},
  series       = {Annals of Mathematics Studies},
  volume       = {76},
  publisher    = {Princeton University Press and University of Tokyo Press},
  address      = {Princeton, NJ and Tokyo},
  year         = {1974},
  isbn         = {978-0691081229},
  note         = {Based on lecture notes of John Milnor},
}

@article{Garcia-Valdecasas:2024cqn,
    author = "Garc{\'\i}a-Valdecasas, Eduardo and Reece, Matthew and Suzuki, Motoo",
    title = "{Monopole breaking of Chern-Weil symmetries}",
    eprint = "2408.00067",
    archivePrefix = "arXiv",
    primaryClass = "hep-th",
    doi = "10.21468/SciPostPhys.18.5.162",
    journal = "SciPost Phys.",
    volume = "18",
    number = "5",
    pages = "162",
    year = "2025"
}

@article{Yonekura:2020ino,
    author = "Yonekura, Kazuya",
    title = "{Topological violation of global symmetries in quantum gravity}",
    eprint = "2011.11868",
    archivePrefix = "arXiv",
    primaryClass = "hep-th",
    doi = "10.1007/JHEP09(2021)036",
    journal = "JHEP",
    volume = "09",
    pages = "036",
    year = "2021"
}

@article{Kapustin:2013uxa,
    author = "Kapustin, Anton and Thorngren, Ryan",
    title = "{Higher Symmetry and Gapped Phases of Gauge Theories}",
    eprint = "1309.4721",
    archivePrefix = "arXiv",
    primaryClass = "hep-th",
    doi = "10.1007/978-3-319-59939-7_5",
    journal = "Prog. Math.",
    volume = "324",
    pages = "177--202",
    year = "2017"
}

@misc{beaudry2018guidecomputingstablehomotopy,
      title={A Guide for Computing Stable Homotopy Groups}, 
      author={Agnes Beaudry and Jonathan A. Campbell},
      year={2018},
      eprint={1801.07530},
      archivePrefix={arXiv},
      primaryClass={math.AT},
      url={https://arxiv.org/abs/1801.07530}, 
}

@inproceedings{Witten:2019bou,
    author = "Witten, Edward and Yonekura, Kazuya",
    title = "{Anomaly Inflow and the $\eta$-Invariant}",
    booktitle = "{The Shoucheng Zhang Memorial Workshop}",
    eprint = "1909.08775",
    archivePrefix = "arXiv",
    primaryClass = "hep-th",
    month = "9",
    year = "2019"
}

@article{Witten:2015aba,
    author = "Witten, Edward",
    title = "{Fermion Path Integrals And Topological Phases}",
    eprint = "1508.04715",
    archivePrefix = "arXiv",
    primaryClass = "cond-mat.mes-hall",
    doi = "10.1103/RevModPhys.88.035001",
    journal = "Rev. Mod. Phys.",
    volume = "88",
    number = "3",
    pages = "035001",
    year = "2016"
}

@article{Yonekura:2016wuc,
    author = "Yonekura, Kazuya",
    title = "{Dai-Freed theorem and topological phases of matter}",
    eprint = "1607.01873",
    archivePrefix = "arXiv",
    primaryClass = "hep-th",
    reportNumber = "IPMU-16-0094",
    doi = "10.1007/JHEP09(2016)022",
    journal = "JHEP",
    volume = "09",
    pages = "022",
    year = "2016"
}

@article{Freed:2004yc,
    author = "Freed, Daniel S. and Moore, Gregory W.",
    title = "{Setting the quantum integrand of M-theory}",
    eprint = "hep-th/0409135",
    archivePrefix = "arXiv",
    doi = "10.1007/s00220-005-1482-7",
    journal = "Commun. Math. Phys.",
    volume = "263",
    pages = "89--132",
    year = "2006"
}

@article{Freed:2014iua,
    author = "Freed, Daniel S.",
    editor = "Donagi, Ron and Douglas, Michael R. and Kamenova, Ljudmila and Rocek, Martin",
    title = "{Anomalies and Invertible Field Theories}",
    eprint = "1404.7224",
    archivePrefix = "arXiv",
    primaryClass = "hep-th",
    doi = "10.1090/pspum/088/01462",
    journal = "Proc. Symp. Pure Math.",
    volume = "88",
    pages = "25--46",
    year = "2014"
}

@article{Dai:1994kq,
    author = "Dai, Xian-Zhe and Freed, Daniel S.",
    title = "{eta invariants and determinant lines}",
    eprint = "hep-th/9405012",
    archivePrefix = "arXiv",
    doi = "10.1063/1.530747",
    journal = "J. Math. Phys.",
    volume = "35",
    pages = "5155--5194",
    year = "1994",
    note = "[Erratum: J.Math.Phys. 42, 2343--2344 (2001)]"
}

@article{Garcia-Etxebarria:2017crf,
    author = "Garc{\'\i}a-Etxebarria, I{\~n}aki and Hayashi, Hirotaka and Ohmori, Kantaro and Tachikawa, Yuji and Yonekura, Kazuya",
    title = "{8d gauge anomalies and the topological Green-Schwarz mechanism}",
    eprint = "1710.04218",
    archivePrefix = "arXiv",
    primaryClass = "hep-th",
    reportNumber = "IPMU-17-0139, MPP-2017-220",
    doi = "10.1007/JHEP11(2017)177",
    journal = "JHEP",
    volume = "11",
    pages = "177",
    year = "2017"
}

@article{Torres:2024sbl,
    author = "Torres, Ethan",
    title = "{Giving a $KO$ to 8D Gauge Anomalies}",
    eprint = "2405.08809",
    archivePrefix = "arXiv",
    primaryClass = "hep-th",
    reportNumber = "CERN-TH-2024-054",
    month = "5",
    year = "2024"
}

@article{Lee:2022spd,
    author = "Lee, Yasunori and Yonekura, Kazuya",
    title = "{Global anomalies in 8d supergravity}",
    eprint = "2203.12631",
    archivePrefix = "arXiv",
    primaryClass = "hep-th",
    reportNumber = "IPMU-22-0011, TU-1148",
    doi = "10.1007/JHEP07(2022)125",
    journal = "JHEP",
    volume = "07",
    pages = "125",
    year = "2022"
}

@article{Benini:2018reh,
    author = "Benini, Francesco and C{\'o}rdova, Clay and Hsin, Po-Shen",
    title = "{On 2-Group Global Symmetries and their Anomalies}",
    eprint = "1803.09336",
    archivePrefix = "arXiv",
    primaryClass = "hep-th",
    reportNumber = "SISSA 10/2018/FISI, SISSA-10-2018-FISI",
    doi = "10.1007/JHEP03(2019)118",
    journal = "JHEP",
    volume = "03",
    pages = "118",
    year = "2019"
}

@article{Cordova:2018cvg,
    author = "C{\'o}rdova, Clay and Dumitrescu, Thomas T. and Intriligator, Kenneth",
    title = "{Exploring 2-Group Global Symmetries}",
    eprint = "1802.04790",
    archivePrefix = "arXiv",
    primaryClass = "hep-th",
    doi = "10.1007/JHEP02(2019)184",
    journal = "JHEP",
    volume = "02",
    pages = "184",
    year = "2019"
}

@article{Tachikawa:2021mvw,
    author = "Tachikawa, Yuji",
    title = "{Topological modular forms and the absence of a heterotic global anomaly}",
    eprint = "2103.12211",
    archivePrefix = "arXiv",
    primaryClass = "hep-th",
    doi = "10.1093/ptep/ptab060",
    journal = "PTEP",
    volume = "2022",
    number = "4",
    pages = "04A107",
    year = "2022"
}

@article{Debray:2021vob,
    author = "Debray, Arun and Dierigl, Markus and Heckman, Jonathan J. and Montero, Miguel",
    title = "{The anomaly that was not meant IIB}",
    eprint = "2107.14227",
    archivePrefix = "arXiv",
    primaryClass = "hep-th",
    reportNumber = "LMU-ASC 24/21",
    doi = "10.1002/prop.202100168",
    journal = "Fortsch. Phys.",
    volume = "70",
    number = "1",
    pages = "2100168",
    year = "2022"
}

@article{Yonekura:2022reu,
    author = "Yonekura, Kazuya",
    title = "{Heterotic global anomalies and torsion Witten index}",
    eprint = "2207.13858",
    archivePrefix = "arXiv",
    primaryClass = "hep-th",
    reportNumber = "TU-1163",
    doi = "10.1007/JHEP10(2022)114",
    journal = "JHEP",
    volume = "10",
    pages = "114",
    year = "2022"
}

@article{Basile:2023knk,
    author = "Basile, Ivano and Debray, Arun and Delgado, Matilda and Montero, Miguel",
    title = "{Global anomalies {\&} bordism of non-supersymmetric strings}",
    eprint = "2310.06895",
    archivePrefix = "arXiv",
    primaryClass = "hep-th",
    reportNumber = "IFT-23-129",
    doi = "10.1007/JHEP02(2024)092",
    journal = "JHEP",
    volume = "02",
    pages = "092",
    year = "2024"
}

@article{Green:1984sg,
    author = "Green, Michael B. and Schwarz, John H.",
    title = "{Anomaly Cancellation in Supersymmetric D=10 Gauge Theory and Superstring Theory}",
    reportNumber = "CALT-68-1182",
    doi = "10.1016/0370-2693(84)91565-X",
    journal = "Phys. Lett. B",
    volume = "149",
    pages = "117--122",
    year = "1984"
}

@article{Saito:2025idl,
    author = "Saito, Shota and Tachikawa, Yuji",
    title = "{Cancelling mod-2 anomalies by Green-Schwarz mechanism with $B_{\mu\nu}$}",
    eprint = "2411.09223",
    archivePrefix = "arXiv",
    primaryClass = "hep-th",
    reportNumber = "IPMU24-0041",
    doi = "10.21468/SciPostPhys.19.1.017",
    journal = "SciPost Phys.",
    volume = "19",
    number = "1",
    pages = "017",
    year = "2025"
}

@article{Gaiotto:2014kfa,
    author = "Gaiotto, Davide and Kapustin, Anton and Seiberg, Nathan and Willett, Brian",
    title = "{Generalized Global Symmetries}",
    eprint = "1412.5148",
    archivePrefix = "arXiv",
    primaryClass = "hep-th",
    doi = "10.1007/JHEP02(2015)172",
    journal = "JHEP",
    volume = "02",
    pages = "172",
    year = "2015"
}

@article{Chen:2023czk,
    author = "Chen, Shi and Tanizaki, Yuya",
    title = "{Solitonic symmetry as non-invertible symmetry: cohomology theories with TQFT coefficients}",
    eprint = "2307.00939",
    archivePrefix = "arXiv",
    primaryClass = "hep-th",
    reportNumber = "YITP-23-59",
    month = "7",
    year = "2023"
}

@book{hatcherspectral,
  author    = {Hatcher, Allen},
  title     = {Spectral Sequences in Algebraic Topology},
  year      = {2004},
  url       = {https://pi.math.cornell.edu/~hatcher/SSAT/SSATpage.html}
}

@book{hatcher2002,
  author    = {Hatcher, Allen},
  title     = {Algebraic Topology},
  publisher = {Cambridge University Press},
  year      = {2002},
  url       = {https://pi.math.cornell.edu/~hatcher/AT/AT.pdf}
}

@book{fomenko-fuchs-2016,
  author    = {Fomenko, Anatoly T. and Fuchs, Dmitry B.},
  title     = {Homotopical Topology},
  publisher = {Springer},
  edition   = {2},
  year      = {2016},
  series    = {Graduate Texts in Contemporary Physics},
  isbn      = {978-3-319-26956-1}
}

@book{mccleary2001,
  author    = {McCleary, John},
  title     = {A User's Guide to Spectral Sequences},
  edition   = {2},
  series    = {Graduate Texts in Mathematics},
  volume    = {58},
  publisher = {Springer},
  year      = {2001},
  isbn      = {978-0387985862}
}

@inproceedings{Cheeger1985DifferentialCA,
  title={Differential characters and geometric invariants},
  author={Jeff Cheeger and James Simons},
  year={1985},
}

@article{Hopkins:2002rd,
    author = "Hopkins, M. J. and Singer, I. M.",
    title = "{Quadratic functions in geometry, topology, and M theory}",
    eprint = "math/0211216",
    archivePrefix = "arXiv",
    journal = "J. Diff. Geom.",
    volume = "70",
    number = "3",
    pages = "329--452",
    year = "2005"
}

@article{Dierigl:2025rfn,
    author = "Dierigl, Markus and Tartaglia, Michelangelo",
    title = "{(Quadratically) Refined discrete anomaly cancellation}",
    eprint = "2504.02934",
    archivePrefix = "arXiv",
    primaryClass = "hep-th",
    reportNumber = "CERN-TH-2025-068, IFT-25-35",
    doi = "10.1007/JHEP08(2025)145",
    journal = "JHEP",
    volume = "08",
    pages = "145",
    year = "2025"
}

@article{Ibanez:1991hv,
    author = "Ibanez, Luis E. and Ross, Graham G.",
    title = "{Discrete gauge symmetry anomalies}",
    doi = "10.1016/0370-2693(91)91614-2",
    journal = "Phys. Lett. B",
    volume = "260",
    pages = "291--295",
    year = "1991"
}

@article{Dierigl:2022zll,
    author = "Dierigl, Markus and Oehlmann, Paul-Konstantin and Schimannek, Thorsten",
    title = "{The discrete Green-Schwarz mechanism in 6D F-theory and elliptic genera of non-critical strings}",
    eprint = "2212.04503",
    archivePrefix = "arXiv",
    primaryClass = "hep-th",
    reportNumber = "LMU-ASC 33/22",
    doi = "10.1007/JHEP03(2023)090",
    journal = "JHEP",
    volume = "03",
    pages = "090",
    year = "2023"
}

@article{Monnier:2018nfs,
    author = "Monnier, Samuel and Moore, Gregory W.",
    title = "{Remarks on the Green{\textendash}Schwarz Terms of Six-Dimensional Supergravity Theories}",
    eprint = "1808.01334",
    archivePrefix = "arXiv",
    primaryClass = "hep-th",
    doi = "10.1007/s00220-019-03341-7",
    journal = "Commun. Math. Phys.",
    volume = "372",
    number = "3",
    pages = "963--1025",
    year = "2019"
}

@article{Hsieh:2018ifc,
    author = "Hsieh, Chang-Tse",
    title = "{Discrete gauge anomalies revisited}",
    eprint = "1808.02881",
    archivePrefix = "arXiv",
    primaryClass = "hep-th",
    reportNumber = "IPMU-18-0130",
    month = "8",
    year = "2018"
}

@article{Cheng:2025ikd,
    author = "Cheng, Peng and Milam, Michael N. and Minasian, Ruben",
    title = "{Integral cubic form of 5D minimal supergravities and non-perturbative anomalies in 6D (1,0) theories}",
    eprint = "2509.18042",
    archivePrefix = "arXiv",
    primaryClass = "hep-th",
    month = "9",
    year = "2025"
}

@article{Davighi:2020uab,
    author = "Davighi, Joe and Lohitsiri, Nakarin",
    title = "{The algebra of anomaly interplay}",
    eprint = "2011.10102",
    archivePrefix = "arXiv",
    primaryClass = "hep-th",
    doi = "10.21468/SciPostPhys.10.3.074",
    journal = "SciPost Phys.",
    volume = "10",
    number = "3",
    pages = "074",
    year = "2021"
}

@article{Davighi:2020bvi,
    author = "Davighi, Joe and Lohitsiri, Nakarin",
    title = "{Anomaly interplay in $U(2)$ gauge theories}",
    eprint = "2001.07731",
    archivePrefix = "arXiv",
    primaryClass = "hep-th",
    doi = "10.1007/JHEP05(2020)098",
    journal = "JHEP",
    volume = "05",
    pages = "098",
    year = "2020"
}

@article{Tachikawa:2021mby,
    author = "Tachikawa, Yuji and Yamashita, Mayuko",
    title = "{Topological Modular Forms and the Absence of All Heterotic Global Anomalies}",
    eprint = "2108.13542",
    archivePrefix = "arXiv",
    primaryClass = "hep-th",
    doi = "10.1007/s00220-023-04761-2",
    journal = "Commun. Math. Phys.",
    volume = "402",
    number = "2",
    pages = "1585--1620",
    year = "2023",
    note = "[Erratum: Commun.Math.Phys. 402, 2131 (2023)]"
}

@article{Garcia-Etxebarria:2018ajm,
    author = "Garc{\'\i}a-Etxebarria, I{\~n}aki and Montero, Miguel",
    title = "{Dai-Freed anomalies in particle physics}",
    eprint = "1808.00009",
    archivePrefix = "arXiv",
    primaryClass = "hep-th",
    reportNumber = "MPP-2018-188",
    doi = "10.1007/JHEP08(2019)003",
    journal = "JHEP",
    volume = "08",
    pages = "003",
    year = "2019"
}

@article{Kapustin:2014zva,
    author = "Kapustin, Anton and Thorngren, Ryan",
    title = "{Anomalies of discrete symmetries in various dimensions and group cohomology}",
    eprint = "1404.3230",
    archivePrefix = "arXiv",
    primaryClass = "hep-th",
    month = "4",
    year = "2014"
}

@article{Hsieh:2020jpj,
    author = "Hsieh, Chang-Tse and Tachikawa, Yuji and Yonekura, Kazuya",
    title = "{Anomaly Inflow and p-Form Gauge Theories}",
    eprint = "2003.11550",
    archivePrefix = "arXiv",
    primaryClass = "hep-th",
    reportNumber = "IPMU-20-0028, TU-1098",
    doi = "10.1007/s00220-022-04333-w",
    journal = "Commun. Math. Phys.",
    volume = "391",
    number = "2",
    pages = "495--608",
    year = "2022"
}

@inproceedings{Moore:2025tmt,
    author = "Moore, Gregory W. and Saxena, Vivek and Freed (contributing an appendix),  Daniel S.",
    title = "{TASI Lectures On Topological Field Theories And Differential Cohomology}",
    booktitle = "{Theoretical Advanced Study Institute in Elementary Particle Physics 2023}: {Aspects of Symmetry}",
    eprint = "2510.07408",
    archivePrefix = "arXiv",
    primaryClass = "hep-th",
    reportNumber = "YITP-SB-2025-04",
    month = "10",
    year = "2025"
}

@article{Pantev:2005rh,
    author = "Pantev, Tony and Sharpe, Eric",
    title = "{Notes on gauging noneffective group actions}",
    eprint = "hep-th/0502027",
    archivePrefix = "arXiv",
    month = "2",
    year = "2005"
}

@article{Pantev:2005wj,
    author = "Pantev, Tony and Sharpe, Eric",
    title = "{String compactifications on Calabi-Yau stacks}",
    eprint = "hep-th/0502044",
    archivePrefix = "arXiv",
    doi = "10.1016/j.nuclphysb.2005.10.035",
    journal = "Nucl. Phys. B",
    volume = "733",
    pages = "233--296",
    year = "2006"
}

@article{Pantev:2005zs,
    author = "Pantev, Tony and Sharpe, Eric",
    title = "{GLSM's for Gerbes (and other toric stacks)}",
    eprint = "hep-th/0502053",
    archivePrefix = "arXiv",
    doi = "10.4310/ATMP.2006.v10.n1.a4",
    journal = "Adv. Theor. Math. Phys.",
    volume = "10",
    number = "1",
    pages = "77--121",
    year = "2006"
}

@phdthesis{Clementthesis,
  author       = {Alain Clément},
  title        = {Integral Cohomology of Finite Postnikov Towers},
  school       = {Université de Lausanne},
  year         = {2002},
  address      = {Lausane},
}

@article{Araki1959Steenrod,
  author  = {Araki, Sh{\^o}r{\^o}},
  title   = {Steenrod reduced powers in the spectral sequence associated with a fiber space},
  journal = {Osaka Mathematical Journal},
  volume  = {11},
  year    = {1959},
  pages   = {1--22}
}

@article{Freed:2016rqq,
    author = "Freed, Daniel S. and Hopkins, Michael J.",
    title = "{Reflection positivity and invertible topological phases}",
    eprint = "1604.06527",
    archivePrefix = "arXiv",
    primaryClass = "hep-th",
    doi = "10.2140/gt.2021.25.1165",
    journal = "Geom. Topol.",
    volume = "25",
    pages = "1165--1330",
    year = "2021"
}

@article{Yonekura:2018ufj,
    author = "Yonekura, Kazuya",
    title = "{On the cobordism classification of symmetry protected topological phases}",
    eprint = "1803.10796",
    archivePrefix = "arXiv",
    primaryClass = "hep-th",
    reportNumber = "IPMU-18-0040",
    doi = "10.1007/s00220-019-03439-y",
    journal = "Commun. Math. Phys.",
    volume = "368",
    number = "3",
    pages = "1121--1173",
    year = "2019"
}

@article{Fiorenza:2016oki,
    author = "Fiorenza, Domenico and Sati, Hisham and Schreiber, Urs",
    title = "{T-Duality from super Lie n-algebra cocycles for super p-branes}",
    eprint = "1611.06536",
    archivePrefix = "arXiv",
    primaryClass = "math-ph",
    doi = "10.4310/ATMP.2018.v22.n5.a3",
    journal = "Adv. Theor. Math. Phys.",
    volume = "22",
    pages = "1209--1270",
    year = "2018"
}

@article{Fiorenza:2012ec,
    author = "Fiorenza, Domenico and Sati, Hisham and Schreiber, Urs",
    title = "{Extended higher cup-product Chern-Simons theories}",
    eprint = "1207.5449",
    archivePrefix = "arXiv",
    primaryClass = "hep-th",
    doi = "10.1016/j.geomphys.2013.07.011",
    journal = "J. Geom. Phys.",
    volume = "74",
    pages = "130--163",
    year = "2013"
}

@article{Fiorenza:2019ckz,
    author = "Fiorenza, Domenico and Sati, Hisham and Schreiber, Urs",
    title = "{The Rational Higher Structure of M-theory}",
    eprint = "1903.02834",
    archivePrefix = "arXiv",
    primaryClass = "hep-th",
    doi = "10.1002/prop.201910017",
    journal = "Fortsch. Phys.",
    volume = "67",
    number = "8-9",
    pages = "1910017",
    year = "2019"
}

@article{Sati:2008kz,
    author = "Sati, Hisham and Schreiber, Urs and Stasheff, Jim",
    title = "{Fivebrane Structures}",
    eprint = "0805.0564",
    archivePrefix = "arXiv",
    primaryClass = "math.AT",
    doi = "10.1142/S0129055X09003840",
    journal = "Rev. Math. Phys.",
    volume = "21",
    pages = "1197--1240",
    year = "2009"
}

@article{Hellerman:2006zs,
    author = "Hellerman, Simeon and Henriques, Andre and Pantev, Tony and Sharpe, Eric and Ando, Matt",
    title = "{Cluster decomposition, T-duality, and gerby CFT's}",
    eprint = "hep-th/0606034",
    archivePrefix = "arXiv",
    doi = "10.4310/ATMP.2007.v11.n5.a2",
    journal = "Adv. Theor. Math. Phys.",
    volume = "11",
    number = "5",
    pages = "751--818",
    year = "2007"
}

@book{ConnerFloyd1964,
  author    = {Conner, P. E. and Floyd, E. E.},
  title     = {Differentiable Periodic Maps},
  year      = {1964},
  publisher = {Princeton University Press},
  address   = {Princeton, NJ}
}

@misc{joyce2025bordismcategoriesorientationsmoduli,
      title={Bordism categories and orientations of moduli spaces}, 
      author={Dominic Joyce and Markus Upmeier},
      year={2025},
      eprint="2503.20456",
      archivePrefix="arXiv",
      primaryClass="math.AT"
}

@article{Teichner:1993aij,
    author = "Teichner, Peter",
    title = "{On the signature of four-manifolds with universal covering spin}",
    doi = "10.1007/BF01444915",
    journal = "Math. Ann.",
    volume = "295",
    number = "1",
    pages = "745--759",
    year = "1993"
}

@article{Zhubr1999,
  author  = {Zhubr, A. V.},
  title   = {Spin bordism of oriented manifolds and the {H}auptvermutung for 6-manifolds},
  journal = {Mathematical Notes},
  volume  = {65},
  number  = {5},
  pages   = {573--582},
  year    = {1999},
  doi     = {10.1007/BF02566068}
}
\bibliographystyle{JHEP} 

\end{document}